\documentclass[fleqn,usenatbib]{mnras}

\usepackage[T1]{fontenc}

\usepackage{enumitem}

\usepackage{soul, xcolor}
\setstcolor{red}

\usepackage{graphicx}
\usepackage{amsmath}	

\usepackage{subcaption}
\usepackage{threeparttable}
\usepackage{booktabs}

\usepackage{changepage}

\usepackage{dirtytalk}

\usepackage{adjustbox}
\usepackage{rotating}

\title[Retrieval study of 51 Eri b]{Retrieval study of cool, directly imaged exoplanet 51 Eri b}
\author[N. Whiteford et al.]
{Niall Whiteford$^{1,2,3}$\thanks{E-mail: niallwhiteford@gmail.com},
Alistair Glasse$^{2,3,4}$,
Katy L. Chubb$^{5,}$,
Daniel Kitzmann$^{6}$,
\newauthor
Shrishmoy Ray$^{7}$,
Mark W. Phillips$^{14}$,
Beth A. Biller$^{2,3}$,
Paul I. Palmer$^{9,3,}$,
\newauthor
Ken Rice$^{1,2}$,
Ingo P. Waldmann$^{8}$,
Quentin Changeat$^{12}$,
Nour Skaf $^{10,11}$,
\newauthor
Jason Wang$^{15}$,
Billy Edwards$^{16}$,
Ahmed Al-Refaie$^{8}$\\
\\
$^{1}$Department of Astrophysics, American Museum of Natural History, Central Park West at 79th Street, NY 10024, USA\\
$^{2}$Institute for Astronomy, University of Edinburgh, Royal Observatory, Blackford Hill, Edinburgh, EH9 3HJ, UK\\
$^{3}$Centre for Exoplanet Science, University of Edinburgh, UK\\
$^{4}$UK Astronomy Technology Centre, Royal Observatory, Edinburgh, Blackford Hill, Edinburgh EH9 3HJ, UK\\
$^{5}$Centre for Exoplanet Science, University of St Andrews, North Haugh, St Andrews, UK \\
$^{6}$Center for Space and Habitability, University of Bern, Gesellschaftsstrasse 6, Bern, 3012, Switzerland\\
$^{7}$Astrophysics Group, University of Exeter, EX4 4QL, Exeter, UK\\
$^{8}$Department of Physics and Astronomy, University College London, Gower Street, WC1E 6BT, UK\\
$^{9}$School of GeoSciences, University of Edinburgh, Edinburgh, UK\\
$^{10}$LESIA, Observatoire de Paris, Université PSL, CNRS, Sorbonne  Universit\'e, Universit\'e de Paris,  5 place Jules \\ Janssen, 92195 Meudon, France\\
$^{11}$ Subaru Telescope, National Astronomical Observatory of Japan, 650 North A'Ohoku Place, Hilo, HI 96720, USA\\
$^{12}$European Space Agency, Space Telescope Science Institute, Baltimore, MD, USA \\
$^{13}$AIM, CEA, CNRS, Universite Paris-Saclay, Universite de Paris, F-91191 Gif-sur-Yvette, France\\
$^{14}$Institute for Astronomy, University of Hawaii, 2680 Woodlawn Drive, Honolulu, HI 96822, USA \\
$^{15}$ Center for Interdisciplinary Exploration and Research in Astrophysics (CIERA) and Department of Physics and Astronomy, \\ Northwestern University, Evanston, IL 60208, USA \\
$^{16}$SRON Netherlands Institute for Space Research, Sorbonnelaan 2, 3584 CA, Utrecht, Netherlands\\
\vspace{-2.cm}
}

\pubyear{2023}

\begin{document}
\label{firstpage}
\pagerange{\pageref{firstpage}--\pageref{lastpage}}
\maketitle

\begin{abstract}
Retrieval methods are a powerful analysis technique for modelling exoplanetary atmospheres by estimating the bulk physical and chemical properties that combine in a forward model to best-fit an observed spectrum, and they are increasingly being applied to observations of directly-imaged exoplanets. We have adapted TauREx3, the Bayesian retrieval suite, for the analysis of near-infrared spectrophotometry from directly-imaged gas giant exoplanets and brown dwarfs. We demonstrate TauREx3's applicability to sub-stellar atmospheres by presenting results for brown dwarf benchmark GJ 570D which are consistent with previous retrieval studies, whilst also exhibiting systematic biases associated with the presence of alkali lines.  We also  present results for the cool exoplanet 51 Eri b, the first application of a free chemistry retrieval analysis to this object, using spectroscopic observations from GPI and SPHERE. While our retrieval analysis is able to explain spectroscopic and photometric observations without employing cloud extinction, we conclude this may be a result of employing a flexible temperature-pressure profile which is able to mimic the presence of clouds. We present Bayesian evidence for an ammonia detection with a 2.7$\sigma$ confidence, the first indication of ammonia in an exoplanetary atmosphere. This is consistent with this molecule being present in brown dwarfs of a similar spectral type. We demonstrate the chemical  similarities between 51 Eri b and GJ 570D in relation to their retrieved molecular abundances. Finally, we show that overall retrieval conclusions for 51 Eri b can vary when employing different spectral data and modelling components, such as temperature-pressure and cloud structures.

\end{abstract}








\section{Introduction}

While over 5000 exoplanets have been confirmed to date, \citep{NASA_Exoplanet_archive, web:NASAExoplanetArchive}, only a very small fraction have been directly imaged due to the significant technical challenge of detecting a signal from an exoplanet many times fainter than its host star. However, extreme coronagraphic spectrometers, including VLT's Spectro-Polarimetric High-contrast Exoplanet REsearch instrument (SPHERE) \citep{Beuzit2008_SPHERE}, the Gemini Planet Imager (GPI) \citep{Macintosh2014_GPI} and VLT's GRAVITY\citep{2017_GRAVITY}, have made it possible to start the characterisation and classification effort of directly imaged exoplanet demographics \citep{Nielsen_2019_GPI, 2020_SHINE_Vigan}. A summary of direct imaging spectroscopy is covered extensively in \cite{Biller_Bonnefoy_2018}.

\begin{figure}
\centering
\includegraphics[width=0.48\textwidth]{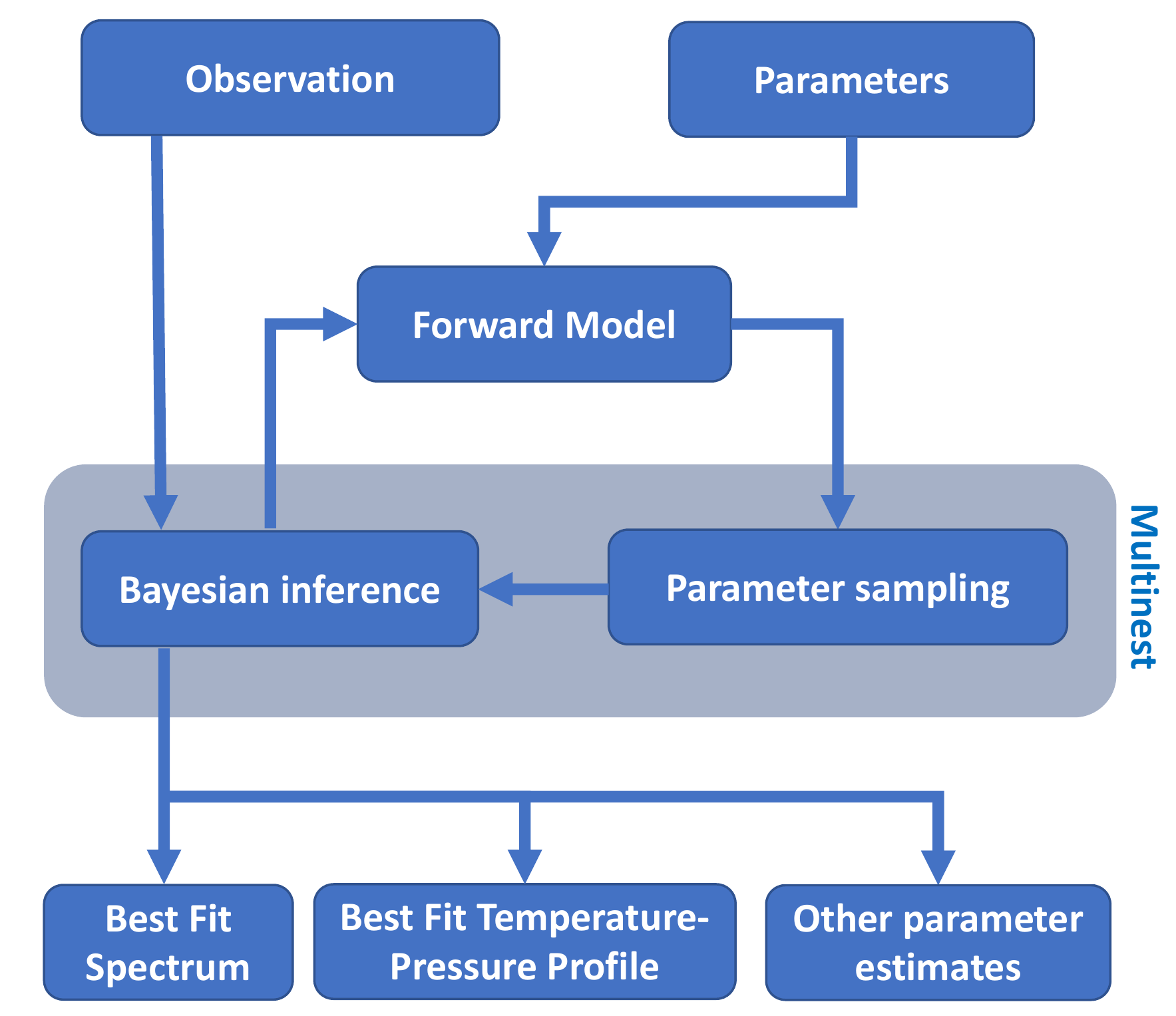}
\caption{Flowchart of the key components used in inverse retrieval techniques.}
\label{flowchaart}
\end{figure}

The development of extrasolar planetary spectroscopy  \citep[see][]{jr:Tinetti2013} has mainly been driven by studies of transiting hot-Jupiters and has allowed for unprecedented insight into the diversity of their atmospheres. This led to the expansion and application of inverse atmospheric modeling techniques (outlined in Figure \ref{flowchaart}) to exoplanetary spectra (see \citealt{jr:LineRetrieval2013} for a review of early exoplanetary retrieval codes). There are now a variety of retrieval codes developed for exoplanet atmospheric characterisation, examples include Nemesis \citep{Irwin_2008}, Chimera \citep{jr:LineRetrieval2013}, BART \citep{Harrington_BART_2016}, SCARLET \citep{2015_Benneke}, POSEIDON \citep{2017_MacDonald_POSEIDON}, Brewster \citep{Burningham2017, Burningham_2021}, HyDRA \citep{2018_Gandhi_Hydra}, petitRADTRANS \citep{2019_Molliere_petitRADTRANS, Nowak_2020_BetaPicb, Molliere_2020}, Platon II \citep{2020PLATONII_Zhang}, Helios-R2 \citep{Kitzmann_2020}, APOLLO \citep{2022_Howe_APOLLO} and TauREx3 \citep{TauREx3, 2022_Refaie_Taurex3p1}. In previous studies, TauREx has been applied to observations of transiting exoplanets \citep{jr:TauRex1,jr:TauREx, 2016_Tsiaras,2016_Rocchetto, 2018_Tsiaras, 2019_Tsiaras, 2019_Changeat, 2020_Edwards, 2020_Skaf, 2020_Pluriel_Whiteford}, with a comparative study of TauREx, CHIMERA and NEMESIS retrieval codes to be found in \citet{2020_Barstow_comparison_paper} with a review of the current state-of-the-art in \citet{2020_barstow_heng_review} and \citet{jr:Madhu2019}. 

There is now an abundance of literature outlining the application of the retrieval approach to directly imaged exoplanet and brown dwarf spectroscopy or photometry. This includes HR8799b \citep{Lee_2013_HR8799b_retrieval}, GJ 570D \citep{Line_2014}, GJ 570D and HD 3651B \citep{Line_2015}, 11 T dwarfs \citep{Line_2017}, HR8799b-e \citep{Lavie_2017_HELIOS}, 2MASS J05002100+0330501 and 2MASS J2224438-015852 \citep{Burningham2017},  GJ 570D and the Epsilon Indi brown dwarf binary system \citep{Kitzmann_2020}, 6 T and 8 Y dwarfs \citep{Zalesky_2019}, $\beta$ Pic b \citep{2020_Nowak}, HR 8799e \citep{Molliere_2020}, HR 8799c \citep{Wang_2020} and the SDSS J1416+1348AB binary \citep{2020_Gonzales}. Most recently it has been employed for analysis of SDSS J125637.13-022452.4 \citep{2021_Gonzales}, a "cloud busting" study of 2MASS 2224-0158, studying the L-T transition \citep{2022_Lueber}, analysing high-resolution observations of HD 4747 B \citep{2022_Xuan} and a population analysis of 50 Late-T Dwarfs \citep{2022_Zalesky}. Here we use the TauREx3 retrieval tool to carry out analysis of directly-imaged exoplanet 51 Eridani b (hereafter 51 Eri b) and brown dwarf benchmark GJ 570D.

Despite the significant development in the field of directly-imaged exoplanet spectroscopy in the last decade, upcoming telescopes will prove essential to further our understanding of these objects. The James Webb Space Telescope (JWST) \citep{jwst2006} and the soon to be constructed Extremely Large Telescope (ELT) \citep{jr:EELT, jr:METIS}, will lead to increased observational capacity, requiring refined and robust analysis techniques.  Retrieval tools will be a corner stone for the analysis of these next generation observations. 

Facilitated by the aforementioned instruments, direct imaging will be a very important technique for the future with the notable benefits that it offers when compared to the currently dominant technique of transmission spectroscopy. These include the ability to view exoplanet and brown dwarf atmospheres as they rotate \citep{jr:Crossfield2014} (as they are not tidally locked) and being able to probe further into the atmosphere, unlocking more spectral features. The currently observed selection of directly imaged exoplanets are limited to young gas-giants which orbit their host stars at large radial distances.  They show similar properties to free-floating planetary mass objects and old, field brown dwarfs. As a result, these three subsets of object can have the same spectral types.  The youngest low surface gravity objects start out as a hot L spectral type, evolving via cooling firstly to a T type \citep{Kirkpatrick2005LTtypes} before finally becoming a very cool Y type \citep{Cushing_2011, Kirkpatrick_2012, 2020_Miles}, at the limits of current observational capabilities.  In this study we will be focusing on T spectral type objects, with their atmospheric signatures dominated by H$_2$O and CH$_4$ absorption. 

The importance of cloud modelling for directly imaged exoplanets and brown dwarfs has been well explored and debated \citep{Marley_2010_clouds, jr:2012_Morley, Marley_2012_HR8799, Lee_2013_HR8799b_retrieval, Morley_2014_Water_clouds, Chilcote2017_BetaPicb, Burningham2017, Charnay_2018, 2020_Bowler_vhs1256b, 2020_Zhou_vhs1256b, 2020_Lew_clouds, Molliere_2020, Burningham_2021}. Previous studies of 51 Eri b, for example, used clouds in their grid modelling \citep{Rajan_2017, Samland_2017} to successfully fit the planet's spectral energy distribution (SED). For a recent and extensive review of exoplanet clouds see \citet{Helling_2018_clouds}. Alternative explanations for the observed SEDs have been explored in \citet{Tremblin_2016} and \citet{Tremblin_2017}, who demonstrated that a reduced atmospheric temperature gradient can reproduce the SEDs of late L and T type brown dwarfs, without the need to invoke clouds. The mechanism reducing the temperature gradient in these atmospheres has been proposed to be diabatic convection triggered by the $\mathrm{CO/CH_4}$ chemical conversion in brown dwarf atmospheres \citep{Tremblin_2019}.

We now describe the specifics of both the retrieval tool and other tools used in our spectral analysis, as applied to spectra of 51 Eri b and GJ 570D. 

\section{Two benchmark T dwarfs: GJ 570D and 51 Eri b observations}

In this section we give a brief overview of our current knowledge and understanding of GJ 570D and 51 Eri b as well as describing the origin of the data used in their model fitting analysis.  We chose to focus on T dwarfs in this first application of TauREx3 to directly-imaged targets as in this temperature regime their SED's are thought to be less influenced by clouds, which are expected to exist below the observable photosphere \citep{1997_Burrows, 1999_Burrows_Sharp, 2006_Lodders_Chemistry_of_Low_Mass_Substellar_Objects}. The inclusion of GJ 570D allows us to benchmark TauREx3 against previous studies using other retrieval codes \citep{Line_2015, Burningham2017, Kitzmann_2020}. 51 Eri b offers a comparable spectral type object but allows us to investigate a completely different mass regime and it has no existing free-chemistry retrieval analysis. We note that clouds seem to be more prominent in the observable atmosphere for low surface gravity objects such as 51 Eri b \citep{Marley_2012_HR8799,Charnay_2018}.

\addtocounter{figure}{-0}
\begin{figure*}
\centering
\begin{subfigure}[b]{0.7\textwidth}
   \includegraphics[width=1\textwidth]{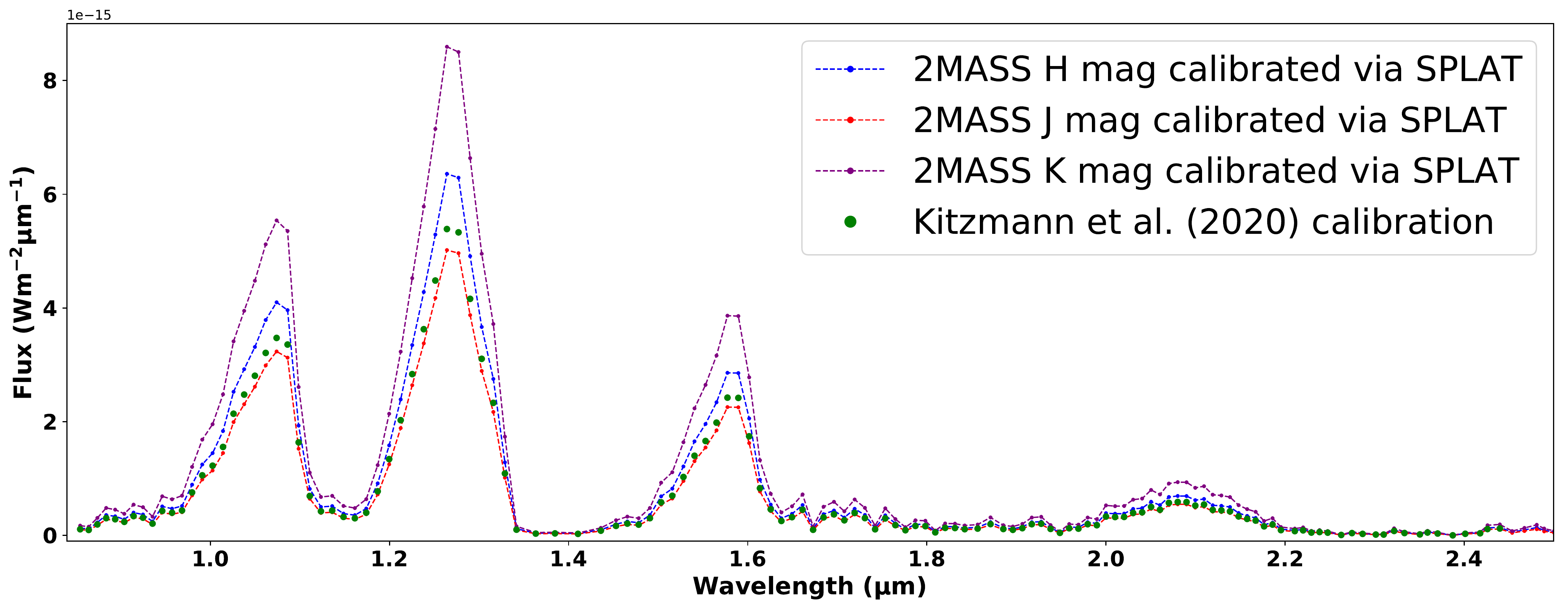}
   \caption{GJ 570D}
   \label{fig:observations_GJ570D} 
\end{subfigure}

\vspace{3mm}

\begin{subfigure}[b]{0.72\textwidth}
   \includegraphics[width=1\textwidth]{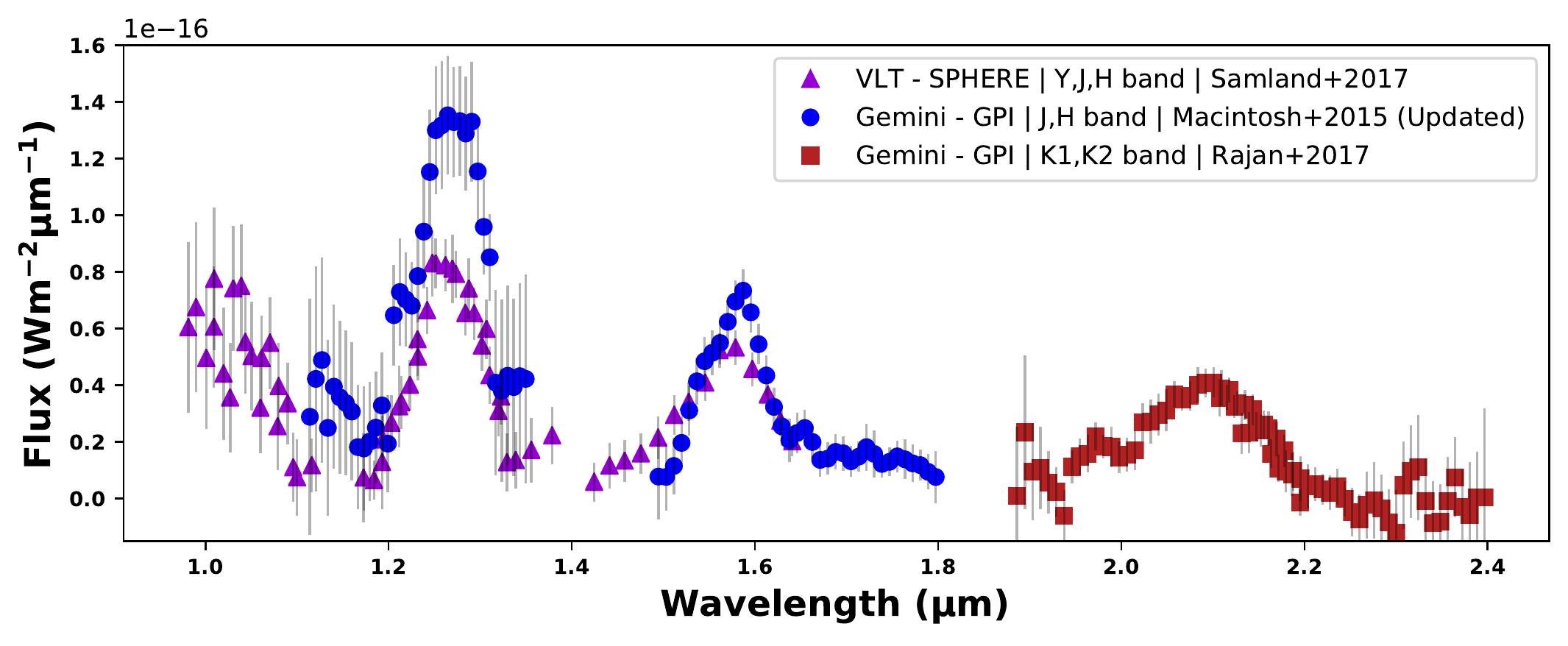}
   \caption{51 Eri b  }
   \label{fig:observations_51Erib}
\end{subfigure}
\caption[]{(a): SpeX prism spectrum of GJ 570D, flux calibrated using SPLAT and the object's 2MASS $J$, $H$ and $K$ band magnitudes. The spectrum used by \citet{Kitzmann_2020}, produced using a different absolute flux calibration approach, is included for comparison. (b): Published data for 51 Eri b. We include $Y$, $J$ and $H$ band SPHERE data from \citet{Samland_2017}, along with the GPI $J$ and $H$ data from \citet{Macintosh2015} (which is updated using a revised stellar flux and presented in \citet{Rajan_2017}) along with GPI $K1$ and $K2$ band data from \citet{Rajan_2017}. There is a clear difference in the $J$ band brightness, and also a difference in the $H$ band brightness, between the GPI and SPHERE observations.}
\end{figure*}

\vspace*{1cm}

\subsection{GJ 570D}

GJ 750D (or 2MASS J14571496-2121477) is a cool T7.5 brown dwarf, with an age of 1-5 Gyr \citep{Liu_2007}, and was among the first T dwarf companions to be discovered by \citet{Burgasser_2004, Burgasser2006}.  It is a very wide component in a hierarchical quadruple system, comprising the inner spectroscopic binary companions GJ 570B and C and the primary GJ 570A from which GJ 570D orbits at a projected separation of 1525 $\pm$ 25 AU \citep{Burgasser_2000}. GJ 570D has been included in order to compare TauREx3 against other retrieval studies as it has become commonly included in novel retrieval approach validations \citep{Line_2014,Line_2015,Line_2017, Burningham2017, Kitzmann_2020, Piette_2020}. It also offers the opportunity to compare retrieval results against studies using grid model fitting. GJ 570D has a comparable spectral type to the exoplanet 51 Eri b, also included in this study.

\subsubsection{Observations and calibration}

We used observations of GJ 570D  taken by the SpeX spectrograph \citep{2003_Rayner_SpeX}, which is mounted on the 3m NASA InfraRed Telescope Facility. The measured spectrum is part of the SpeX Prism Library \citep{Burgasser_2014_SpeX_Library} and was first published in \cite{Burgasser_2004}. The data were reduced using the pipeline described in \cite{CushingIRTF}, with the spectrum spanning 0.65 to 2.56 $\mu$m at an average spectral resolving power of 120. Using the SpeX Prism Library\footnote{SpeX Prism Library: \url{ http://pono.ucsd.edu/~adam/browndwarfs/spexprism/library.html}} data analysis toolkit (SPLAT\footnote{SPLAT: \url{ http://pono.ucsd.edu/~adam/browndwarfs/splat/}}) (see \cite{Burgasser_SPLAT} for details), we flux calibrated the data using photometry from the 2MASS survey\footnote{2MASS Survey Archive: \url{https://irsa.ipac.caltech.edu/Missions/2mass.html}} \citep{2006_2MASS_Survey}. 

The spectra shown in Figure \ref{fig:observations_GJ570D} have then been calibrated using $J$ (15.324 $\pm$ 0.05 mag), $H$ (15.268 $\pm$ 0.09 mag) and $K$-band (15.242 $\pm$ 0.16 mag) fluxes. In the following analysis we used the spectrum calibrated using the $H$-band magnitude.  As outlined in \citet{Line_2015}, neighbouring pixels may not be statistically independent, due to the duplication of flux information. Therefore, when analysing this data set we only include every third data point (pixel) in our model fitting.   

\label{sec:GJ570D_observations}

\subsection{51 Eri b}
\label{sec:51Erib}

51 Eri b was the first exoplanet discovered by GPI \citep{Macintosh2014_GPI}, and has one of the smallest angular and physical separations ($\sim$0.5", $\sim$13 AU) of any directly imaged exoplanet.  It orbits a young F0-type host, with age estimates of $20\pm6$Myrs from \citet{Macintosh2014_GPI} and 26$\pm$3Myrs from \citet{2016_Nielsen}.  With a spectral type of T6.5$\pm$1.5 \citep{Rajan_2017}, 51 Eri b is notably the latest spectral type planet yet imaged.  

Exhibiting methane absorption (a first for directly-imaged exoplanets) with its lower effective temperature ($\sim$700K) and low mass ($< 10$ M$_{\rm Jup}$), 51 Eri b defined a new category of directly-imaged exoplanets.  Further, its SED indicates that the L/T transition occurs at lower temperatures for these lower surface gravity objects compared to the higher surface gravity brown dwarfs \citep{Rajan_2017}. 

Studies of this exoplanet have included clouds in order to fit the spectroscopic and photometric data.  \citet{Rajan_2017} used two self-consistent grid models, one with a patchy iron/silicate cloud component, and the other with sulfide/salt cloud to explain the spectral profile, while \citet{Samland_2017} used grid models produced using petitCODE \citep{2015_Molliere, 2017_Molliere}  which employed a slightly modified version of the \citet{AckermanMarley2001} prescription in their cloud modelling. \citet{Samland_2017} also tested the \citet{jr:2012_Morley} cloud models against their observations. \citet{Samland_2017} couldn't differentiate between patchy and uniform clouds while \citet{Rajan_2017} found a preference for patchy iron/silicate clouds in the model fitting. Both studies concluded that clouds were needed to fit the spectrum well.  We outline previous model fitting results from previous studies in Table \ref{table_bulk_51_Eri_b}. Neither \citet{Samland_2017} or \citet{Rajan_2017}, however, employed a free chemistry model as we have done in this study.


\subsubsection{Observations}

In this study we used a combination of observations of 51 Eri b from 2015-2016. These included spectroscopic data taken with GEMINI-GPI's Integral Field Spectrograph \citep{Macintosh2014_GPI} (IFS) in the $J$, $H$, $K1$ and $K2$ bands \citep{Rajan_2017} (where $J$ and $H$ band observations are updated from \citet{Macintosh2015}) and VLT-SPHERE's IFS \citep{Beuzit2008_SPHERE, 2019_Beuzit_SPHERE} using its $YJ$, $YH$ filters \citep{Samland_2017}. The spectra are shown in Figure \ref{fig:observations_51Erib}, calibrated as outlined in \citep{Samland_2017} and \citep{Rajan_2017}. 

We also employed photometric measurements from KECK-NIRC2's \citep{2003_NIRC2_McLean_Sprayberry} Lp and Ms filters \citep{Rajan_2017}, where we used two combinations of data for our analyses: one which combined SPHERE's $Y$, $J$ and $H$ bands along with GPI's $K1$ and $K2$ band data and the other which combined only the GPI bands. We used this approach as the aforementioned GPI and SPHERE observations differed significantly in brightness in both the $J$ and $H$ bands.

Unlike with the GJ 570D data, we did not exclude any data from the analysis. This was motivated by the data's already low spectral resolution, combined with the relatively large errors, where exclusion of data would severely impact the retrievals ability to constrain parameters. We note that the potential for correlated noise to impact the retrieval is more prominent when using these full data sets.

\section{Modelling}

\subsection{Retrieval method overview}

In this study we employ the inverse retrieval method. This approach is reviewed in \cite{jr:FortneyRA2018} and \cite{Madhusudhan_2018} and is outlined in Figure \ref{flowchaart}. In its simplest form, the technique obtains a best fit to observed spectra using a varying forward model defined by a handful of constraining parameters. Variations in the forward model explore the permitted parameter space while statistically deriving the best-fit to an observation. Inverse techniques calculate the posterior distributions of these forward model parameters that best fit the observed data. 

The forward model contains a set of input parameters and converts them into the observable format;  typically a spectrum at a specified spectral resolution. For the purposes of this paper, we replicate observations of thermal emission in the near and mid-infrared. The model combines fundamental quantities such as the molecular chemistry and the temperature-pressure profile present in the observed atmosphere, the radius and mass of the object, and the existence of clouds or aerosol opacities.  

Retrieval models generally use a Bayesian sample approach to select the best fit model. As outlined below, the use of Bayesian retrievals allows the formal inclusion of prior knowledge and full exploration of the likelihood probability distribution of the data. Bayesian retrievals have become the norm in atmospheric analyses of transmission and secondary eclipse spectra of transiting exoplanets. However, Bayesian parameter inference has now regularly proven itself as an effective tool in the pursuit of statistically rigorous exoplanet and brown dwarf atmospheric characterisations in the field of direct imaging.

\subsection{TauREx3}

TauREx3 (Tau Retrieval of Exoplanets) is a publicly available\footnote{TauREx3: \url{ https://github.com/ucl-exoplanets/TauREx3_public}} Bayesian retrieval code designed to be applied to spectroscopic observations of extrasolar atmospheres \citep{jr:TauRex1, jr:TauREx, TauREx3}. It can be employed to analyse emission, transmission and phase-curve spectroscopic data. Figure \ref{flowchaart} gives an overview of the TauREx code for emission retrieval. The following subsections outline the key components of TauREx3.

\subsubsection{Atomic and Molecular cross sections}

TauREx has its own purpose built molecular and atomic opacities, which can be accessed from the publicly available ExoMolOP database \citep{20ChRoAl.exo}\footnote{ExoMolOP: \url{http://exomol.com/data/data-types/opacity/}}. The line lists used for this database originate mainly from the ExoMol project \citep{Tennyson_Yurchenko_2012_ExoMol} but also HITEMP \citep{HITEMP_2010_Rothman}, HITRAN \citep{HITRAN} and MoLLIST \citep{MOLLIST}. This includes the latest line lists for TiO \citep{McKemmish_TiO_ExoMol}, H$_2$O \citep{jt734}, CO \citep{15LiGoRo.CO}, CO$_2$ \citep{20YuMeFr.co2}, CH$_4$ \citep{jt698}, VO \citep{jt644}, H$_2$S \citep{jt640} and NH$_3$ \citep{jt771}. ExoMol provides line lists for extended temperature ranges for a variety of molecules. The pressure and temperature broadened profiles for the resonance doublets of Na and K are computed using methods described in \cite{16AlSpKi.broad} and \cite{19AlSpLe.broad}. All other line data for these atomic species are taken from either the NIST~\citep{NISTWebsite} or Kurucz~\citep{KURonline} database (see Section~\ref{sec:Na_K}). We note here that all the results (Tables and Figures) presented in this study were retrieved using the broadening parameters of Allard et al. and non-resonance lines from the Kurucz database, unless stated otherwise.The TauREx3 cross-sections used in this work were sampled at $R$~=~$\frac{\lambda}{\Delta \lambda}$~=~15,000 across the 0.3~-~50$\mu$m wavelength region. For a more detailed discussion of TauREx's line list library see \cite{20ChRoAl.exo}. During the molecular and atomic radiative transfer calculations performed by TauREx, the model is produced at a much higher resolution than that of the observed spectrum. These high resolution spectra are then binned down to the data resolution in order to calculate the log-likelihood. 

\subsubsection{Temperature-pressure profiles}

\begin{figure}
\centering
\begin{subfigure}[b]{0.485\textwidth}
   \includegraphics[width=1\linewidth]{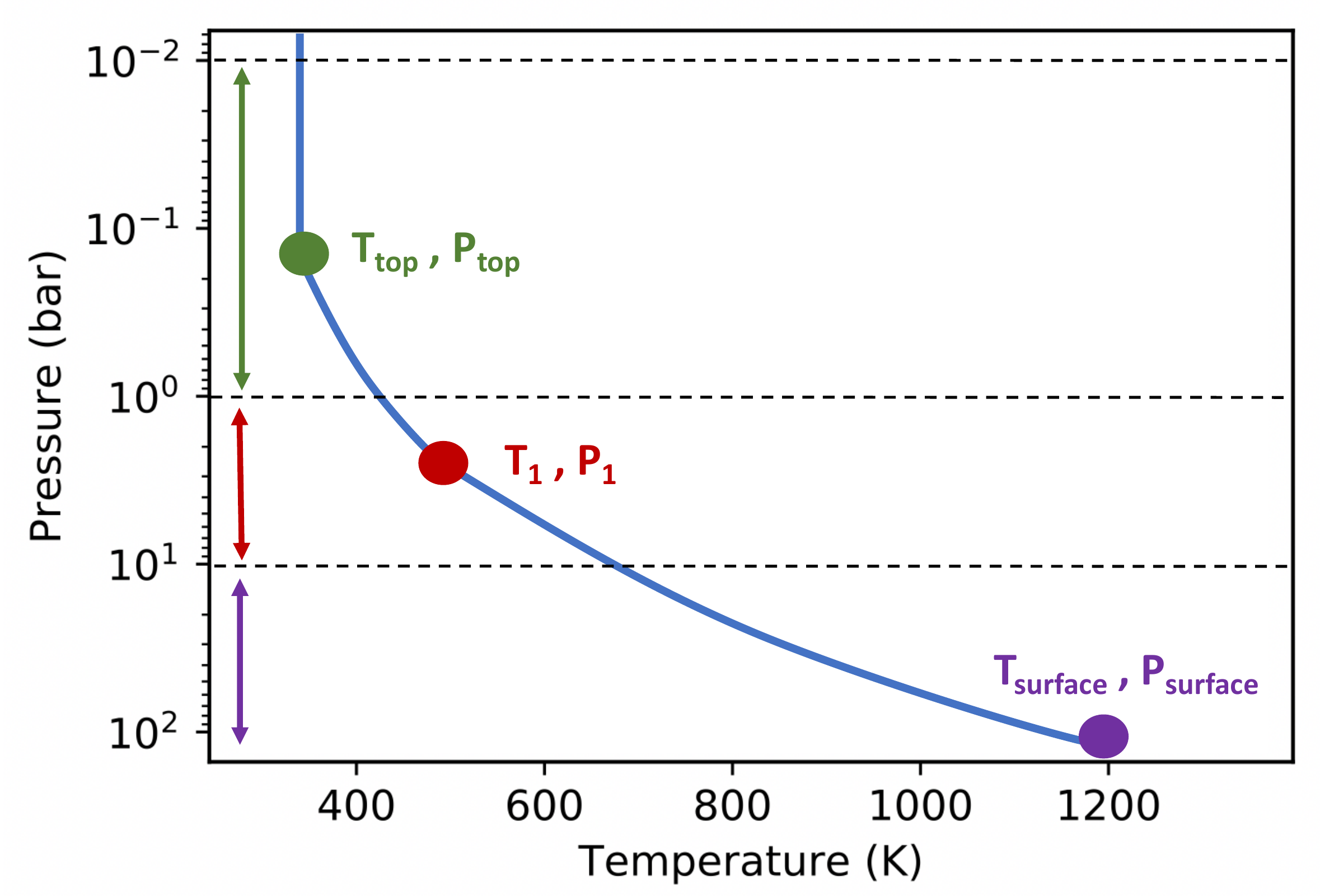}
   \label{fig:TP1} 
\end{subfigure}

\vspace{3mm}

\caption[]{$npoint$ temperature-pressure profile used in our analysis. The figure outlines the structure of the n-point profile, with n = 3 in this example.}
\label{fig:TP_profile_demo}
\end{figure}

To accurately model directly-imaged emission spectroscopy, an appropriate temperature-pressure parameterisation must be adopted. TauREx offers a variety of temperature-pressure profile options, ranging from radiative two-stream modelling such as the \cite{Guillot_2010} prescription (for highly-irradiated planets) to more ad-hoc geometric approaches in which temperature-pressure nodes are allowed to vary freely. In Figure \ref{fig:TP_profile_demo}, we illustrate the $n-point$ temperature-pressure profile adopted in our analysis. 

The $n-point$ profile is determined by several parameters including the top of atmosphere temperature, $T\textsubscript{top}$, and top of atmosphere pressure, $P\textsubscript{top}$ (set at 10$^{-3}$ bar in this study). The other parameters include the tropopause temperature and pressure, $T\textsubscript{1}$ and $P\textsubscript{1}$, as well as the surface pressure $P\textsubscript{surf}$ (set at 500 bar for this study) and temperature $T\textsubscript{surf}$.Temperatures are then linearly interpolated between these temperature-pressure nodes in log space.

In order to be able to compare retrievals with temperature-pressure profiles of differing degrees of freedom we also added with a less flexible profile following a simple parameterisation employed in \citet{Lavie_2017_HELIOS}. This originated from a reduced version of equation 126 in \citet{Heng_2014}:

\begin{equation}
T^{4} = \frac {T_{int}^{4}}{4} \left( \frac{8}{3} + \widetilde{m} \kappa_{0} \right)
\end{equation}

\noindent where $T_{int}$ is the internal temperature and $\kappa_{0}$ the constant component of the infrared opacity. $\widetilde{m}$ is column density determined via $P_{0}=\widetilde{m} \cdot g$ with $g$ being the surface gravity at the bottom of our model atmosphere (500 bar). This simpler profile paramterisation only has two free parameters within our retrievals: $\kappa_{0}$ and $T_{int}$.

\subsubsection{Effective Temperature Calculation}

We have included the effective temperature $T_{\rm eff}$ as a derived parameter which is useful for comparing retrieval results to grid models and evolutionary tracks. For this, we followed the same approach as adopted in \citet{Line_2015}, integrating the spectrum from 0.1 to 50 ${ \mu}$m (at the native resolution of the input cross sections) to calculate the total emission flux.  The effective temperature, as associated uncertainties, is then derived using the Stefan-Boltzmann law and a random sampling of 10$\%$ of models ran.

\subsubsection{Bayesian Analysis}

TauREx employs Bayesian statistics as the cornerstone for the retrieval analysis. Bayes' theorem states that:

\begin{equation}
 P(\theta\mid x, \mathcal{M}) = \frac{P(x\mid\theta, \mathcal{M}) \, P(\theta, \mathcal{M})}{P(x\mid \mathcal{M})},
\end{equation}

\noindent where $P(\theta, \mathcal{M})$ is the Bayesian prior, and $\mathcal{M}$ is the forward model. $P(\theta\mid x, \mathcal{M})$ is the posterior probability of the model parameters $ \theta$ given the data, \textit{x}, assuming the forward model $\mathcal{M}$. The likelihood, $P(x\mid\theta, \mathcal{M})$ is given by:

\begin{equation}
 P(x\mid\theta, \mathcal{M}) = \frac{ 1 }{\mathcal{E} \sqrt{2\pi}} {\rm exp} \left[     -\frac{1}{2} \sum_{\lambda}^{N} \left( \frac{  x_{\lambda} - \mathcal{M}_\lambda} {\mathcal{E}_\lambda} \right) ^2  \right],    
\end{equation}

\noindent where $\mathcal{E}$ is the error on the input spectral data. This is defined via:

\begin{equation}
\mathcal{E}^{2}_{\lambda} = \sigma_{\lambda}^{2} + 10^{b},    
\end{equation}

\noindent where $\sigma_{\lambda}$ is the measured error for the ${\lambda}$th flux and \textit{b} is a tolerance factor which is included as a free parameter in the retrieval analysis \citep{Tremaine_2002, 2010_Hogg, 2013_Foreman-Mackey}. This $10^{b}$ factor has been used extensively throughout the literature within various retrieval frameworks across many data sets \citep{Line_2015, Line_2017, Burningham2017, Burningham_2021}. 

The $10^{b}$ error inflation term can account for imperfections in the forward model's capability to fit the observed emission spectrum and/or account for underestimated uncertainties. It also, more demonstrably, allows for the down-weighting of sections of a spectrum (in our case the $K$ band) where the spectral resolution is highest as well as possessing the smallest error bars. Such sections of data can lead to the neglect of other important parameters driving regions of a spectrum, such as in the case of our GJ 570D data. Including the error inflation can therefore allow for a more equally weighted consideration of the whole spectrum when performing the Bayesian evidence calculations. This is discussed more in section \ref{sec:model_setup}.

\label{sec:Bayesian_Analysis}

\subsubsection{Nested Sampling via Multinest}

TauREx includes the implementation of Bayesian statistics via nested sampling (NS) using Multinest \citep{Feroz_Hobson_2008, Feroz_2009, Feroz_2013} via PyMultinest \citep{PyMultiNest}. NS derives the Bayesian Evidence given by:    

\begin{equation}
E = \int P(\theta \mid \mathcal{M})P(x\mid\theta, \mathcal{M})d\theta, 
\end{equation}

\noindent where $E = P(x \mid \mathcal{M})$ is the Bayesian Evidence which allows for formal model selection. The statistical results from MultiNest are then used to derive the parameter estimates which combine to produce the highest Log-Evidence. Using MultiNest, we sampled the parameter space using 3000-5000 live points at a sampling efficiency of 0.8 which is the default for parameter estimation.

Via the nested sampling Log-Evidence, we can compare model results using the Bayes Factor B: 

\begin{equation}
 log(b) = \Delta log(Ev) = Log(Ev2) - log(Ev1), 
\end{equation}

This is a ratio of evidence of two competing models (Ev1 and Ev2), allowing for comparison. Table \ref{KassRafteryTable}, from \citet{Kass_Raftery_1995}, outlines how $log(B)$ can be interpreted.

\begin{table}
	\centering
	\caption{Interpreation of the Bayes ratio outlined in \citealp{Kass_Raftery_1995}}
	\label{tab:abundances}
	\begin{tabular}{ |c|c| } 
		\hline 
		$log(b)$ &  Interpretation\\
		\hline 
	    0 - 0.5 & $No \ Evidence$ \\
		0.5 - 1 & $Some \ Evidence$ \\
		1 - 2 & $Strong \ Evidence$ \\
		>2 & $Decisive$ \\
		\hline 
    	\end{tabular}
		\label{KassRafteryTable}
\end{table}

\label{section:Nested_Sampling_via_Multines}

\subsubsection{Priors}

TauREx has preset default priors set for all the possible free parameters. This includes all that are necessary for the forward model such as mass, radius, temperature-pressure prescription and atmospheric trace gases considered. By default TauREx employs uninformative priors with large prior ranges (e.g. trace-gas abundance priors are log-uniform, log(abundances) = $1.0 - 1.0 \times 10^{-12}$). The default values can be manually over-ridden, allowing the user to limit or open-up the parameter space. Narrowly defined bounds have the benefit of reducing computational expense but run the risk of being overly restrictive. The priors and prior bounds set for the retrieval analysis performed in this paper were either uniform, log-uniform or Gaussian priors based on values from previous published studies (when such values were available). See Table \ref{tab:retrievalpriors} for a full overview of the priors set.

\begin{figure}
\centering
\includegraphics[width=0.45\textwidth]{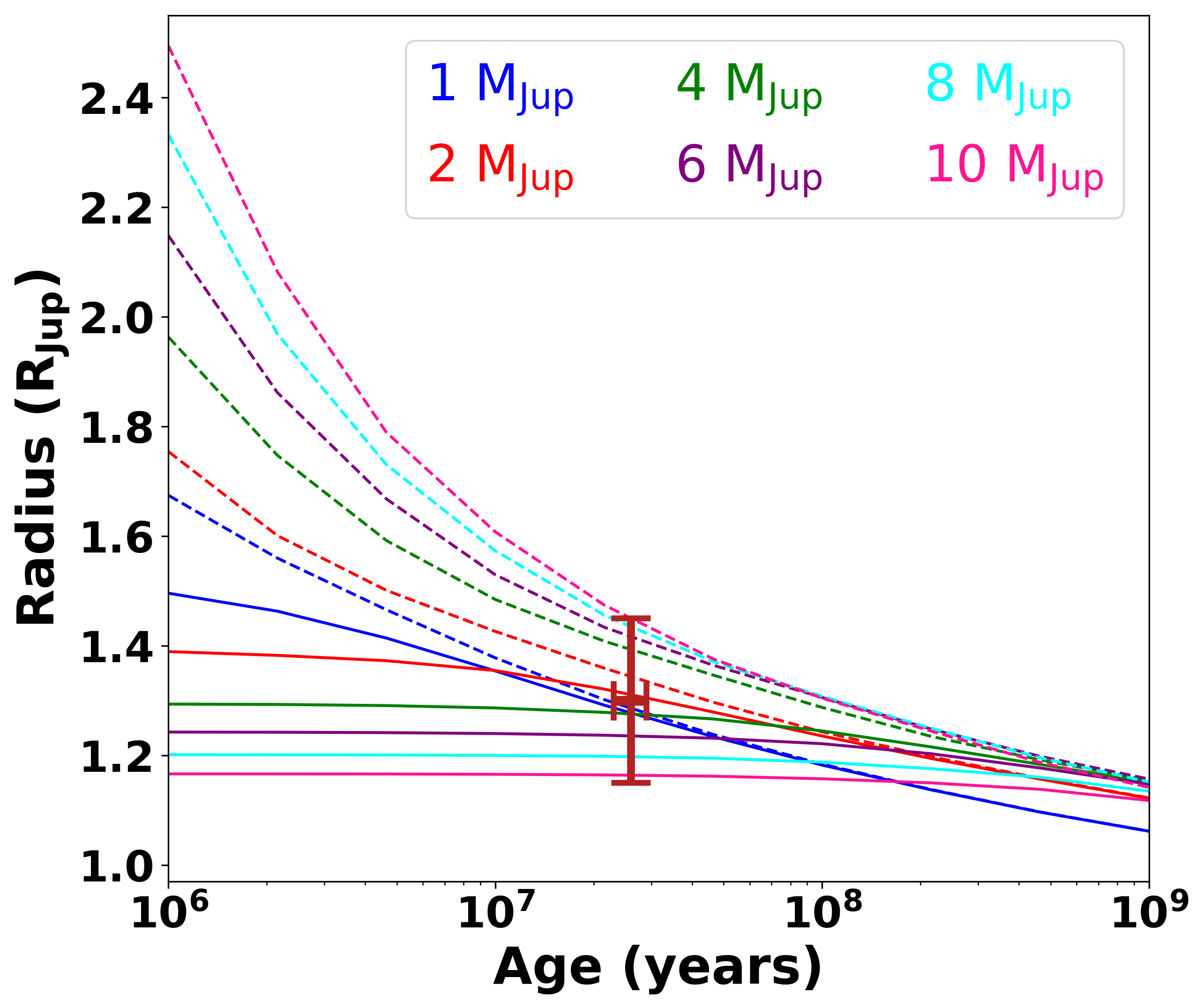}
\caption{Evolutionary tracks from \citet{2008_Fortney} with age uncertainty of 51 Eri system from \citet{Rajan_2017} indicated. The Gaussian radius prior we adopt for the 51 Eri b analysis is also indicated.}
\label{RadVsAge_Fortney2008}
\end{figure}

Given the lower quality of the 51 Eri b data we adopted an informative Gaussian prior for our retrievals. This was based on the system age estimate from \citet{Rajan_2017} and the evolutionary tracks from \citet{2008_Fortney} as shown in Figure \ref{RadVsAge_Fortney2008}. We didn't adopt a Gaussian prior on the mass as the reported values in the literature \citep{Macintosh2015, Rajan_2017, Samland_2017, Nielsen_2019_GPI} have a large spread in the planetary mass regime. 

\begin{table*}
    \begin{threeparttable}
    \centering
    \caption{Table of retrieval priors. The middle sections list the parameters used for the $n-point$ and \citet{Lavie_2017_HELIOS} temperature-pressure profiles, while the bottom section outlines the parameters used in the deck and slab cloud schemes}.
    \begin{tabular}{cccc}
    \toprule
  
    Retrieved parameter & Distribution Type  & GJ 570D Bounds & 51 Eri b Bounds  \\ 
    \hline
    Mixing Ratio  &  Log-Uniform & 1e-12 - 1e-1 & 1e-12 - 1e-1  \\ 
    Radius  &  Uniform | Gaussian & 0.5 - 2.0 R$_{\rm Jup}$ &  1.3 $\pm$ 0.15 R$_{\rm Jup}$ \tnote{1,2} \\ 
    Mass & Uniform & 13 - 80 M$_{\rm Jup}$ & 1 - 13 M$_{\rm Jup}$\tnote{3}\\
    Distance & Gaussian & 5.8819 $\pm$ 0.0029 pc\tnote{4} & 29.4 $\pm$ 0.3 pc\tnote{5} \\
    $S_{\rm cal}$ & Gaussian & 1.0, 0.1 STD &  1.0, 0.1 STD \\
    $10^{b}$ & Uniform & 0.01 x min($\sigma_{\lambda}^2$) $\leq 10^{b}$ $\leq$ 100 x min($\sigma_{\lambda}^2$) & -\\
    \hline
    $T_{surf}$ & Uniform & 1250 - 2500 K & 1250 - 2500 K\\
    $P_{surf}$ & Log-Uniform & 5e2 - 1e1 bar & 5e2 - 1e1 bar\\
    $T_{1}$ & Uniform & 100 - 2000 K & 100 - 2000 K\\
    $P_{1}$ & Log-Uniform & 1e1 - 1e-1 bar & 1e1 - 1e-1 bar\\
    $T_{top}$ & Uniform & 0 - 1000 K & 0 - 1000 K \\
    $P_{top}$ & Log-Uniform & 1e-1 - 1e-3 bar & 1e-1 - 1e-3 bar \\
    \hline
    $T_{int}$ & Uniform & - & 10 - 1500 K \\
    $\kappa_{0}$ & Log-Uniform & - & 1e-15 - 1e1 \\
    \hline
    $\tau_{0}$& Uniform & - & 0.01 - 100 \\
    $\alpha$& Log-Uniform & - & -10 - 10 \\
    $P_{top}$ & Log-Uniform & - & 1e2 - 5e7  bar\\
    $P_{bottom}$ & Log-Uniform & - & 1e2 - 5e7 bar\\
    $C_{frac} $ & Uniform & - & 0 - 1 \\
    \bottomrule
    \end{tabular}
    \begin{tablenotes}
    \item[1] Note: Gaussian prior. 
    \item[2] Note: We use a Gaussian prior informed using evolutionary models from \citet{2008_Fortney} combined with the age presented in \citet{Rajan_2017} .  
    \item[3] Note: We make the assumption of a planetary mass object.
    \item[4] Note: GJ 570D distance comes from Gaia Archive: https://gea.esac.esa.int/archive/.
    \item[5] Note: The 51 Eri b distance comes from \citet{Macintosh2015}.
    \end{tablenotes}
    \label{tab:retrievalpriors}
    \end{threeparttable}
\end{table*}


\subsubsection{Mode for direct imaging}

We modified TauREx3 to allow us to model directly imaged targets.  First, we removed stellar emission from the forward model and added an inverse square law scaling for the exoplanet or brown dwarf emission:

\begin{equation}
 \mathrm{Absolute \ Flux} = {F_{\mathrm{emission}}} \cdot {S_{\mathrm{cal}}} \cdot \frac{R^2} {D^2}, 
\end{equation}

\noindent where \textit{${F_{\rm emission}}$} is the emission flux from the forward model and \textit{R} is the object radius and \textit{D} its distance from the Earth. \textit{${S_{\rm cal}}$} is a scaling calibration factor, to account for imperfect absolute flux calibration assumptions (see Figure \ref{fig:observations_GJ570D}). A calibration factor such as this was used in \citet{2020_Oreshenko}.
Within the retrieval, $S_{cal}$ can also be inversely considered as scaling the observed data to the model derived flux via $Obs_{cal}$:

\begin{equation}
 \mathrm{Obs_{cal}} = \frac{1}{S_{\mathrm{cal}}}, 
\end{equation}

\noindent We have added surface gravity $log(g)$ as an inferred parameter, determined via Newton's Law of Universal Gravity: 

\begin{equation}
\log(g) = \log\left[\frac{GM}{R^2}\right], 
\end{equation}

\noindent where $G$ is the gravitational constant, $M$ is the object's mass and $R$ is the object's radius. We have also included the calculation of the carbon to oxygen (hereafter C/O) ratio, which for the brown dwarf and exoplanet we study in this paper, is driven predominantly by the relative abundances of H$_2$O and CH$_4$. Therefore, this ratio should really be considered as a CH$_4$ to H$_2$O ratio.  In the case of our T dwarf analysis the inferred C/O ratio is calculated via:

\begin{equation}
C/O = \frac{\chi_{CH_{4}}+\chi_{CO}}{\chi_{H_{2}O}+\chi_{CO}}, 
\end{equation}

\noindent where $\chi$ is the mixing ratio of the relative molecules.We also add an inferred metallicity via the retrieved abundances. This is approximated by summing metal-containing molecules weighted by the number of metal atoms which is then divided by the abundance of neutral hydrogen. This value is then compared to the summation of the retrieval-traced solar metals relative to hydrogen, using values from \citet{Asplund2009}. Metallicity is therefore calculated via:

\begin{equation}
M_{object} = \sum_{molecules}{\frac{f_{m} \cdot n }{\chi_{H2} \cdot 2}}, 
\end{equation}

\noindent where $f_{m}$ is the gas fraction of a particular molecule m, n is the number of "metal" atoms in a given molecule (eg. for CO2 n=3), $\chi_{H2}$ is the gas fraction of neutral hydrogen. Therefore [M/H] is determined via:

\begin{equation}
[M/H] = \frac{M_{object}}{M_{solar}}, 
\end{equation}

\noindent where $M_{solar}$ is determined via all relevant solar elemental abundances relative to solar H.

\subsubsection{Clouds}

\begin{figure}
     \centering
     \begin{subfigure}[b]{0.5\textwidth}
         \centering
         \includegraphics[width=\textwidth]{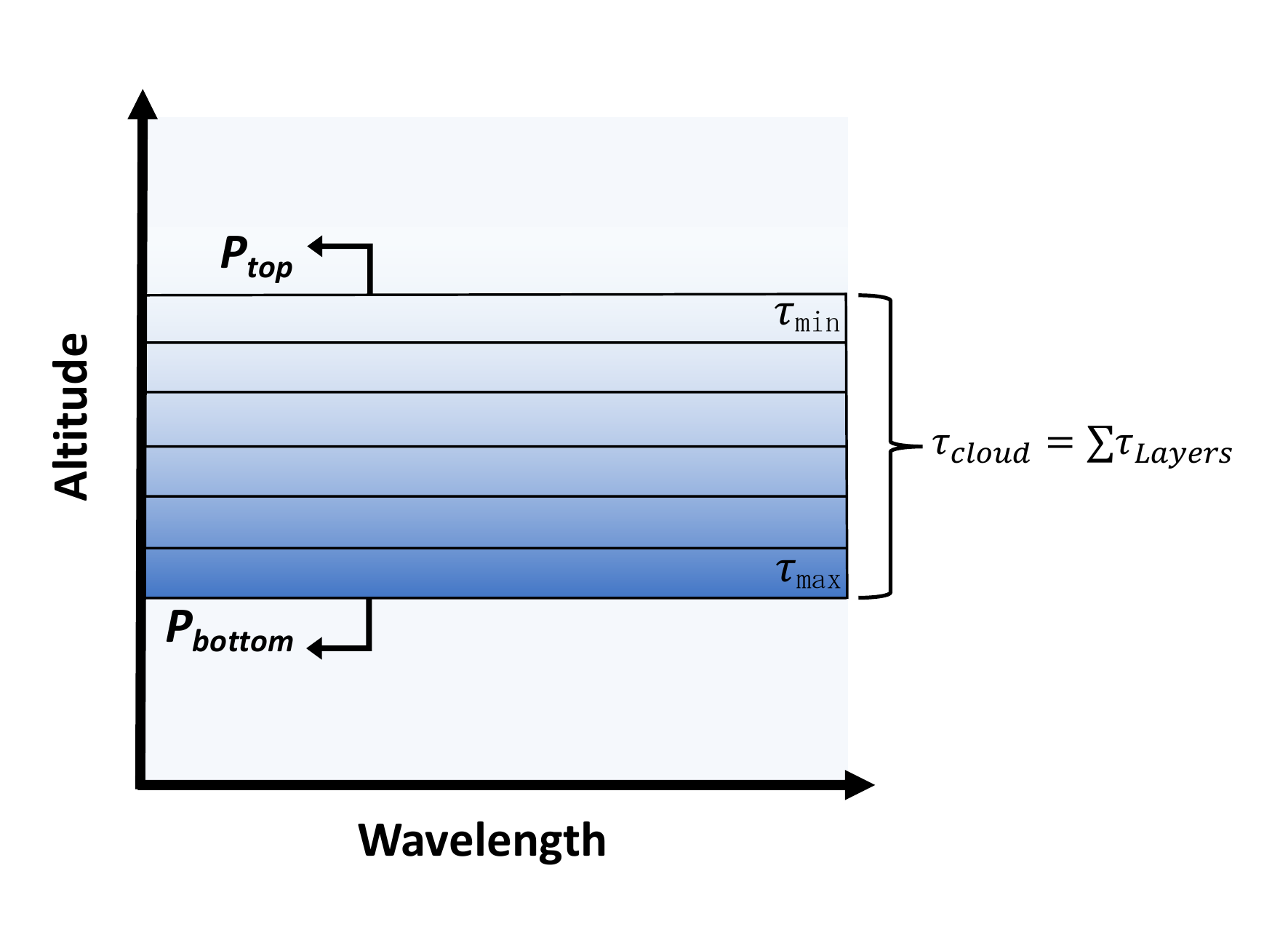}
         \caption{Slab cloud}
         \label{fig:___}
     \end{subfigure}
     \hfill
     \begin{subfigure}[b]{0.5\textwidth}
         \centering
         \includegraphics[width=\textwidth]{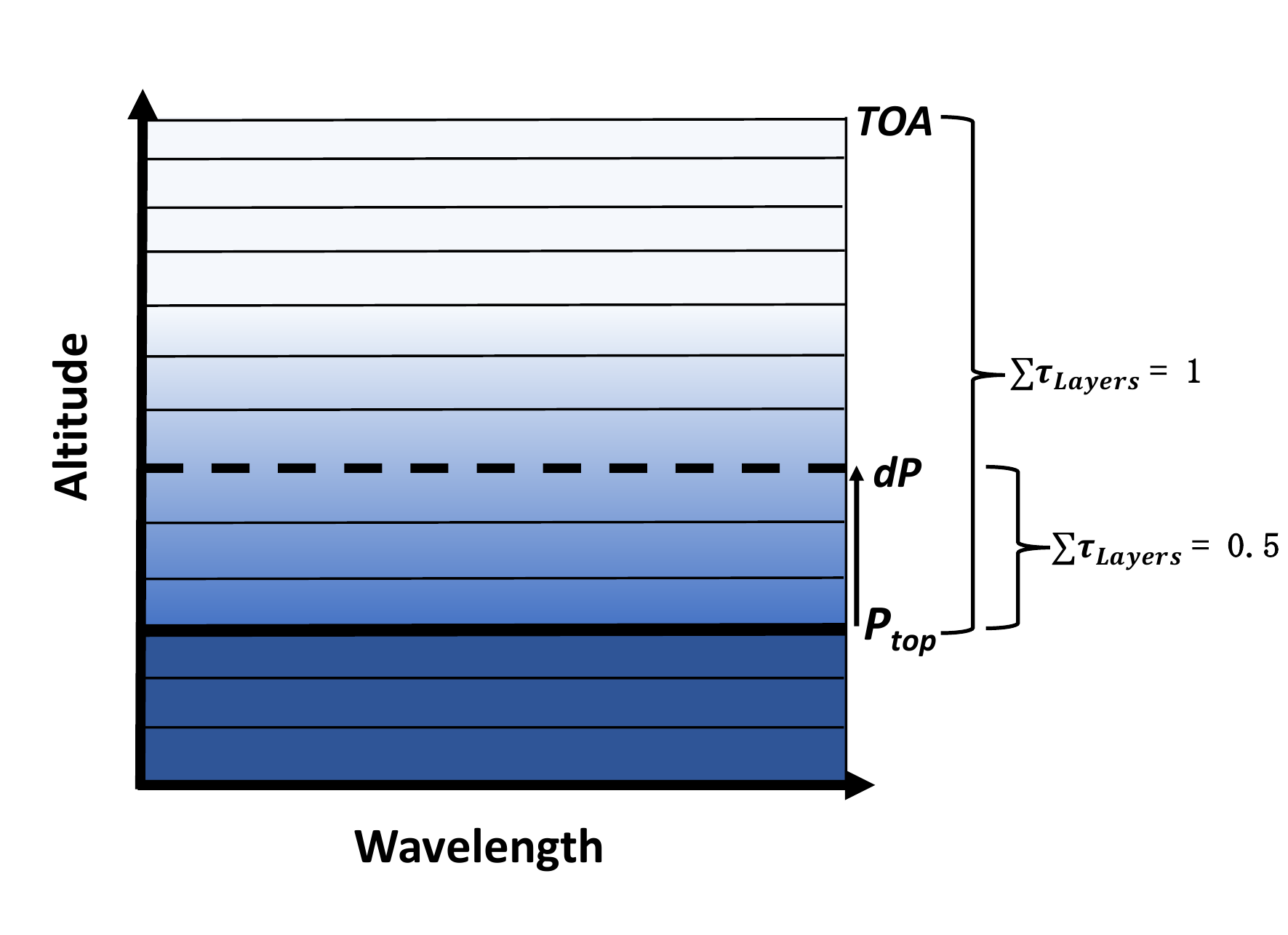}
         \caption{Deck cloud}
         \label{fig:_____}
     \end{subfigure}
\caption{Diagram outlining cloud structures employ in this study. (a) outlines the slab cloud structure. (b) outlined the deck cloud structure.}
\label{powerlawcloud}
\end{figure}

\noindent In order to explore the impact of cloud extinction we employed the power law deck and slab cloud parameterisations from \citep{Burningham2017} which are illustrated in Figure \ref{powerlawcloud}. The optical depth of the power law deck cloud is set by $P_{top}$ pressure where the cloud becomes optically thick ($\tau$=1). The cloud opacity drops off above this pressure via d$\tau$/ dP $\propto$ exp($\Delta$P/$\Phi$) where $\Delta$P is the height above and below the $P_{top}$ pressure and $\phi$ is:

\begin{equation}
\Phi = \frac{P_{top}\cdot(10^ {\Delta logP}-1)}{10^ {\Delta logP}}
\end{equation}{}

\noindent The decay is parameterised by $\Delta logP$ and the cloud is made non grey with the optical depth following $\tau \propto \lambda^{\alpha} $. Therefore, in total, the power law slab is retrieved via 3 parameters: $P_{top}$, $\Delta logP$ and $\alpha$. The total optical depth of the slab cloud is determined via:

\begin{equation}
\tau_{cloud} = \tau_{0} \left( \frac{\lambda}{\lambda_{0}} \right) ^{\alpha } = \sum \tau_{Layers}
\end{equation}{}

\noindent where $\lambda_{0}$=1$\mu$m. $\tau_{0}$ and $\alpha$ are the two retrievable components of this cloud prescription as well as the top pressure boundary of $P_{top}$ and the bottom pressure boundary of the cloud $P_{bottom}$. $\tau_{cloud}$ is distributed throughout the layers in the cloud slab pressure boundaries, weighted by d$\tau$/ dP $\propto$ P where dP is relative to $P_{bottom}$. Therefore, the total optical depth is distributed such that the bottom layer has the maximum optical depth while the top layer has the minimum optical depth present. In total, the power law slab is retrieved via 4 parameters: $\tau_{0}$, $\alpha$, $P_{top}$ and $P_{bottom}$.

\noindent These are flexible but simplistic cloud approach that lack the rigour of the more physically motivated approaches included in \citet{Molliere_2020} and \citet{Burningham_2021}. For example, these parameterisations do not allow us to probe specific cloud species or particle sizes but is still suitable for this study. In an effort to investigate the potential presence and impact of patchy clouds on exoplanet and brown dwarf spectra, the basic patchy cloud consideration from \citealp{Marley_2010_clouds} has been added:

\begin{equation}
F_{Tot} = C_{frac}\cdot F_{Cloudy} + (1-C_{frac})\cdot F_{Clear}
\end{equation}

\noindent where $F_{Tot}$ is the total flux, $F_{Clear}$ is the flux from regions without clouds, $F_{Cloudy}$ is the flux from regions with clouds and $C_{frac}$ is the fraction of surface area with clouds. The non-cloud properties are identical for the two forward models which are linearly combined. Employing this fractional cloud consideration therefore acts to add an additional retrieved parameter to both the slab and deck cloud parameterisations.

\label{cloud_description}

\subsubsection{Forward model setup}
\label{sec:model_setup}


The atmospheric forward model is described in \citet{TauREx3}. We assume a hydrogen dominated atmosphere with a H$_{2}$ and He mixing ratio He/H$_{2}$ = 0.17567.
We set molecular trace-gas volume mixing ratios to be constant in pressure, which we refer to as isoprofiles.  TauREx3 does allow for pressure dependent  abundance profiles \citep{2019_Changeat} but this comes at a significant increase in model complexity and so will be explored in later work. 

We use error inflation (as outlined in Section \ref{sec:Bayesian_Analysis}) for our analysis of GJ 570D but not for our 51 Eri b analysis. This is because 51 Eri b's observations have much larger error bars and the observation's resolution is more uniform throughout the spectrum, negating the spectral band weighting issue experienced with the GJ 570D data set (again, see  Section \ref{sec:Bayesian_Analysis}). In our GJ 570D data it is apparent that that $K$ band errors are much smaller than the $J$ band data (see Figure \ref{fig:observations_GJ570D}). The impact of adding this error inflation parameter acted to allow the fit to the $J$ band data to improve without affecting the goodness of fit in the $K$ band and also allowed for the overall Log Evidence to increase slightly. This was interpreted as an increase in error size in the $K$ band, while negligible in the $J$ band, allowing for a better overall fit by de-weighting the small error bars found predominately in the $K$ band when performing the Bayesian likelihood calculation.

In this study, a plane-parallel approximation is used to model the atmosphere, with the pressure ranging from $10^{-3}$ to 500 bar, uniformly sampled in log-space with 100 atmospheric layers. Collision induced absorption (CIA) of H$_2$--H$_2$ and H$_2$--He \citep{abel_h2-h2, fletcher_h2-h2, abel_h2-he} is included. 

In the case of the 51 Eri b data analysis we use multiple scaling factors S$_{cal}$ to account for the inclusion of observations from different instruments. This is employed in the case of the SPHERE $Y$, $J$ and $H$ data (S$_{cal \ SPH}$) being combined with the GPI $K1$ and $K2$ band data (S$_{cal \ GPI}$). When employing data from a single instrument we simply use one scaling S$_{cal}$ factor.

\subsection{ATMO 2020}

We compare our cloudless retrievals to self-consistent radiative-convective grid models. For this we use the recently published \texttt{ATMO} 2020 set of atmosphere and evolutionary models for cool brown dwarfs and self-luminous giant exoplanets \citep{Phillips_2020_ATMO}.

The \texttt{ATMO} code is a 1D radiative-convective equilibrium model, and has been most recently described in \citet{Phillips_2020_ATMO} and \citet{Goyal_2020}. Briefly, \texttt{ATMO} defines the TP-profile of an atmosphere on a logarithmic optical depth grid with 100 model levels. The outer boundary condition in the first model level is fixed at a pressure of $10^{-5}\,$bar and is given an optical depth of $\tau\sim10^{-4}-10^{-7}$ depending on surface gravity. The inner boundary condition in the last model level is not fixed in pressure and is given an optical depth of $\tau=1000$. The model then iterates the pressure and temperature in each model level towards radiative-convective and hydrostatic equilibrium using a Newton-Raphson solver. On each iteration chemical equilibrium abundances are calculated for the current TP-profile using a Gibbs energy minimisation scheme based on that of \citet{Gordon1994}. \texttt{ATMO} also has the ability to calculate non-equilibrium chemical abundances self-consistently with the TP-profile, using kinetic networks or relaxation schemes \citep{2016_Drummond, Phillips_2020_ATMO}. Once the chemical abundances have been computed, the opacities used by \texttt{ATMO} can be obtained from pre-computed correlated-$k$ tables for individual gases \citep{2014_Amundsen}, and are combined within the code using the random overlap to obtain the total mixture opacity consistently with the pressure, temperature and abundances in each iteration \citep{Amundsen_2017}. The radiative flux is computed by solving the integral form of the radiative transfer equation in 1D plane-parallel geometry including isotropic scattering following \citet{Bueno_1995}. The convective flux is computed using mixing length theory using the same method as \citet{Gustafsson_2008}, with the adiabatic gradient computed using equation of state tables from \citet{Saumon_1995}.

This grid includes solar metallicity atmosphere models spanning $T_\mathrm{eff}=200-3000\,$K and $\log(g)=2.5-5.5$ ($g$ in units of $\mathrm{cm/s^2}$), with steps of $100\,$K for $T_\mathrm{eff}>600\,$K, $50\,$K for $T_\mathrm{eff}<600\,$K, and 0.5 in $\log(g)$. The \texttt{ATMO} 2020 model set consists of three atmosphere model grids spanning this parameter range. The first is calculated assuming chemical equilibrium, and the second and third are calculated assuming non-equilibrium chemistry with different strengths of vertical mixing. Each model in the grid is generated with the \texttt{ATMO} code and consists of a TP-profile, chemical abundance profiles, and a spectrum of the emergent flux from the top of the atmosphere, which are publicly available for download\footnote{ATMO 2020: \url{http://opendata.erc- atmo.eu}}.

\subsubsection{Sampling using Markov Chain Monte Carlo}
To calculate the best fits from the \texttt{ATMO} 2020 grid to the spectrophotometry of 51 Eri b (see section \ref{sec:51Erib}), we used a
Markov chain Monte Carlo (MCMC) method utilising the $emcee$ python package \citep{2013_Foreman-Mackey}. We generated each independent model using an interpolation to the \texttt{ATMO} 2020 grid with temperatures ranging from $200\,\rm{K}$ to $3000\,\rm{K}$ and $\log(g)$ from $2.5\,\rm{cm/s^{2}}$ to $5.5\,\rm{cm/s^{2}}$ for models assuming chemical equilibrium, and temperature ranging from $350\,\rm{K}$ to $1800\,\rm{K}$ and $\log(g)$ from $3.0\,\rm{cm/s^{2}}$ to $5.5\,\rm{cm/s^{2}}$ for models assuming non-equilibrium chemistry due to vertical mixing. The radius was constrained between $0.07\,\rm{R_{\odot}}$ ($\sim 0.7\,\rm{R_{\rm Jup}}$) and $0.2\,\rm{R_{\odot}}$  ($\sim 2\,\rm{R_{\rm Jup}}$) for both cases, using a rough estimation from the \texttt{ATMO} evolutionary tracks, given the system's age. With this grid, the MCMC was set up with 100 walkers and was executed for 500 steps. The posteriors were constructed after discarding the first 200 steps, to account for the `burn-in'. This eliminates any bias caused by the initial values supplied to the MCMC as a starting point in the parameter space. All results are reported with an uncertainty of 1$\sigma$.



\section{Results: GJ 570D}

\vspace{2cm}

\begin{figure}
\centering
\includegraphics[width=0.48\textwidth]{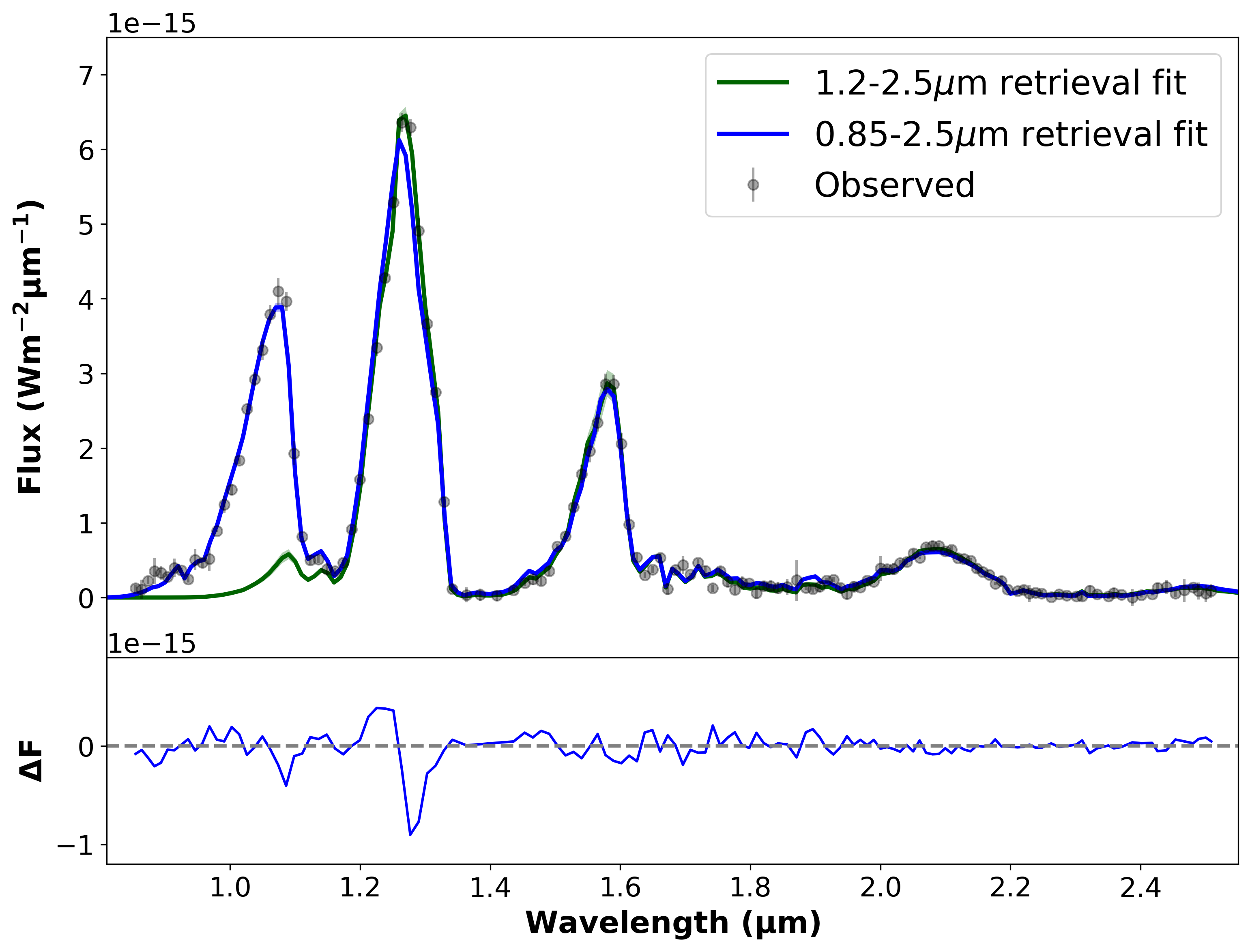}
\caption{GJ 570D retrieval spectral fit}
\label{fig:GJ570D_chem_spectral_fit}
\end{figure}

\begin{table*}
    \begin{adjustbox}{angle=90}
    \begin{threeparttable}
	\centering
	\caption{Summary of  retrieval bulk parameters for GJ 570D along with values from previous studies.}
	\label{tab:bulk_parameters}
	\begin{tabular}{lccccccr} 
		\toprule
		& Mass (M$_{\rm Jup}$) & Radius (R$_{\rm Jup}$) & log(g) (cm/s$^{2}$) & T$_{\rm eff}$ (K) & C/O & [M/H] \\
		\hline
		This work (TauREx3)  & $48.00^{+13.03}_{-11.87}$ &  $1.17^{+0.08}_{-0.08}$  & $4.93^{+0.11}_{-0.12}$ & $722^{+23}_{-26}$ & $0.87^{+0.08}_{-0.07}$  & $-0.19^{+0.05}_{-0.03}$ \\
		This work (ATMO 2020 - EC FM)  & - & $0.71^{+0.04}_{-0.02}$ & $4.64^{+0.34}_{-0.30}$ & $826.34^{+12.88}_{-17.21}$ & - & - \\
		This work (ATMO 2020 - NEC FM)  & - & $0.72^{+0.06}_{-0.03}$ & $4.63^{+0.16}_{-0.10}$ & $813.33^{+14.01}_{-27.19}$ & - & - \\
		\hline
		\citealp{Kitzmann_2020} (FCR) & $53^{+24}_{-20}$ & $1.13^{+0.05}_{-0.06}$ & $5.01^{+0.13}_{-0.19}$ & $703^{+17}_{-30}$ & $1.11^{+0.09}_{-0.09}$ & $-0.13^{+0.06}_{-0.08}$   \\
		\citealp{Kitzmann_2020} (ECR) & $17^{+3.8}_{-3.0}$ & $1.00^{+0.10}_{-0.09}$ & $4.61^{+0.08}_{-0.08}$ & $730^{+18}_{-17}$ & $0.83^{+0.09}_{-0.08}$ & $-0.15^{+0.05}_{-0.04}$\\
		\citealp{Burningham2017} (FCR) & $19.80^{+28.60}_{-15.96}$ & $0.96^{+0.80}_{-0.11}$ & $4.73^{+0.31}_{-1.17}$ & $752.25^{+35.51}_{-82.10}$ & - & -  \\
		\citealp{Line_2015} (FCR) & $30.90^{+26.64}_{-15.76}$ & $1.14^{+0.10}_{-0.09}$ & $4.76^{+0.27}_{-0.28}$ & $714.11^{+20.19}_{-23.15}$ & $1.09^{+0.16}_{-0.14}$ & $-0.25^{+0.13}_{-0.12}$ \\
		
		\hline
		
		\citealp{2020_Oreshenko}: Sonora (SML) & - & - &  $4.93^{+0.38}_{-0.55}$  & $808^{+43}_{-27}$ & - & - &\\
		\citealp{2020_Oreshenko}: AMES-cond (SML) & - & - & $5.27^{+0.43}_{-0.67}$ & $878^{+23}_{-78}$ & - & - &\\
		\citealp{2020_Oreshenko}: HELIOS (SML) & - & - & $5.08^{+0.62}_{-0.68 }$ & $800^{+14}_{-100}$ & - & - &\\
		
		\hline
		\citealp{Samland_2017} (FM) & -  &  $0.94^{+0.04}_{-0.04}$  &  $4.67^{+0.04}_{-0.04}$ &  $769^{+14}_{-13}$ & - &  - \\
		\citealp{2015_Filippazzo} (EM) & $37.28^{+24.05}_{-24.05}$ & $0.94^{+0.16}_{-0.16}$ & $4.90^{+0.50}_{-0.50}$ & $759^{+63}_{-63}$ & - & - \\
		\citealp{2009_Testi} (FM) & - & - &  5.0 & 900 & - & - &\\
		\citealp{2009_Del_Burgo} (FM) & - & - &  $4.5^{+0.5}_{-0.5}$  & $948^{+58}_{-58}$ & - & - &\\
		\citealp{Saumon2006} (EM, FM) & $42.5^{+4.5}_{-4.5}$ & $0.855^{+0.023}_{-0.023}$ & 5.09-5.23 & 800-820 & - & - \\
		\citealp{Burgasser2006} (EM) & - & - & 5.1 & 780-820 & - & - &\\
		\bottomrule
	\end{tabular}
	\begin{tablenotes}
	\item[1] EC FM is Equilibrium Chemistry Forward Model 
	\item[2] NEC FM is Non-Equilibrium Chemistry Forward Model
    \item[3] FCR is Free Chemistry Retrieval
    \item[4] ECR is Equilibrium Chemistry Model
    \item[5] SML is Supervised Machine Learning
    \item[6] EM = Evolutionary Model
    \item[7] FM = Forward Model
    \end{tablenotes}
	\label{table_bulk_parameters_GJ570D}
	\end{threeparttable}
	\end{adjustbox}
\end{table*}

\begin{table*}
	\centering
	\caption{Summary of GJ 570D retrieved molecular abundances along with a comparison to previous studies. TW = This work. }
	\label{tab:abundances}
	\begin{tabular}{lcccccccccr} 
		\hline
		& TW,0.85-2.5$\mu$m & TW, 1.2-2.5$\mu$m & \citet{Kitzmann_2020}  & \citet{Burningham2017} & \citet{Line_2015} \\
		\hline 
		
		log(H$_{2}$0)  & $-3.33^{+0.03}_{-0.03}$ & $-3.11^{+0.05}_{-0.05}$ &$-3.33^{+0.05}_{-0.06}$ &  $-3.42^{+0.16}_{-0.22}$ & $-3.40^{+0.13}_{-0.13}$ & \\
		\hline
		log(CH$_{4}$) & $-3.39^{+0.03}_{-0.03}$ & $-3.34^{+0.05}_{-0.06}$ &  $-3.28^{+0.06}_{-0.09}$ & $-3.44^{+0.20}_{-0.31}$ &    $-3.45^{+0.10}_{-0.10}$ & \\
		\hline
		log(NH$_{3}$) & $-4.58^{+0.04}_{-0.04}$ & $-4.69^{+0.07}_{-0.09}$ & $-4.38^{+0.07}_{-0.10}$ &  $-4.82^{+0.26}_{-2.47}$ &  $-4.64^{+0.15}_{-0.15}$ & \\
		\hline 
		log(CO) & $-7.66^{+3.22}_{-2.86}$ & $-7.06^{+4.12}_{-3.12}$ & $-7.70^{+2.7}_{-2.4}$ & $-7.47^{+3.05}_{-3.04}$ &  $-7.53^{+2.65}_{-3.07}$ & \\ 
		\hline
		log(CO$_{2}$) & $-8.35^{+2.50}_{-2.39}$ & $-8.58^{+2.12}_{-2.09}$ &  $-7.70^{+2.7}_{-2.4}$ &  $-7.86^{+2.67}_{-2.66}$ & $-7.76^{+2.23}_{-2.89}$ & \\ 
		\hline
		log(H$_{2}$S) & $-8.59^{+2.42}_{-2.26}$ & $-3.86^{+0.12}_{-2.26}$ &  $-8.47^{+2}_{-2}$ &  $-8.74^{+2.68}_{-2.20}$ &  $-8.94^{+2.22}_{-2.11}$ & \\
		\hline
		log(Na+K)& $-5.99^{+0.03}_{-0.03}$ & $-4.37^{+0.06}_{-0.06}$ & $-5.86^{+0.04}_{-0.03}$ &  $-5.47^{+0.09}_{-0.30}$ &  $-5.45^{+0.06}_{-0.06}$ & \\
		\hline
    	\end{tabular}
		\label{table_abundances_GJ570D}
\end{table*}	

In order to evaluate TauREx3's emission model against brown dwarf observations, we perform retrieval analysis on the Spex observations of GJ 570D. We compare our results with previous studies which employed other retrieval codes, with the aim of determining if the results were consistent with these previous studies. The results of the comparison are shown in Table \ref{table_bulk_parameters_GJ570D}, with the retrieval priors used in the analysis listed in Table \ref{tab:retrievalpriors}.

\subsection{Na+K systematic model bias}\label{sec:Na_K}

\begin{figure}
\centering
\includegraphics[width=0.48\textwidth]{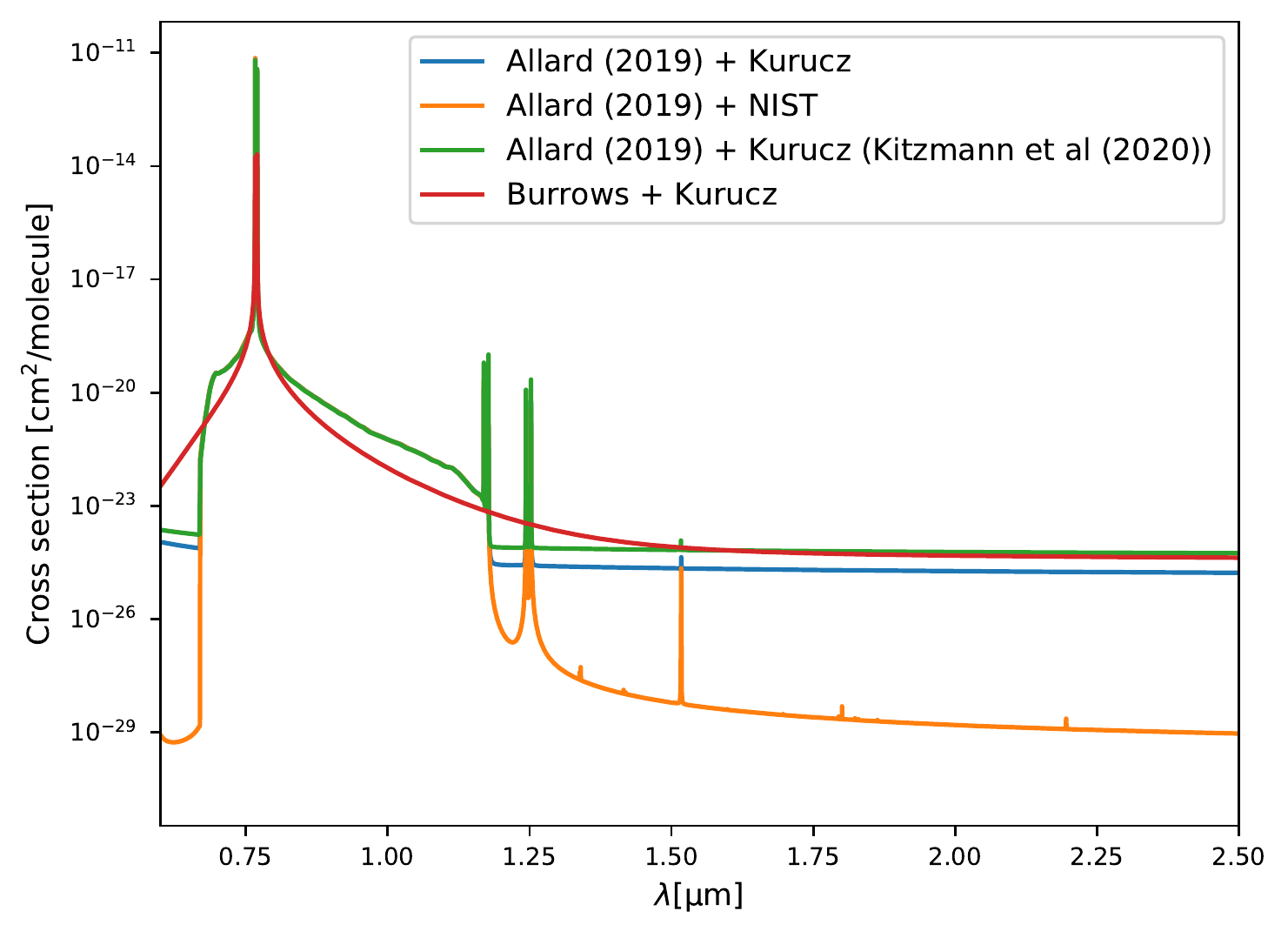}
\includegraphics[width=0.48\textwidth]{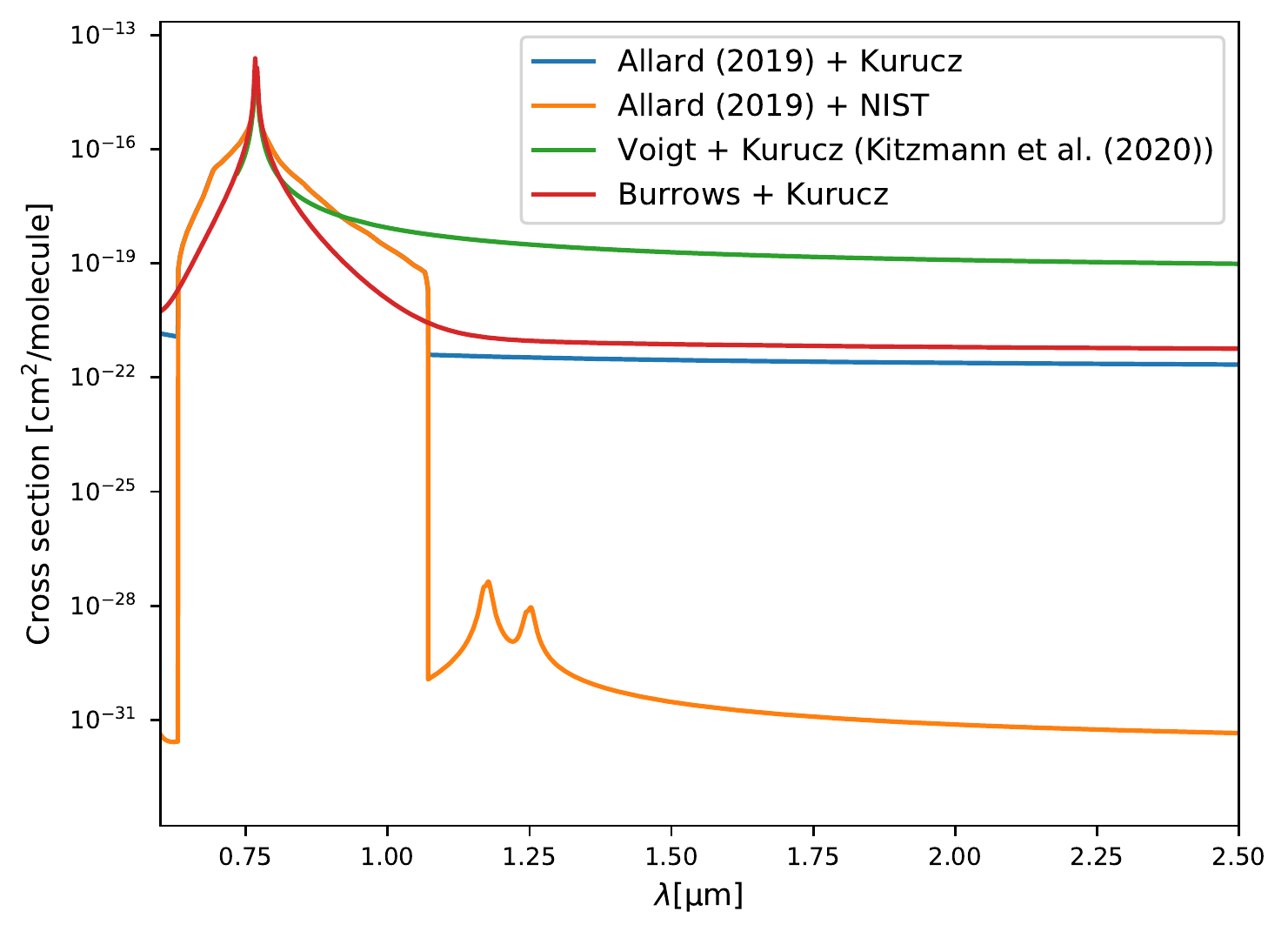}
\caption{A comparison of different methods used to compute the resonance doublet and non-resonance lines of Na and K. The first panel gives cross sections for K computed at $T~=~1000$~K, $P~=~0.1$~bar, and the second panel the cross sections for K computed at $T~=~600$~K, $P~=~10$~bar. The cross sections in green are those which were used in \citet{Kitzmann_2020}. All other combinations shown (using either Burrows et al. \citep{03BuVo.broad} or Allard et al. \citep{16AlSpKi.broad,19AlSpLe.broad} for the computation of the resonance doublets, and either NIST~\citep{NISTWebsite} or Kurucz~\citep{KURonline} for the non-resonance lines) were tested in the present study.
}
\label{fig:cross-section}
\end{figure}

We encountered a systematic bias in the retrieved estimates for mass and radius when attempting to fit the 0.85-2.5$\mu$m spectrum of GJ 570D using a flat prior. This bias resulted in a non-physical radius (see \citealp{2008_Fortney} and \citealp{2009_Chabrier_mass_radius_relationship} for typical radii) of 1.4 to 1.55 R$_{\rm Jup}$, along with a mass value that converged to the prior's upper boundary.  The mass was also found to increase with the radius, likely in an effort to maintain the best fit surface gravity. Both retrievals and grid-modelling approaches have encountered the problem of mass values converging to the upper boundary of the model prior space \citep{Line_2015, Zalesky_2019, 2015_Schneider}.

To investigate this effect, we ran retrievals with varying absolute flux calibrations and also employed the \citet{jr:Mad&Seager2009} temperature-pressure profile. Neither of these approaches negated the systematic bias. The application of a tight Gaussian prior on the radius was also tested, but in this case the mass was still seen to converge to the upper boundary of its flat prior. We find this systematic issue to be sensitive to the sodium and potassium (Na+K) cross sections, a dominating source of contribution in near-infrared model fitting as shown in \citet{Line_2015}, \citet{Burningham2017} and \citet{2020_Oreshenko}.

It is noteworthy that this issue seems most prevalent when fitting the whole 0.85 - 2.5 $\rm{\mu}$m spectrum. The resonance doublets of K and Na are at $\sim$0.77  $\mu$m and $\sim$0.59  $\mu$m respectively. We encountered examples when the bias issue would not be present when fitting only 0.85 - 1.2 $\rm{\mu}$m ($\sim$0.77 $\mu$m K / $\sim$0.59 $\mu$m Na resonance doublet impacted region), or 1.2 - 2.5 $\rm{\mu}$m (non-resonance lines/resonance doublet line wings region).  This indicates a potential issue with either the combination of the resonance doublets and non-resonance lines within the Na+K cross sections, or with the extent of the broadening of the resonance doublets. Some different combinations of computing the cross sections of the resonance doublet and non-resonance lines of K are illustrated in Figure \ref{fig:cross-section}. The resonance doublets tested in the present study were either treated using the broadening parameters of Burrows et al. \citep{03BuVo.broad} or Allard et al. \citep{16AlSpKi.broad,19AlSpLe.broad}. The non-resonance lines from both  the NIST~\citep{NISTWebsite} or Kurucz~\citep{KURonline} databases were also tested. Testing various combinations didn't negate the aforementioned bias. We again note that the results presented in this study (Tables and Figures) were retrieved using the broadening parameters of Allard et al. and non-resonance lines from the Kurucz database. The issues related to the Na and K cross sections are discussed further in Section \ref{sec:discussion}.

We therefore present two separate retrieval analyses for GJ 570D. First, to avoid the impact of this systematic bias but to still attain a set of values for the scaling factors (radius, distance and $S_{\rm cal}$) along with the mass (and by extension the inferred surface gravity) we ran a retrieval fitting only the 1.2 - 2.5 $\rm{\mu}$m part of the spectrum. This cut-off of the potassium resonance doublet impacted region of the spectrum allowed for physically credible results for the mass and radius using flat priors. We then used these values as fixed (non-fitted) priors in a subsequent retrieval to infer the chemical properties of the atmosphere. This was necessary as extending the fit of the 1.2 - 2.5 $\rm{\mu}$m retrieval to the 0.85 - 1.2 $\mu$m data showed a significant mismatch between the model fit and the observed SED in this region, as shown in Figure \ref{fig:GJ570D_chem_spectral_fit}. This two-step approach leads to the most literature consistent values for the retrieved parameters but does lead to a very slightly lower Bayesian Evidence value   (see section \ref{section:Nested_Sampling_via_Multines}) due to a slightly worse fit of the $J$ band peak. 

While this strategy did derive results consistent with previous studies it does have its limitations and imperfections. Firstly, the assumptions of flat priors while also truncating the data is not an ideal approach. The temperature pressure profile, which is fit in the second retrieval, is significantly constrained as the scaling factors, with which it is intricately linked, are fixed. The same can be said for the alkali abundance, which is strongly correlated to surface gravity. While our approach derives an alkali abundance consistent with previous studies, likely as a result of being able to make use of the alkali dominated wavelength region, we acknowledge this has been driven to an extent by our constraint on this parameter.

\subsection{Scaling factors and bulk parameters}

\begin{figure*}
\centering
\includegraphics[width=0.7\textwidth]{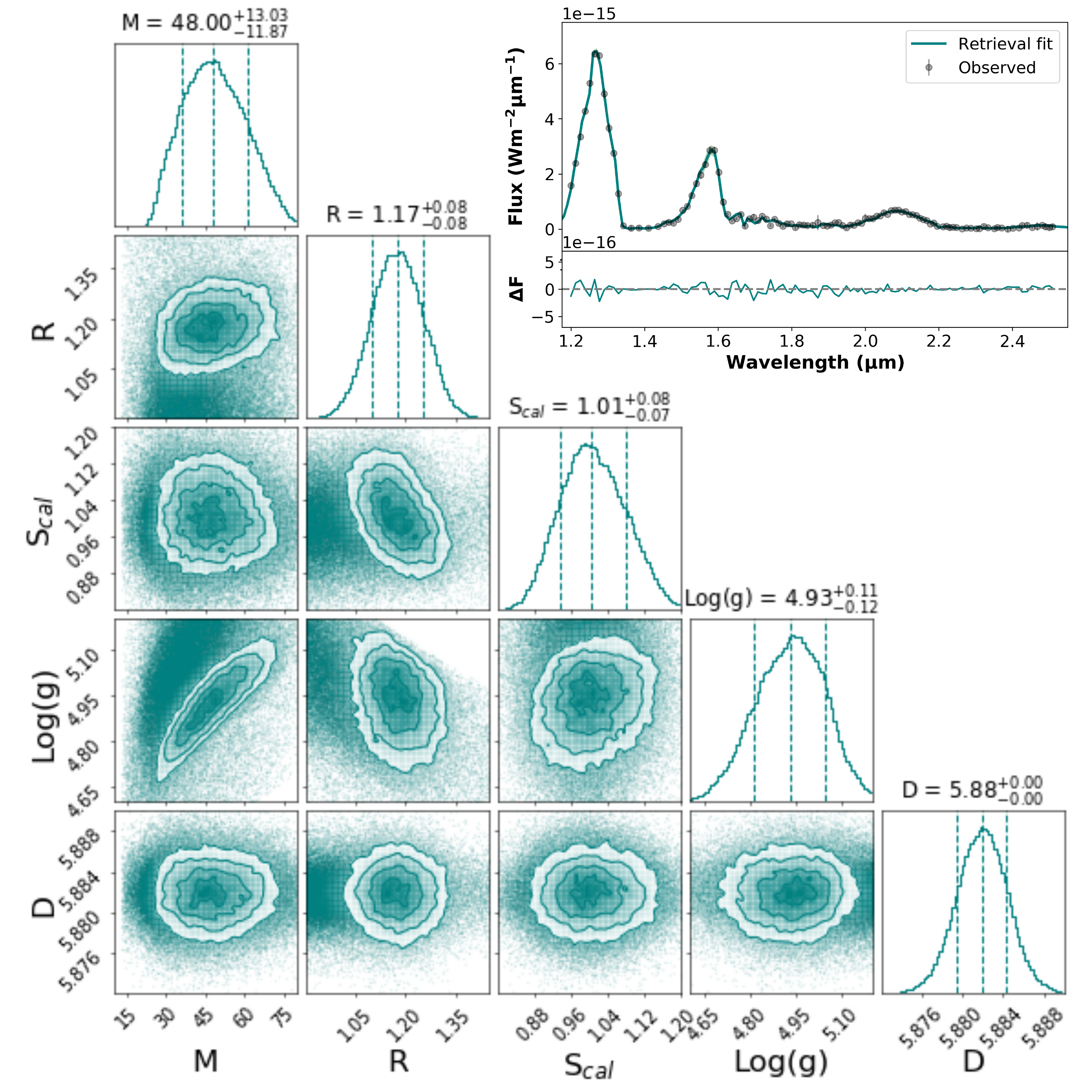}
\caption{GJ 570D bulk parameter posterior probability distributions for the spectral fit of the 1.2-2.5$\mu$m data used.}
\label{fig:bulk_posteriors_GJ570D}
\end{figure*}

The model posteriors for the mass, radius, $S_{cal}$ and distance, along with the inferred surface gravity, can see seen in Figure \ref{fig:bulk_posteriors_GJ570D}, along with the spectral fit to the data used. These results are also summarised in Table \ref{table_bulk_parameters_GJ570D}. In general, the retrieved parameter values are consistent with previous studies. Values from previous studies can also be seen in Table \ref{table_bulk_parameters_GJ570D}. Mass is consistent with all previous studies outlined in the table, apart from the equilibrium chemistry retrieval presented in \citet{Kitzmann_2020}. Radius is consistent with all previous free chemistry retrievals quoted in the table, and is 2$\sigma$ consistent with the slightly lower radii presented in the equilibrium chemistry retrieval from \citet{Kitzmann_2020} and non-retrieval analysis conducted in previous studies. As mass and radius are largely consistent with previous studies, so too is the inferred surface gravity. As the distance prior is so tightly constrained because of the precise Gaia measurements \citep{2016_Gaia, 2018_Gaia}, the distance parameter does not play a significant role in the scaling of the spectrum. Our retrieved effective temperature matches well with all previously conducted retrieval studies, whilst some other studies such as \citet{Saumon2006} and \citet{Burgasser2006} have obtained slightly higher values for this parameter.

\subsection{Abundances}

The posterior distributions for the retrieved abundances are shown in Figure \ref{fig:posteriors_chem_GJ570D}, and listed in Table \ref{table_abundances_GJ570D}. The resulting SED fit, derived combining these retrieved abundances along with the locked scaling parameters outlined previously, is shown in Figure \ref{fig:GJ570D_chem_spectral_fit}.  These show that the three most abundant molecules are H$_{2}$O, CH$_{4}$ and NH$_{3}$, whilst Na+K is also well constrained.

The abundance for Na+K that we retrieve is similar to that from \citet{Kitzmann_2020} but noticeably different from the values presented in \citet{Line_2015} and \citet{Burningham2017}, which we ascribe to the use of the broadening coefficients from \cite{16AlSpKi.broad} and \cite{19AlSpLe.broad} in our analysis and that from \citet{Kitzmann_2020}.

Overall, these abundances (and by extension the C/O and [M/H] ratio) are similar to those from previous retrieval studies of this object presented in \citet{Line_2015}, \citet{Burningham2017} and \citet{Kitzmann_2020}. Our super-solar $0.87^{+0.08}_{-0.07}$ C/O ratio for GJ 570D is in good agreement with the reported $0.65-0.97$ C/O for its host star presented in \cite{Line_2015}. Our value is slightly lower than that derived in \citet{Line_2015}'s and \citet{Kitzmann_2020}'s free chemistry retrievals, but is consistent with \cite{Kitzmann_2020}'s equilibrium chemistry model. We do note however that this comparison is imperfect, as our inferred C/O value only considers the pure gas phase and this neglects the elemental losses dues to condensation.

\subsection{Temperature-Pressure profile}

\begin{figure*}
\centering
\includegraphics[width=0.8\textwidth]{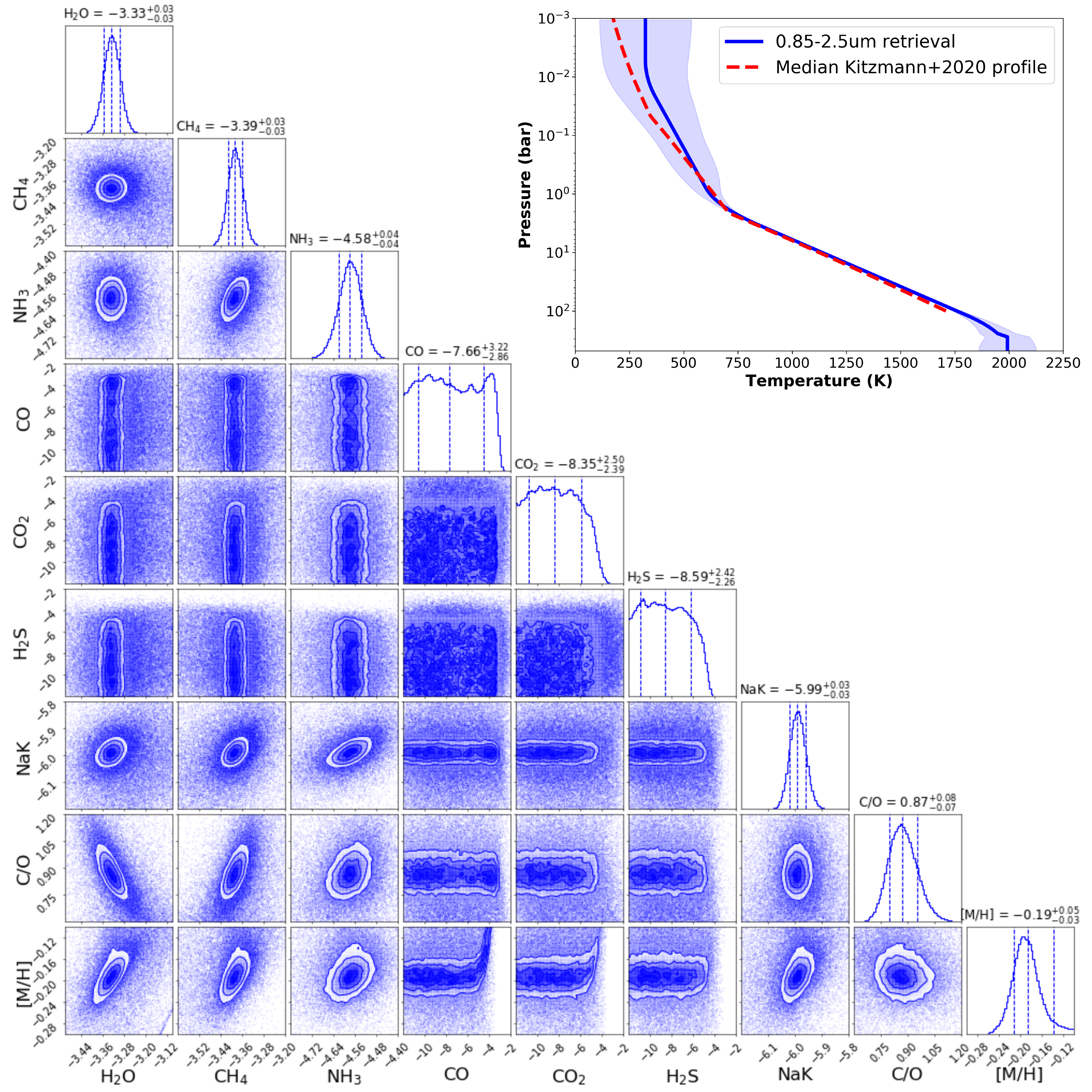}
\caption{GJ 570D mixing ratio posteriors. C/O and [M/H] posteriors are inferred parameters, while all the other parameters are sampled as part of the retrieval. The retrieved temperature-pressure profile is also shown along with a comparison to the median profile retrieved in the \citet{Kitzmann_2020} study}.
\label{fig:posteriors_chem_GJ570D}
\end{figure*}

Our retrieved temperature-pressure profile is very similar to that obtained in the \citet{Kitzmann_2020} study (see Figure \ref{fig:posteriors_chem_GJ570D} for comparison, where the blue band marks the one sigma error on our derived profile). The agreement  in the $1-10$ bar pressure region is particularly close, as expected in this region which contributes most to the spectral emission profile. We are further encouraged that this good agreement continues up into the stratospheric region where the constraining influence of the spectral emission is smaller.

\section{Results: 51 Eri b}

In this section we outline our retrieval results for 51 Eri b, compared to previous studies, all of which required clouds to produce the observed SED. Here, we find that inverse retrieval methods can recreate the observed SED with cloud-free atmospheres but using a more flexible ($npoint)$ temperature-pressure profile. These results include evidence of a tentative ammonia detection. We also compare our results for 51 Eri b to those for GJ 570 D (which is a close match in spectral type). Finally, we outline retrievals which included a power law deck or slab cloud combined with a less flexible temperature-pressure profile \citep{Lavie_2017_HELIOS}.

We ran retrievals on the SPHERE $Y$, $J$, and $H$ band data and separately on the GPI $J$ and $H$ band data.  For both retrievals, we adopted the GPI $K1$ and $K2$ data. We present comparison posteriors in Figure \ref{fig:51Erib_posteriors_combo}, with the results from the individual data sets shown in Figures \ref{fig:51Erib_GPI_posteriors} and \ref{fig:51Erib_SPHERE_GPI_posteriors}. As \citet{Samland_2017} did not fit the GPI $K1$ and $K2$ data in their study, we also present retrieval results using only the SPHERE $Y$, $J$ and $H$ band data. The posteriors for these results are presented in figure \ref{fig:51Erib_SPHERE_only_posteriors}. We do note here, however, that \citet{Samland_2017} included SPHERE and GPI photometry in their fitting which was a driving component of the high metallicity they derive.

The retrieval priors used in this analysis are presented in Table \ref{tab:retrievalpriors}, with an overall summary of the retrieval results in Tables \ref{table_bulk_51_Eri_b} and \ref{table_abundances_51_Eri_b}. The following subsections focus on the retrievals which used the $npoint$ temperature-pressure profile and omitted clouds as these derived the highest (or comparably indistinguishable) Log(Ev). We then discuss the cloudy retrievals in a subsequent subsection.

\begin{table*}

\begin{adjustbox}{angle=90}

\begin{threeparttable}

\centering
\caption{Summary of retrived bulk parameters for 51 Eri b. Values from previous studies are also included for comparison. L17: \citet{Lavie_2017_HELIOS}. GPI all: GPI $J,H,K1,K2$ data. SPHERE \& GPI: SPHERE $Y,J,H$ $\&$ GPI $K1,K2$ data. SPHERE only: SPHERE $Y,J,H$ data. U: Uniform clouds coverage. P: Patchy cloud coverage }
\label{tab:bulk_parameters}
\begin{tabular}{lcccccccccccr} 
		\toprule
		& TP Profile Type & Log(Ev) & Mass (M$_{\rm Jup})$ & Radius (R$_{\rm Jup}$) & log(g) & T$_{\mathrm{eff}}$ (K) & C/O & [M/H] \\
		\hline
        TW, Cloudless (GPI $J,H,K1,K2$ data) & $npoint$, flexible & 5596.02 & $8.50^{+2.86}_{-3.44}$ & $1.09^{+0.11}_{-0.11}$ & $4.26^{+0.16}_{-0.23}$ & $769^{+37}_{-41}$ & $0.92^{+0.19}_{-0.27}$ & $-0.26^{+0.66}_{-0.18}$ \\
        TW, Cloudless (SPHERE $Y,J,H$ $\&$ GPI $K1,K2$ data) & $npoint$, flexible & 5141.33 & $7.93^{+3.22}_{-3.54}$ & $1.18^{+0.12}_{-0.12}$	& $4.16^{+0.18}_{-0.26}$ & $700^{+42}_{-45}$ & $0.97^{+0.09}_{-0.20}$ & $-0.04^{+0.95}_{-0.49}$ \\
        TW, Cloudless (SPHERE $Y,J,H$ data) & $npoint$, flexible & 2330.45 & $8.25^{+3.01}_{-3.33}$ & $1.31^{+0.12}_{-0.11}$ & $4.09^{+0.15}_{-0.23}$ & $909^{+37}_{-50}$ & $0.40^{+0.26}_{-0.15}$ & $-0.66^{+0.14}_{-0.11}$ \\

  

		
		\hline
		\citealp{Nielsen_2019_GPI} & & & $2.6^{+0.3}_{-0.3}$ & - & - & - & - & - \\
		\citealp{Samland_2017} (PTC-uniform clouds)\tnote{1}  & & & $9.1^{+4.9}_{-3.3}$ & $1.11^{+0.16}_{-0.14}$ & $4.26^{+0.25}_{-0.25}$  & $760^{+20}_{-20}$ & - & -  \\
		\citealp{Samland_2017}  (PTC-patchy clouds)\tnote{1} & & & $14.5^{+8.7}_{-5.6}$ & $1.11^{+0.16}_{-0.14}$ & $4.47^{+0.24}_{-0.26}$ & $757^{+24}_{-24}$ & -  & - \\
		\citealp{Samland_2017}  (PTC-clear)\tnote{1} & & & $14.5^{+4.7}_{-3.1}$ & $0.40^{+0.02}_{-0.02}$ & $5.35^{+0.15}_{-0.12}$ & $982^{+18}_{-15}$ & -  & - \\
		\citealp{Samland_2017}  (\citealp{jr:2012_Morley} clouds) & & &  $64.9^{+19.1}_{-15.6}$ & $1.01^{+0.07}_{-0.06}$ & $5.19^{+0.10}_{-0.11}$ & $684^{+16}_{-20}$ & - & $1.03^{+0.10}_{-0.11}$\tnote{2} \\
		\citealp{Rajan_2017} (Iron-silicate, patchy clouds)  & & & - & $0.68^{+0.13}_{-0.14}$ & 3.25 & $737^{+39}_{-46}$ & - & -  \\
		\citealp{Rajan_2017} (Sulfide, salt, uniform clouds) & & & - & $0.90^{+0.23}_{-0.26}$ & $4.05^{+0.36}_{-0.35}$ & $605^{+61}_{-66}$ & - & - \\
		\citealp{Macintosh2015} (cloud-free) & & & 67 & 0.76 & 5.5 & 750 & -  & - \\
		\citealp{Macintosh2015} (partial-cloud) & & & 2 & 1 & 3.5 & 700 & - & -\\
		\bottomrule
        \end{tabular}
        \label{table_bulk_51_Eri_b}

\end{threeparttable}

\end{adjustbox}
\end{table*}

\begin{table*}
    \begin{threeparttable}
	\centering
	\caption{Summary of 51 Eri b retrieved molecular abundances , C/O ratio and metallicity [M/H] with a comparison to our retrieved values for GJ 570D. These are the values from the highest Log(Ev) (cloudlesss, $npoint$ TP) retrievals. }
	\label{tab:abundances}
	\begin{tabular}{lcccr} 
		\toprule
		 & 51 Eri b (1)\tnote{1} & 51 Eri b (2)\tnote{2} & GJ 570D  \\
		\hline 
		
		log(H$_{2}$0) & $-3.52^{+0.16}_{-0.16}$ & $-3.50^{+0.16}_{-0.19}$  & $-3.33^{+0.03}_{-0.03}$ \\
	
		log(CH$_{4}$) & $-3.63^{+0.12}_{-0.13}$ & $-3.60^{+0.09}_{-0.11}$ & $-3.39^{+0.03}_{-0.03}$ \\

		log(NH$_{3}$) & $-4.85^{+0.15}_{-0.18}$ & $-4.61^{+0.11}_{-0.14}$ & $-4.58^{+0.04}_{-0.04}$ \\

		log(CO) & $-3.32^{+1.13}_{-5.68}$ & $-5.10^{+2.27}_{-4.54}$ & $-7.66^{+3.22}_{-2.86}$ \\ 

		log(Na+K)& $-9.52^{+1.69}_{-1.59}$ & $-7.65^{+2.58}_{-2.83}$ & $-5.99^{+0.03}_{-0.03}$ \\
		
		C/O &  \ \: $0.97^{+0.09}_{-0.20}$ &  \ \:  $0.92^{+0.19}_{-0.27}$ & \ \: $0.87^{+0.08}_{-0.07}$ & \\
		
		[M/H] & $-0.04^{+0.95}_{-0.49}$ &  $-0.26^{+0.66}_{-0.18}$ &  $-0.19^{+0.05}_{-0.03}$ \\

		\bottomrule
    	\end{tabular}
		\label{table_abundances_51_Eri_b}
	    \begin{tablenotes}
        \item[1] 51 Eri b (1) refers to results retrieved using SPHERE $Y$, $J$, $H$ and GPI $K1$, $K2$ band data.
        \item[2] 51 Eri b (2) refers to results retrieved using GPI $J$, $H$, $K1$ and $K2$ band data
        \end{tablenotes}
		\end{threeparttable}
\end{table*}

\addtocounter{figure}{-1}
\begin{figure*}
\centering
\begin{subfigure}[b]{0.8\textwidth}
   \includegraphics[width=1\linewidth]{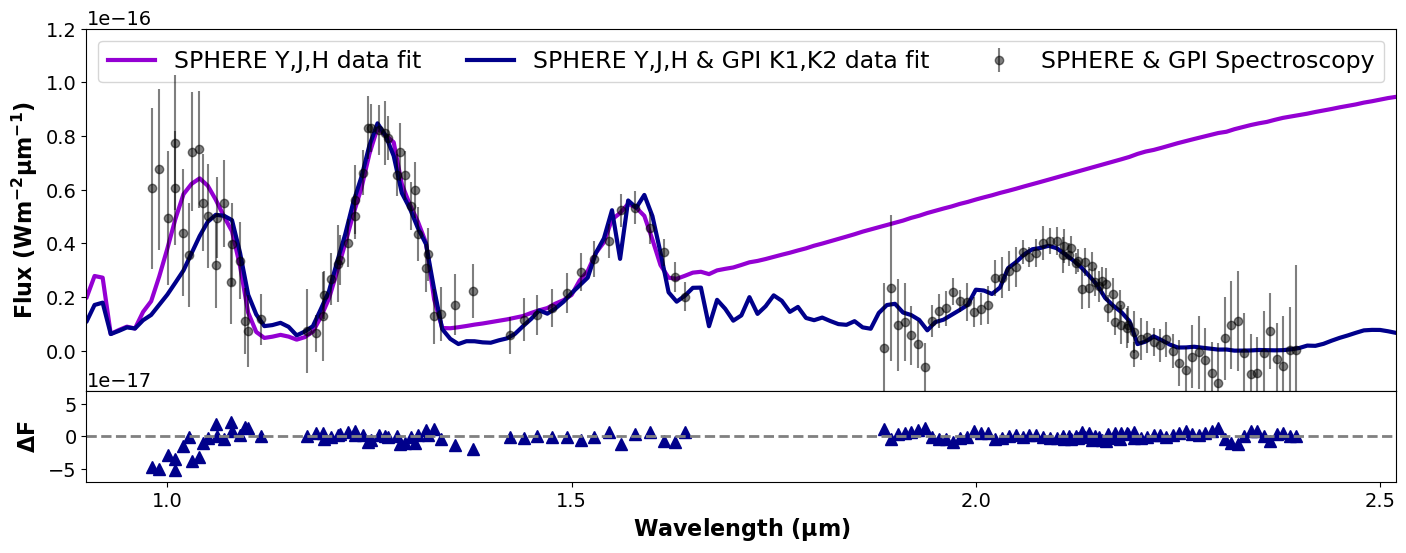}
   \caption{}
   \label{fig:spec_fit_1} 
\end{subfigure}

\begin{subfigure}[b]{0.8\textwidth}
   \includegraphics[width=1\linewidth]{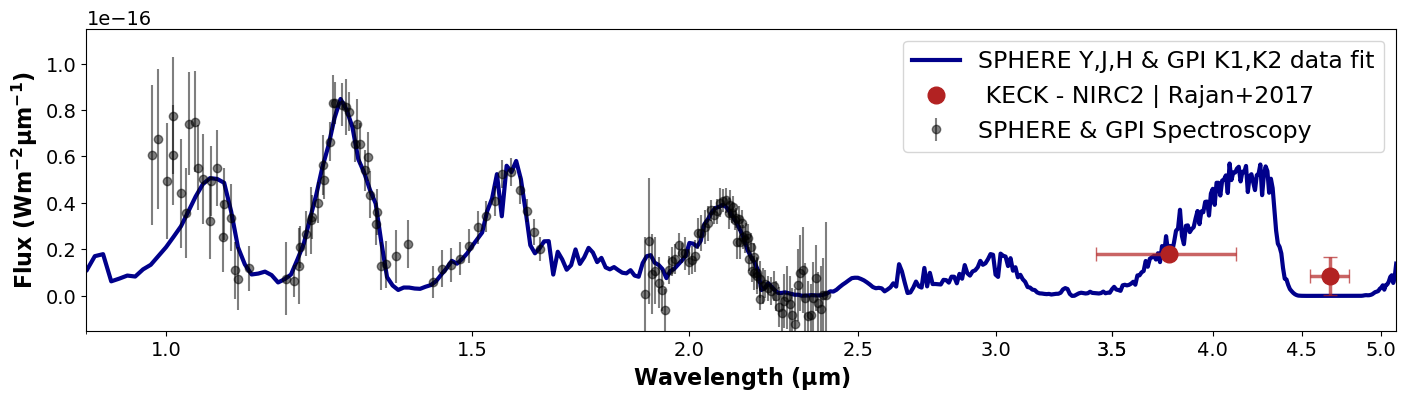}
   \caption{}
   \label{fig:spec_fit_2}
\end{subfigure}

\begin{subfigure}[b]{0.8\textwidth}
   \includegraphics[width=1\linewidth]{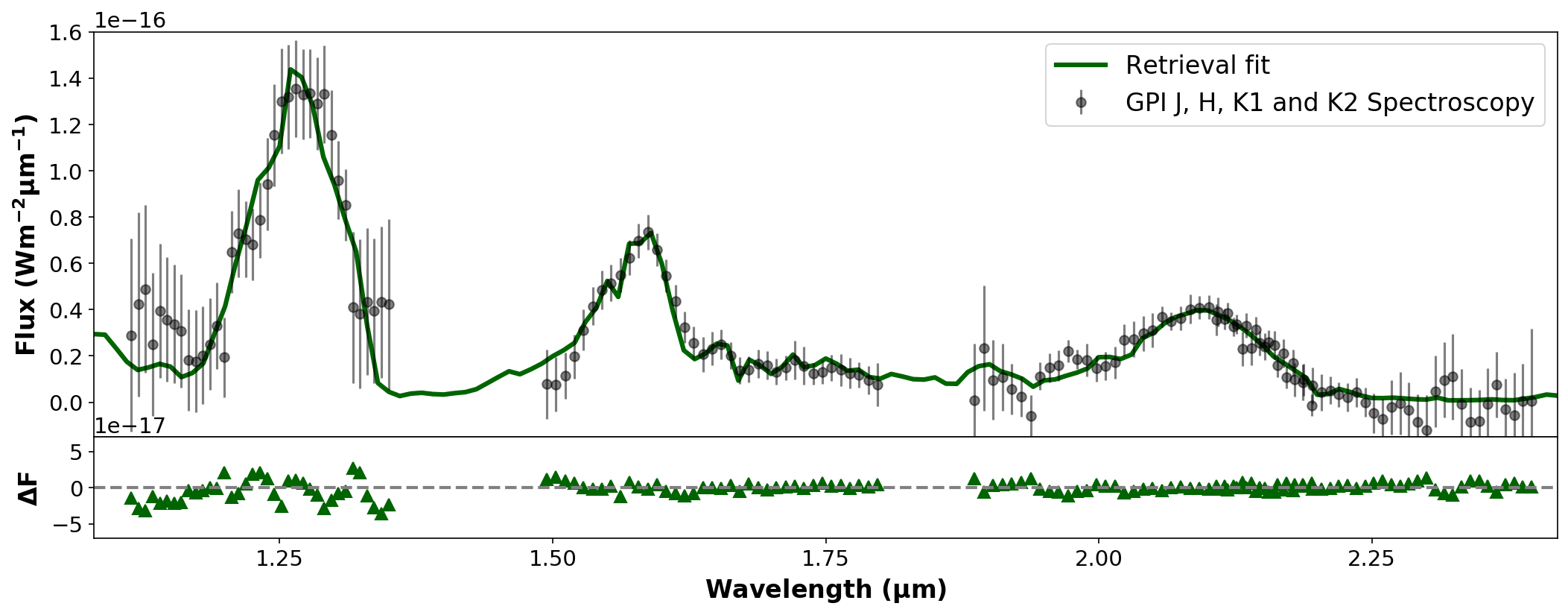}
   \caption{}
   \label{fig:spec_fit_3}
\end{subfigure}

\begin{subfigure}[b]{0.8\textwidth}
   \includegraphics[width=1\linewidth]{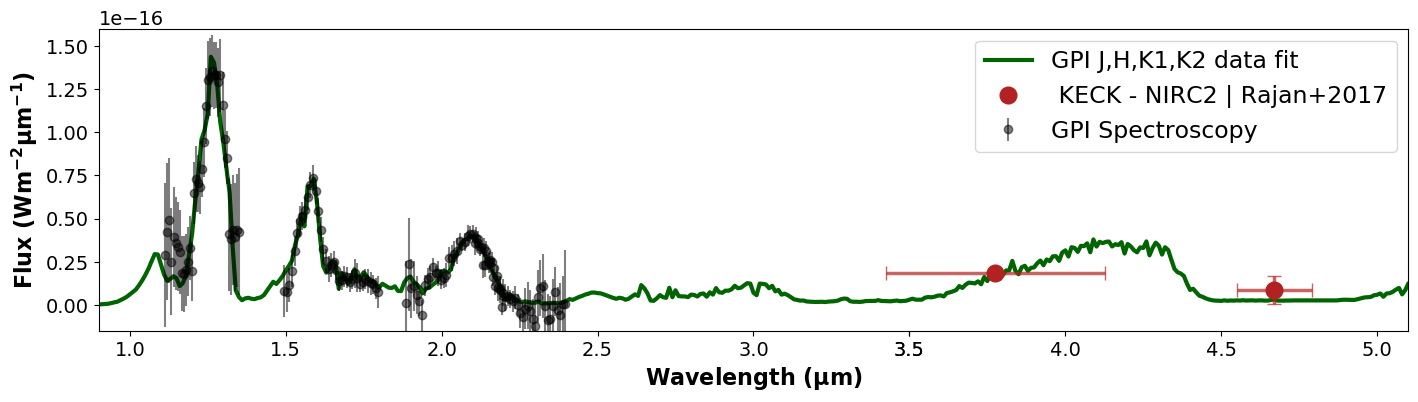}
   \caption{}
   \label{fig:spec_fit_4}
\end{subfigure}

\caption[]{51 Eri b SED fits via cloudless retrievals. (a) illustrates the model fit for retrievals including the SPHERE data where the dark violet fit shows the retrieval fit to only the SPHERE $Y$, $J$ and $K$ band data and the dark blue fit shows the retrieval fit when the SPHERE data is combined with the GPI $K1$ and $K2$ data. (b) shows the SPHERE $Y$,$J$,$H$ and GPI $K1$ and $K2$ data fit extrapolated to longer wavelengths, with the inclusion of KECK-NIRC2 photometry. (c) illustrates the model fit for the retrieval using the GPI $J$, $H$, $K1$ and $K2$ data. (b) shows the GPI $J$, $H$, $K1$, $K2$ data fit extrapolated to longer wavelengths, with the inclusion of KECK-NIRC2 photometry.}
\label{fig:spec_fit_all}
\end{figure*}

\begin{figure}
\centering
\includegraphics[width=0.48\textwidth]{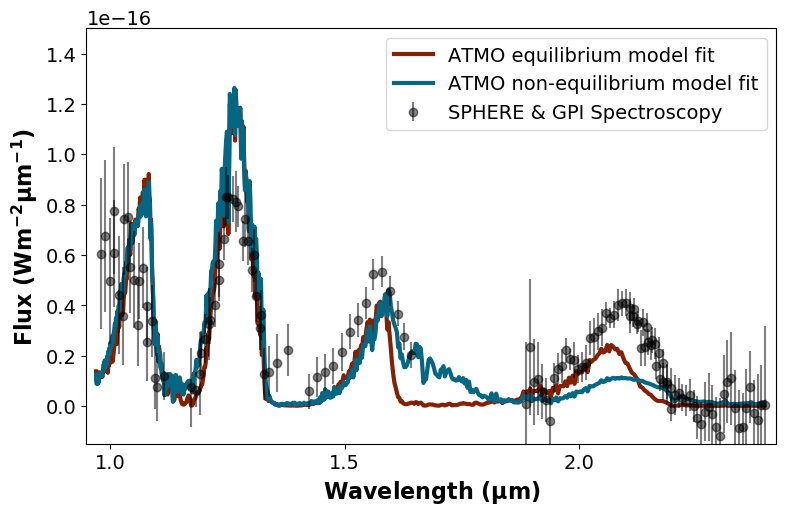}
\caption{51 Eri b spectral fit using ATMO}
\label{fig:_spectral_atmo_fits}
\end{figure}

\begin{figure*}
\centering
\includegraphics[width=1.0\textwidth]{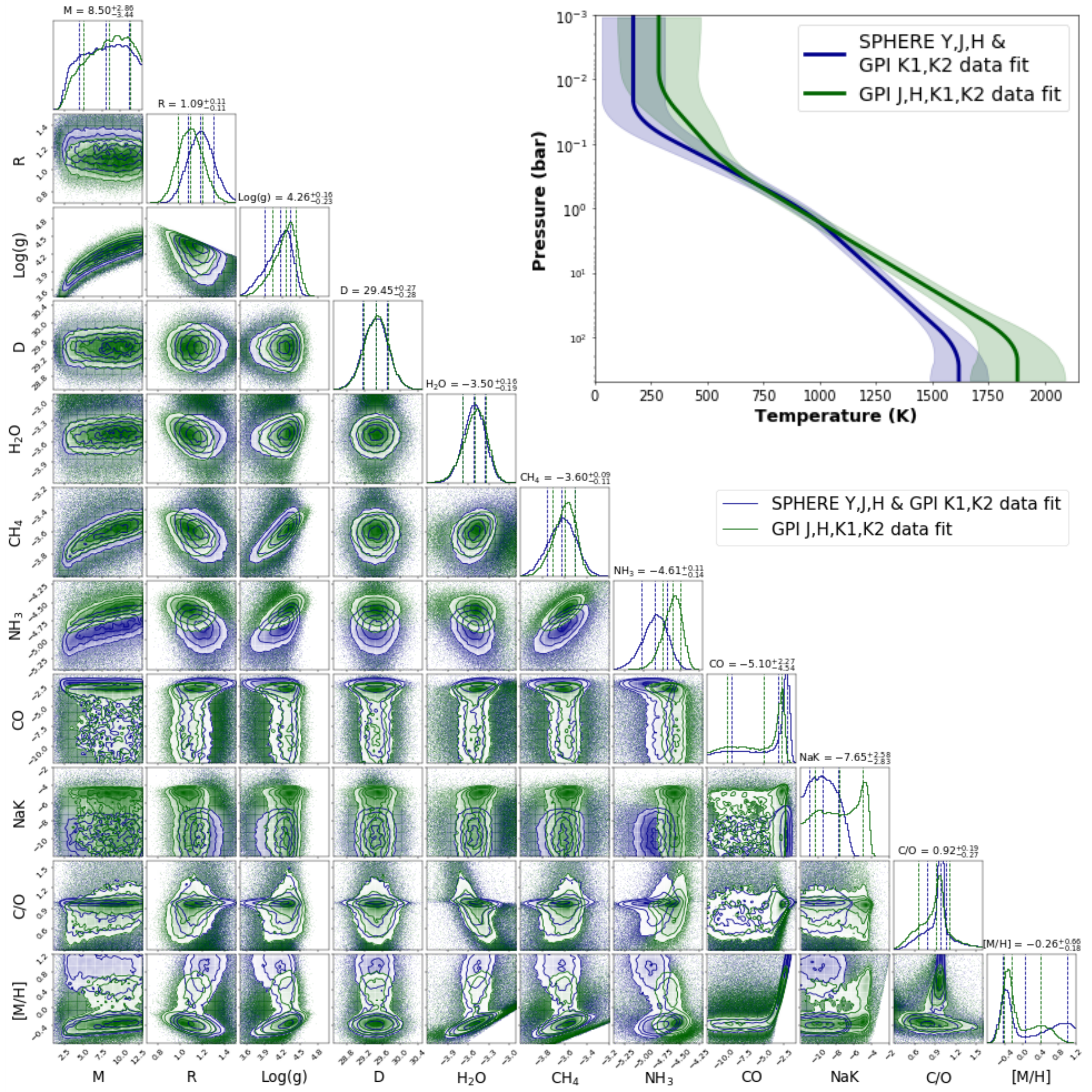}
\caption{51 Eri b posteriors. Blue indicates the retrieved values from the SPHERE $Y$,$J$,$H$ and GPI $K1$,$K2$ data set. Green indicates the retrieved values from the GPI $J$,$H$,$K1$,$K2$ data set. Log($g$), C/O and [M/H] posteriors are inferred parameters, while all the other parameters are sample as part of the retrieval. A comparison of the retrieved temperature-pressure profiles from each respective data set are also shown.}
\label{fig:51Erib_posteriors_combo}
\end{figure*}

\subsection{Scaling factors and bulk parameters}

Our highest Log(Ev) posterior probability distributions for the Mass, Radius, $S_{\rm cal}$ and Distance, along with the inferred surface gravity are presented in Figure \ref{fig:51Erib_posteriors_combo}. We find that our retrieval analysis is able to produce excellent fits to 51 Eri b's observed SED (see Figure \ref{fig:spec_fit_all}) while deriving physically credible mass and radius values. This is the case when analysing each data set as outlined previously. 

The $S_{cal}$ factors derived indicate a preference for a brighter $K$ band absolute flux calibration in both retrievals where this data is employed. In the cases of the SPHERE data being employed within the retrieval, an $S_{cal}\sim$1 is derived, indicating a model preference for this absolute flux calibration given the priors set. All the derived $S_{cal}$ values can be found in Figure \ref{fig:51Erib_GPI_posteriors}, \ref{fig:51Erib_SPHERE_GPI_posteriors} and \ref{fig:51Erib_SPHERE_only_posteriors}.

We note that the cloudless models used in the previous studies did not fit the SED particularly well. The cloudless models \citep{2008_Saumon_Marley_Evolution_models} in \citet{Macintosh2015} derived a barely sub-stellar mass of 67 M$_{\rm Jup}$ with a low radius of 0.76 R$_{\rm Jup}$ while \citet{Samland_2017}'s cloudless model \citep{2015_Molliere, 2017_Molliere} derived a mass that was 1\,$\sigma$ consistent with that of a planetary mass object, but had an improbably small radius of 0.40 R$_{\rm Jup}$ for a Jovian exoplanet, violating electron degeneracy pressure laws for an object such as this \citep{2009_Chabrier_mass_radius_relationship}. We attempted to fit the SED of 51 Eri b using a cloudless ATMO grid model as shown in Figure \ref{fig:_spectral_atmo_fits} and Figure \ref{fig:ATMO_corner_plots_51Erib}, illustrating that these grid models are unable to explain the SED of this object or to constrain its surface gravity, radius or effective temperature, using both chemical equilibrium and chemical disequilibrium assumptions as shown in Figures \ref{fig:ATMO_corner_plots_51Erib}.  

Our retrieved effective temperature values are consistent with expectations for a T dwarf except in the case of the retrieval using only the SPHERE data as longer wavelength data is neglected in this instance.  This resulting SED fit is, however, inaccurate when extrapolated to the $K$ band as shown in Figure \ref{fig:spec_fit_all}a. 

\subsection{Abundances, tentative ammonia detection}

The higest Log(Ev) posterior distributions for the retrieved abundances are shown in Figure \ref{fig:51Erib_posteriors_combo} while retrieved abundances are shown in Table \ref{table_abundances_51_Eri_b}. Comparison with GJ 570D shows that the abundances of 51 Eri b and GJ 570D match to within 1$\sigma$, not unexpected given their similar spectral types (51 Eri b: T6.5$\pm$1.5, GJ 570D: T7.5). We see that the derived [M/H] values for 51 Eri b, while consistent with GJ 570 D, have large uncertainties. This can also be seen in the large posterior tails for [M/H] shown in Figure \ref{fig:51Erib_posteriors_combo}. This appears to be a result of the large uncertainties seen in the (unconstrained) retrieval CO abundances. 

As presented in Table \ref{table_abundances_51_Eri_b}, the retrieved Na+K abundance for 51 Eri b is the only abundance which is not 1-$\sigma$ consistent with that retrieved for GJ 570D. This could either be a physical effect due to 51 Eri b's much lower surface gravity, or it could be related to the Na and K cross sections used in our retrievals. It could also be an impact of absent data below $\sim$1$\mu$m where this species plays a key role in contribution. We combine the Na and K cross sections together at solar abundance ratios, which could be an incorrect assumption for one or both of these objects. However, we found a minimal change in the retrieval results when separate Na and K cross sections (not combined at solar ratios) were used. 


We report a tentative detection of ammonia in the atmosphere of 51 Eri b. This is another example of similar characteristics between 51 Eri b and GJ 570 D. This detection is at a confidence of $\sim $ 2.7$\sigma$ ($log(b)$ = 2.36) for the data set combining SPHERE and GPI obervations, and at 2.5$\sigma$ confidence ($log(b)$ = 1.95) for the data set employing only GPI observations. This was done using a Bayes factor to sigma conversion \citep{2008_Trotta}. If verified, this would be the first detection and constraint on the presence of ammonia in a directly imaged exoplanet. This molecular species is present in planet forming, or protoplanetary disks \citep{2016_Salinas_ammonia_planet-forming_disk} and has long been included in models of substellar atmosphere \citep{AckermanMarley2001, Saumon_2012}. It is also shown to be present in Jupiter's atmosphere \citep{2020_Becker_Ammonia_Jupiter_lightning}.


We note, however, that this detection in both retrievals presented in Figure \ref{fig:51Erib_posteriors_combo} and Table \ref{table_abundances_51_Eri_b} are driven by the GPI $K1$ and $K2$ band data. We do not detect ammonia when only analysing the $Y$, $J$ and $H$ band data as shown in Figure \ref{fig:51Erib_SPHERE_only_posteriors}. Also, as noted previously, this analysis does not account for potential cross correlated noise, which could reduce the confidence of this detection.

\subsection{Temperature-Pressure profile}

The derived $npoint$ temperature-pressure profiles, retrieved with each data set are shown in Figure \ref{fig:51Erib_posteriors_combo}. Despite the differing spectral data inputs, the results are similar and are consistent at the 2-sigma level. We attribute the hotter profile between $\sim$20 to 100 bar when using only GPI data due to this having a brighter $J$ band peak compared to the SPHERE $J$ band peak as shown in Figure \ref{fig:observations_51Erib}.

\begin{figure}
\centering
\includegraphics[width=0.46\textwidth]{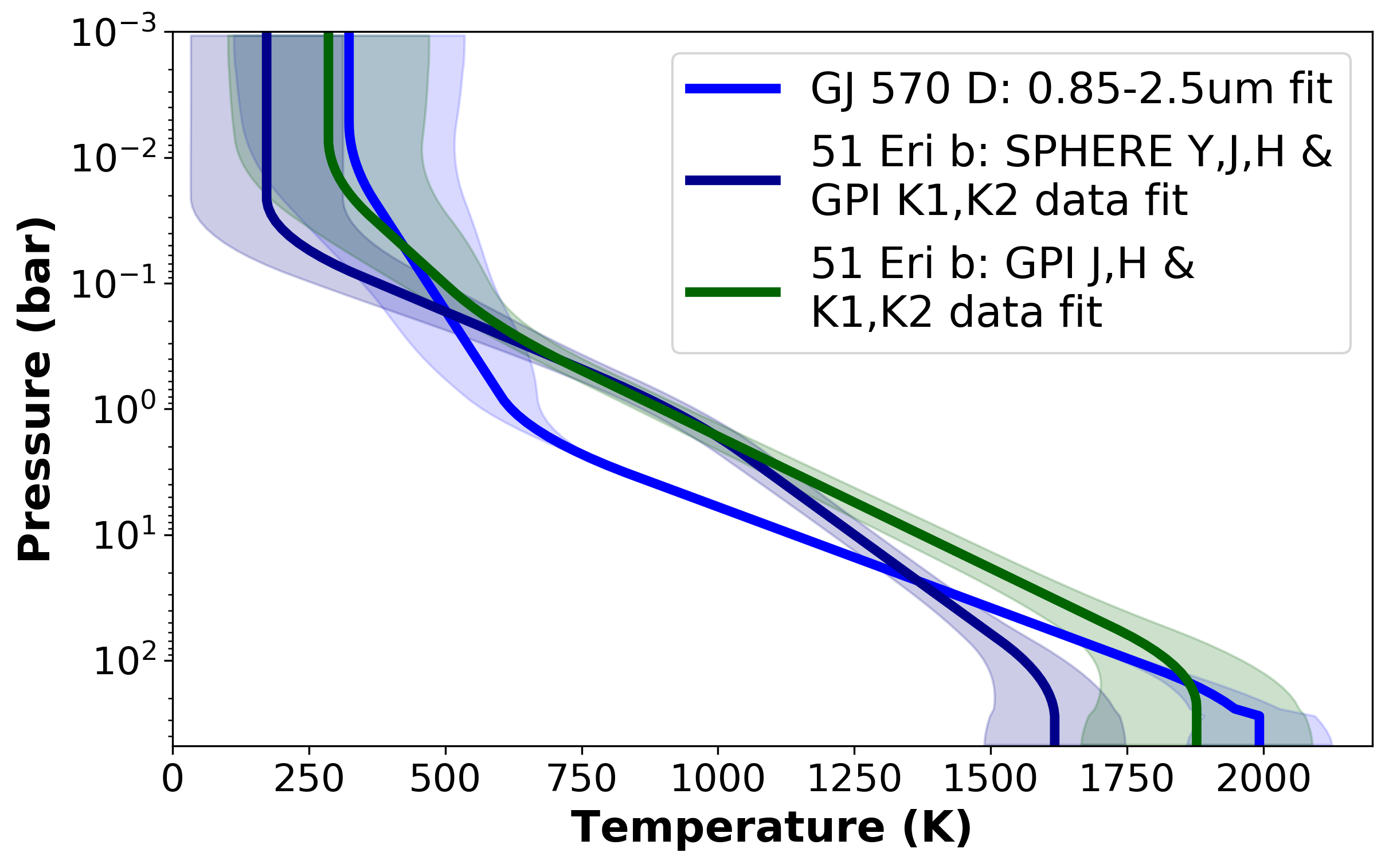}
\caption{Retrived 51 Eri b TP profile compared to the GJ 570D TP profile.}
\label{fig:TP_profile_51EribVsGJ570D_3point}
\end{figure}

We do not include the temperature-pressure profile retrieved using only the SPHERE data as this is an imperfect solution, as mentioned previously (given its inability to explain the GPI $K1$ and $K2$ data). This is a symptom of neglecting data, photometric and spectroscopic, at the longer wavelengths in the case of this particular retrieval. \citet{Samland_2017} avoided such an issue by employing photometric data points at longer wavelengths. The similarities in atmospheric properties between 51 Eri b and GJ 570D, as highlighted in the previous subsection, also encompass the temperature-pressure profile. This is shown in Figure \ref{fig:TP_profile_51EribVsGJ570D_3point}, where the temperature gradients of both objects are similar but 51 Eri b has a slightly steeper, and thus more isothermal, temperature gradient in the photosphere.

\begin{figure}
\centering
\includegraphics[width=0.46\textwidth]{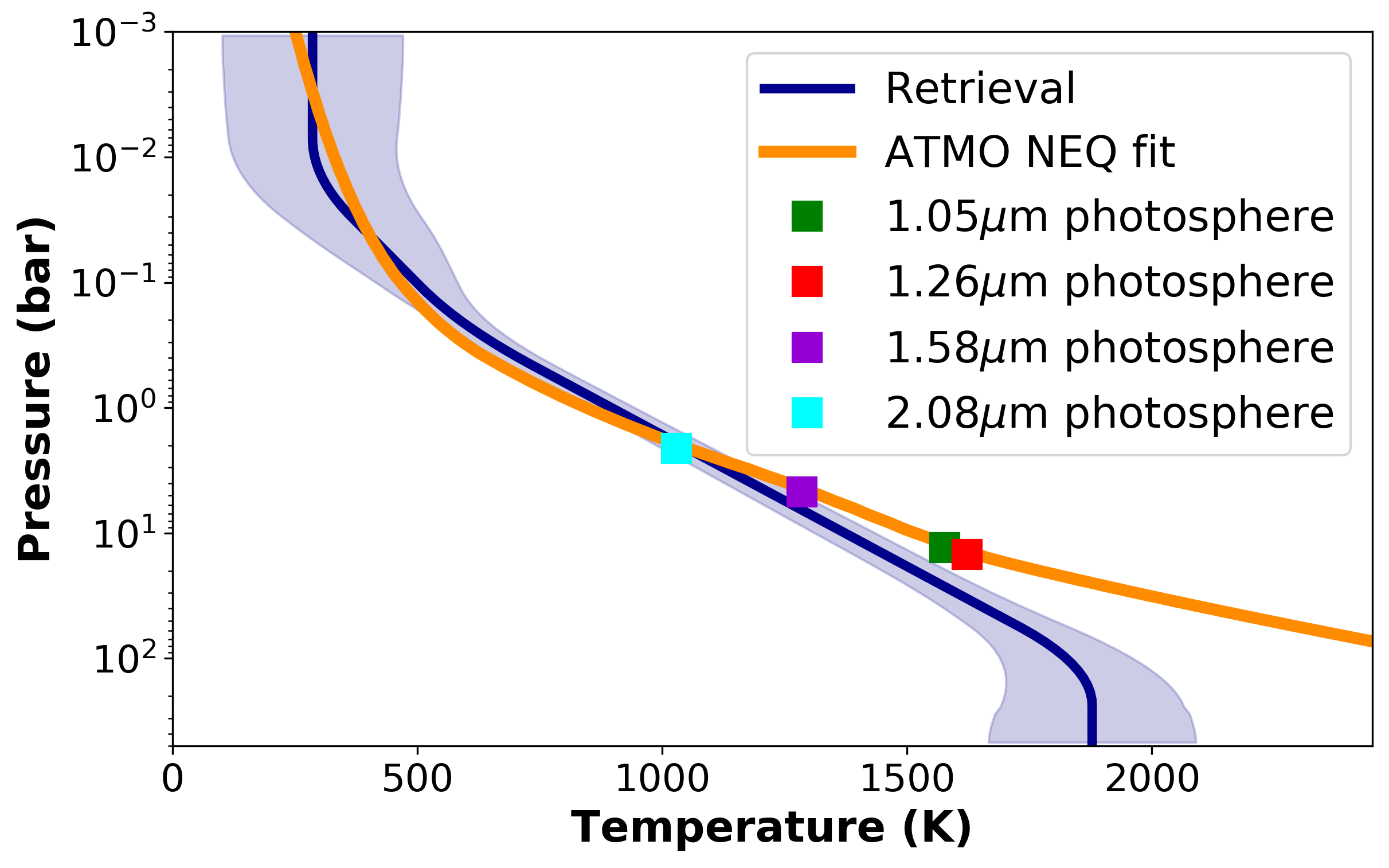}
\caption{Temperature-pressure profile comparison between profiles derived by TauREx3 and ATMO 2020 for 51 Eri b.}
\label{atmo_vs_taurex_TP}
\end{figure}

In Figure \ref{atmo_vs_taurex_TP} we show how our retrieved temperature-pressure profile differs mainly in the $Y$ and $J$-band photospheric contributions regions when compared to the radiative-convective equilibrium profile from the \texttt{ATMO} 2020 grid models. In other words, our retrieval analysis derives cooler $Y$ and $J$-band photospheric temperatures. 

The differences between profiles derived for 51 Eri b compared to the GJ 570 D retrieval (see Figure \ref{fig:TP_profile_51EribVsGJ570D_3point}) and \texttt{ATMO} 2020 fitting (see Figure \ref{atmo_vs_taurex_TP}) may indicate the presence of an unmodelled cloud as the profile departs from an adiabatic gradient and becomes more isothermal. This kind of behaviour has been noted in previous retrieval studies such as \citet{Burningham2017} and \citet{Molliere_2020} when a cloudless retrieval attempted to account for clouds included in mock data by making the profile more isothermal. As we noted above, our retrieved profile acts to produce a cooler $Y$ and $J$ photosphere and as such may be inadvertently mimicking the presence of a cloud layer. Alternatively, reduced, non-adiabatic temperature gradients triggered by chemical transitions have been suggested as an explanation for the SEDs of brown dwarfs \citep{Tremblin_2016, Tremblin_2019}. Thus, the retrieved non-adiabatic temperature profile could also be indicative of thermo-compositional convection taking place in the atmosphere of 51 Eri b.

\begin{figure}
\centering
\includegraphics[width=.46\textwidth]{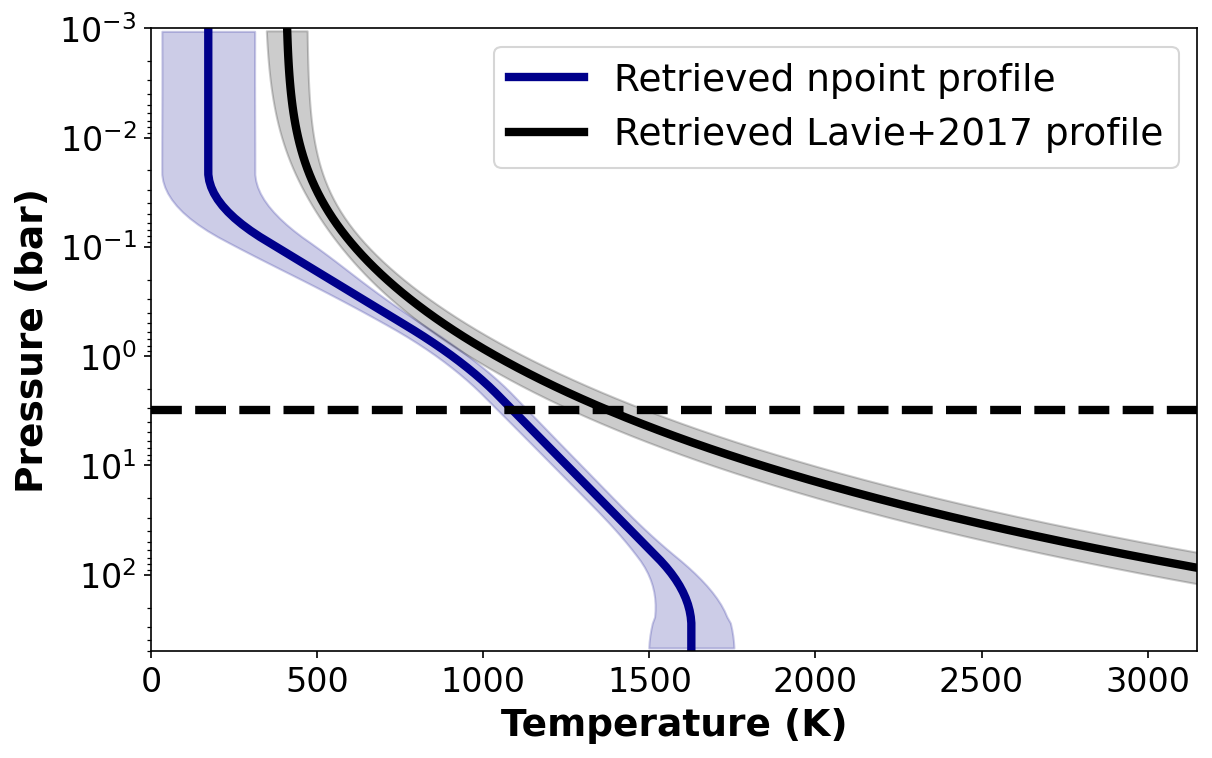}
\includegraphics[width=.46\textwidth]{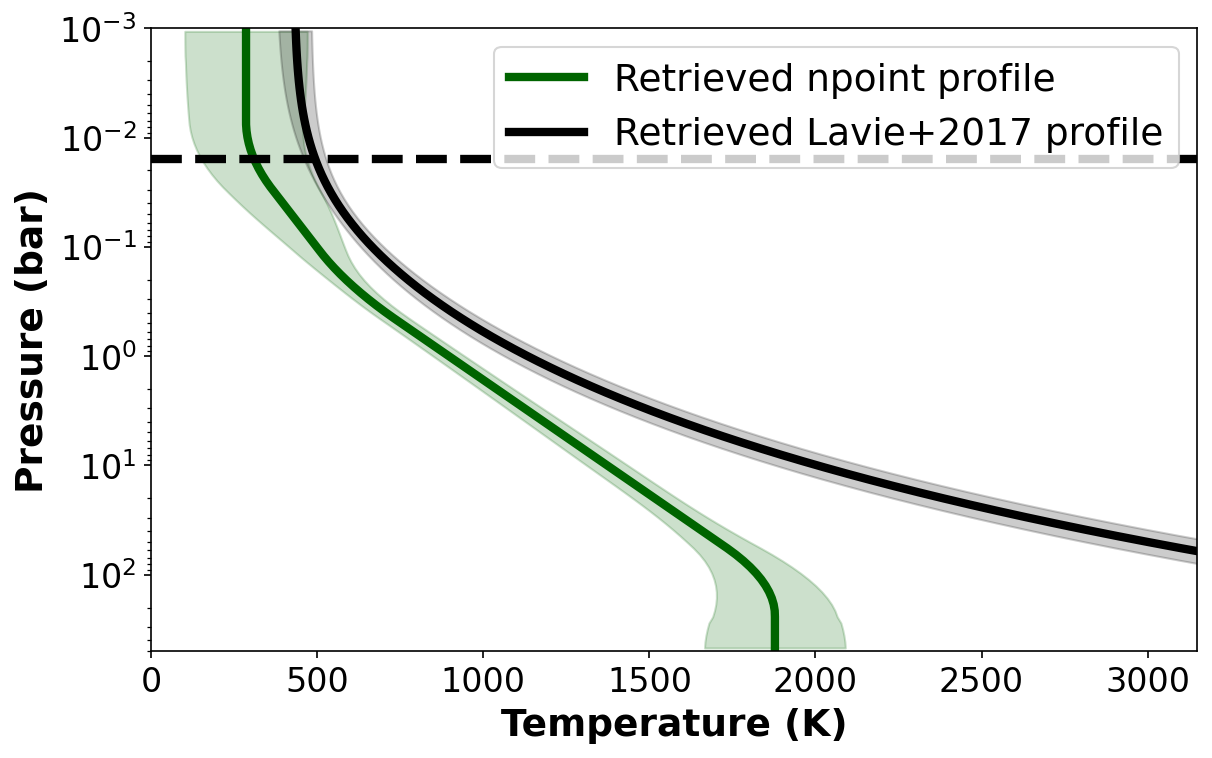}
\caption{Retrieved temperature pressure profiles using the npoint and \citet{Lavie_2017_HELIOS} prescriptions. Also indicated is the retrieved cloud position when using the \citet{Lavie_2017_HELIOS} profile. Top: Using SPHERE $Y$, $J$, $H$ and GPI $K1$, $K2$ band data. Bottom: Using GPI $J$, $H$, $K1$ and $K2$ band data. }
\label{fig:51Erib_TP_comparisons_clouds}
\end{figure}

\newpage

\subsection{The question of formation}

Parameters derived from retrieval analysis can allow us to peer into the formation history of exoplanets \citep{2020_Nowak, 2022_Molliere_formation}. Here, one can attempt to differentiate between possible formation mechanisms for 51 Eri b, primarily gravitational instability (GI) \citep{1974_Bodenheimer_GI,jr:Boss1997,jr:Durisen2007} or core accretion \citep{jr:Pollack1996,2007_Lissauer_Stevenson_FormationofGiantPlanets}. GI is a rapid mechanism that has similarities with the general star formation process.  When the system is very young, the disk may become massive enough to become gravitationally unstable, producing spiral density waves that may collapse to form bound objects, which could then slowly contract to produce planetary-mass bodies. Core accretion occurs when an initial solid core forms, then slowly accretes gas from the surrounding disk. If the mass gets high enough, it may enter a phase of \say{runaway gas accretion} where the protoplanet very rapidly gains a significant amount of gas. Overall, the timescale of core accretion is much longer than that of GI.

\begin{figure}
\centering
\includegraphics[width=0.35\textwidth]{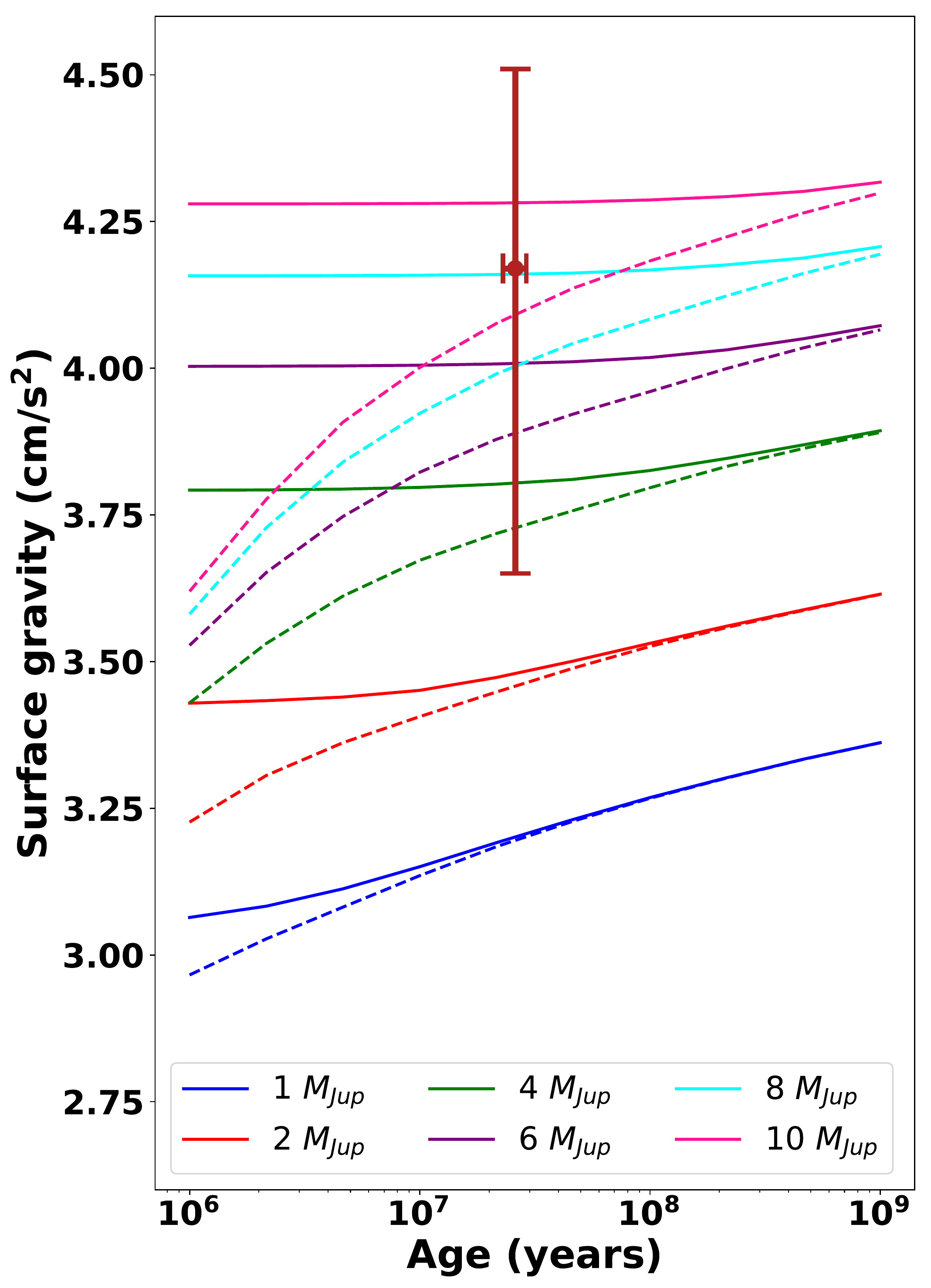}
\caption{Planetary thermal evolution tracks for different planet masses from Fortney et al. (2008), updated from Marley et al. (2007). Dotted lines indicate hot start planets. Solid lines indicate cold start planets. Purple highlighted region indicates the age of 51 Eri b, as stated in Rajan et al. (2017). The error bar indicates the retrieved 2 sigma confidence boundary for surface gravity from the SPHERE $Y$, $J$, $H$ and GPI $K1$, $K2$ data set with the age estimate from \citet{Rajan_2017}. }
\label{fig:Fortney_tracks}
\end{figure}

All our derived radii values are consistent with both the classical cold start and and hot start planetary thermal evolution models from \citet{2008_Fortney} (updated models from \citet{2007_Marley}) as outlined in Figure \ref{fig:Fortney_tracks}, using age estimates from \citet{Macintosh2015} or \citet{Rajan_2017}. Figure \ref{fig:Fortney_tracks} also shows that our derived surface gravity values are consistent with both classical cold start and hot start model predictions at the 2-sigma level. As such, with the current uncertainties derived from retrievals such as that presented in this study, we are unable to differentiate between formation pathways using these models.

However, using carbon and oxygen abundances from \citet{2017_Luck_star_abundances} (Identifier: c Eri, Carbon log $\varepsilon$ = 8.41, Oxygen log $\varepsilon$ = 8.80) we derive a C/O ratio of $\sim$0.41 for 51 Eri. Therefore, the large mass retrieved for 51 Eri b, its $\sim$13 AU separation, both coupled with a super-stellar C/O ratio could point towards formation via GI \citep[e.g.,][]{jr:Vigan2017}. A core accretion pathway would happen on a much longer timescale resulting in planetesimal enrichment \citep{2016_Mordasini}, thus lowering the initial C/O ratio \citep{2017_Espinoza}. However, \citet{jr:Ilee2017} illustrate that even GI could produce a wide range of possible atmospheric abundances and so one should interpret the C/O ratio with caution. We also note that a super-stellar C/O ratio for a T dwarf could also be due to oxygen depletion via condensate processes and the formation of clouds below the photosphere \citep{1999_Burrows_Sharp, 2006_Lodders_Chemistry_of_Low_Mass_Substellar_Objects}. Therefore, the use of inferred C/O ratio informing on possible formation pathways should be approached with caution for T dwarf exoplanets.

\subsection{Cloudy vs cloudless retrievals}

In order to perform a cloudy vs cloudless comparison we used the simpler but less inflexible temperature-pressure profile parameterisation from \citet{Lavie_2017_HELIOS}. This was combined with both the uniform and patchy power law deck and slab clouds outlined in Section \ref{cloud_description}. This analysis was motivated by the cloudy results of \citet{Samland_2017} and \citet{Rajan_2017} as well as the the well documented trend of flexible temperature-pressure structures being able to mimic spectral imprint of cloud opacity. An overview of out cloudy retrieval evidences are outlined in Table \ref{table_51_Eri_b_cloud_evs}.

When considering the use of the \citet{Lavie_2017_HELIOS} profile only, the Log(Ev) derived a strong preference for clouds to be included \ref{table_51_Eri_b_cloud_evs}, matching the results of \citet{Samland_2017} and \citet{Rajan_2017}. We also match \citet{Samland_2017} and \citet{Rajan_2017} in relation to no preference for patchy clouds vs a preference for patchy clouds, relative to the data they used when performing model fitting analysis. However, in the cases of both broadband spectroscopy combinations used in our analysis, the retrievals determined a higher or comparable Log(Ev) when clouds were omitted and the more flexible $npoint$ temperature-pressure profile was employed (see Table \ref{table_51_Eri_b_cloud_evs}). These results, however, do not mean that 51 Eri b is cloud free. In fact, many studies \citep{Burningham2017, Molliere_2020} have documented that retrievals can often use flexible profile to mimic the presence and spectral contributions of clouds, even when retrieving on synthetic data where clouds were included. As can be seen in Figure \ref{fig:51Erib_TP_comparisons_clouds}, the retrieved temperature-pressure structure is significantly different in the contributing photosphere where the cloud contribution is needed to counteract the enforcement of a more adiabatic, less isothermal and hotter profile.  \citet{Burningham_2021} has shown that longer wavelength data, particularly in the mid-IR, is crucial for accurately constraining the temperature-pressure profile when using a reasonably flexible temperature-pressure profile parameterisation. It is also very likely that the simple and flexible power law cloud parameterisations included in our analysis were not the most suitable compared to more physically motivated and condensate specific cloud modelling.


\begin{table*}

\begin{threeparttable}

\centering
\caption{Comparison of 51 Eri b retrieval Log Evidences when employing different model setups.  L17: \citet{Lavie_2017_HELIOS}. GPI all: GPI $J,H,K1,K2$ data. SPHERE \& GPI: SPHERE $Y,J,H$ $\&$ GPI $K1,K2$ data. SPHERE only: SPHERE $Y,J,H$ data. Uniform: Uniform clouds coverage. Patchy: Patchy cloud coverage }
\label{tab:bulk_parameters}
\begin{tabular}{lcccccccccccr} 
		\toprule
		Cloud Type & TP Profile Type & Log(Ev) & \\
		\hline
		
        Deck cloud, Patchy (GPI all) & L17, inflexible & 5596.29  \\
        Cloudless (GPI all) & $npoint$, flexible & 5596.02 \\
        Slab cloud, Patchy (GPI all) & L17, inflexible & 5593.38 \\
        Deck cloud, Uniform (GPI all) & L17, inflexible & 5591.43  \\
        cloudless (GPI all) & L17, inflexible & 5588.28 \\
        Slab cloud, Uniform (GPI all) & L17, inflexible & 5587.58 \\

		\hline
		Cloudless (SPHERE \& GPI) & $npoint$, flexible & 5141.33  \\
		Deck cloud, Uniform (SPHERE \& GPI) &   L17, inflexible & 5132.55  \\
		Slab cloud, Uniform (SPHERE \& GPI) &   L17, inflexible &  5130.97\\
		Deck cloud, Patchy (SPHERE \& GPI) &  L17, inflexible & 5130.05  \\
		Slab cloud, Patchy (SPHERE \& GPI) &   L17, inflexible & 5128.25 \\
		cloudless (SPHERE \& GPI) & L17, inflexible & 5105.07 \\
		\bottomrule
        \end{tabular}
        \label{table_51_Eri_b_cloud_evs}

\end{threeparttable}

\end{table*}

\section{Discussion}\label{sec:discussion}

We have presented retrieval results which are consistent with previous studies and often provide improvements relative to forward models used in non-retrieval studies. This can mainly be attributed to the increased flexibility of model parameters, especially in the free chemistry retrievals. However, consistency between retrieval studies is encouraging when taking into account the use of different samplers, temperature-pressure prescriptions and differing cross-section inputs. The success of these studies demonstrates the scope for application of these tools to both the extensive archival data and future planned observations of brown dwarfs and directly imaged giant exoplanets.

There are, of course, limitations and imperfections in our retrieval analysis as we make assumptions such as isoprofile (constant) mixing ratios, something not expected to be the case in real atmospheres. However, adding additional capabilities to existing retrieval frameworks, such as non-isoprofile mixing ratios, will certainly be probed in future work using both archival and future observation of directly-imaged exoplanets and brown dwarfs. In fact, the current quality and quantity of brown dwarf observations offer a perfect testbed for new modelling parameterisations.

Retrieval analysis is also quite computationally intensive, often requiring computing clusters to run within a reasonable time frame when compared to simply iterating over a grid of forward models. This could become increasingly problematic when significantly higher resolution and increased spectral coverage observations from JWST allow for further parameters to be probed, increasing the overall parameter space and, hence, the computational expense. Recently, however, efforts have been made to use machine learning for the model selection, showing the possibilities for significant gains in computational efficiency \citep{jr:ExoGAN}.

Based on our analysis of the GJ~570D spectrum, we found that different approaches when considering the Na+K line broadening can have a significant effect on the retrieved abundances. This seems to have knock-on effects with other retrieved parameters, such as radius and mass, seemingly in an attempt to preserve surface gravity for a larger object whilst driving up the metallicity. These parameters have been found to be degenerate in other studies such as \cite{Kitzmann_2020}. Some potential reasons for the issues caused by the Na and K cross sections are outlined below.
\begin{itemize}
    \item The profiles of \cite{16AlSpKi.broad} and \cite{19AlSpLe.broad} are only considered valid up to a H$_2$ density of 10$^{21}$~cm$^{-3}$. They therefore break-down at pressures above 10-100~bar. \cite{Kitzmann_2020} took the approach here of switching back to Voigt profiles at these high pressures. The difference in cross sections computed using these varying approaches is illustrated in Figure~\ref{fig:cross-section}. It can be seen that the divergence between the computed cross sections used in this work and those of \cite{Kitzmann_2020} are much higher at larger pressures for this reason. We would not expect the deviations at such larger pressures to have such an impact, but it is worth looking into this more in the future. 
    \item The line cores of Na and K, based on the data of \cite{16AlSpKi.broad} and \cite{19AlSpLe.broad}, are computed considering Lorentzian broadening only, with the effects of Doppler broadening  not taken into consideration. It is possible this has some effect and is worthy of further investigation.
    \item We considered the effects due to using both different sources and different line-wing cutoffs for the non-resonance lines of Na and K. We compared using lines from the NIST~\citep{NISTWebsite} and the Kurucz~\citep{KURonline} databases, and found Kurucz contains more lines for both Na and K. The effects of these various approaches for different pressures and temperatures can be seen in Figure~\ref{fig:cross-section}. The larger number of non-resonance lines in the Kurucz database leads to a larger overall opacity when pressure-broadening is taken into account, with more pronounced effects at higher pressures. However, the use of the different sources for the non-resonance data was found to have negligible effect on the retrieval results.
    \item We tried using a completely different scheme for treating the line profiles of the Na and K resonance doublets; that of \cite{03BuVo.broad}. The use of these cross sections did show some effects in terms of the retrieved parameters of Na+K abundance, radius, and mass. However, they still did not give physically plausible radius and mass values, which led us to proceed with the method of splitting the spectra into two regions, as outlined in Section~\ref{sec:Na_K}
    \item We only use H$_2$-broadening and not He-broadening for the Na and K resonance doublets. As the contribution from H$_2$-broadening is much higher than from He, this is thought to be a good approximation. However, it would be worth looking into He-broadening in the future, as outlined in \cite{20PeYuCh.broad} and \cite{17Peach.broad}.
    We did not implement an instrument profile within our analysis as used in \citet{Kitzmann_2020}. Such a profile can account for flux being spread across instrument pixels. TauREx3 does not account for such a spread when binning the higher resolution forward model to the resolution of the observation. Such an instrument profile could explain why Helios-R \citep{Kitzmann_2020} was able to fit the heavily alkali influenced $J$ band peak more successfully than TauREx3 and thus may have helped negate the bias issue we experienced in this study. This was also one of the few differences in our retrieval approach and that outlined in the \citep{Kitzmann_2020} study.
    \end{itemize}

    We could not identify the exact source of the issue causing unrealistic radius and mass values to be retrieved when the full wavelength coverage spectrum is used. However, it is apparent that the Na+K cross sections used in the retrievals can have a significant impact on the retrieved parameters. This suggests that the $Y$ band sodium and potassium lines could be driving both retrieval and grid-modelling approaches to derive masses which converge towards the upper bounds of the prior space \citep{Line_2015, Zalesky_2019, 2015_Schneider}.
    

We note there are now several studies that have used the updated broadening coefficients from \cite{16AlSpKi.broad} and \cite{19AlSpLe.broad} in analysis of T dwarf spectra. One of these studies, \citet{Kitzmann_2020}, did not encounter this issue for GJ 570D. \citet{2020_Oreshenko} negated these known issues when modeling the 0.85 - 1.2 $\mu$m region by neglecting this wavelength region in their analysis. \citet{Piette_2020} modulated their K cross sections with a multiplicative factor within their retrieval while only analysing data $>$1.1 $\mu$m. This topic warrants further investigation in the future, but we note that it may be less prevalent when studying data from JWST, which will benefit from having wider wavelength spectral coverage, down-weighting the problematic Na+K dominated region when carrying out retrieval analysis. Further studies of the broadening behaviour of Na and K lines in laboratory settings would likely prove invaluable.

More dynamical (model independent) constraints for directly imaged exoplanets and brown dwarfs will help reduce the volume of parameter space explored by the retrieval method. Such measurements have been carried out for Gl~229B \citep{2019_Brandt_Dynamical_Mass_Gl_229B}, ultracool binaries \citep{2017_Dupuy_Dynamical_Masses} and Beta Pic b \citep{2018_Snellen_BetaPicb_mass, 2019_Dupuy}, with HST monitoring campaigns also underway for cool brown dwarfs \citep{2018Dupuy_hst_prop, 2020Dupuy_hst_prop}. This would significantly improve constraints on retrieval mass priors, and may also help constrain the radius values retrieved in various studies as surface gravity plays a key role in shaping the SED. Retrieval analyses have, quite often, returned physically improbable radius values, both in the results presented here (which we attribute to the issues of the Na+K opacities) and in other studies \citep{Kitzmann_2020}. Additionally, the temperature-pressure structure would also likely be better constrained, as radius and effective temperature are inversely correlated. During this study we have seen examples of pressure-temperature profiles changing as a result of varying radii whilst maintaining a similar surface gravity, demonstrating a significant and problematic degeneracy. Dynamical and model independent mass measurements for objects in the directly-imaged regime will help constrain the parameter space significantly. Better parallax measurements, such as from Gaia, help constrain the scaling factor (radius)  further. For example, both our and the \cite{Kitzmann_2020} study benefited from better distance constraints versus that of \citet{Line_2015} and \citet{Burningham2017}. The narrowing of parameter space for these model drivers may result in the ability to better probe other, more elusive, properties, and will also reduce the computational expense of retrieval analyses.

We retrieved very similar effective temperatures and abundances for both 51 Eri b and GJ 570D. This further supports the use of brown dwarfs as proxies for the harder-to-observe cohort of planetary mass companions. Another example of a close exoplanet analogue is PSO J318.5-22, a free-floating planetary mass brown dwarf with a spectrum which closely matches those of the atmospheres of the HR 8799 planets \citep{Liu_2013_PSOJ318, Bonnefoy_2016_HR8977, Miles_2018_PSOJ318}. These free floating objects are much easier to observe and can offer a window into their characteristic counterpart exoplanets, as we can make use of the superior quality of spectral data availability for these objects. Therefore, in the same way PSO 318 has long been documented to have overlapping properties with the same spectral type HR 8799 planets, 51 Eri b also has striking chemical similarities to the benchmark T dwarf GJ 570D and other late T dwarfs from the \citet{Line_2015} and \citet{Line_2017} studies.

The atmospheric similarities between the bona fide exoplanet 51 Eri b and late-T field brown dwarfs extends to mixing ratios, most notably that of ammonia. We acknowledge, though, that such a tentative detection, motivated by the GPI $K$ band data, needs further observations to provide a higher confidence detection. This could, perhaps, be achieved using VLT-GRAVITY \citep{2017_GRAVITY}, Subaru-REACH \citep{SCExAO_2018, 2018_IRD_Subaru} or KECK-KPIC \citep{2019_KPIC}. These instruments deliver higher-resolution observations than that provided by SPHERE and GPI. This would allow us to detect more subtle features. The high-resolution data from REACH and KPIC would allow us to probe individual lines using both retrievals and cross-correlation methods \citep{2019_Brogi_Line, 2018_Hoeijmakers}.

We only employed a single scaling S$_{cal}$ factor for 51 Eri b when considering data take from a single instrument and two in the case of the SPHERE plus GPI combination. However, this may be an imperfect approach in the case of using only the GPI data as this spectrum is stitched together from different bands which can employ different data reduction pipelines and photometric calibrations. Such an approach of allowing each band to scale independently was a successfully strategy adopted in \citet{Nowak_2020_BetaPicb} when combining observations of Beta Pic b. Crucially, such flexibly appears employable when using a high quality data set, as in the case of the GRAVITY Beta Pic b data used in the \citet{Nowak_2020_BetaPicb}, with this data appearing to anchor the model and deriving a very small uncertainty for the GRAVITY data scaling factor.  Our 51 Eri b data quality from GPI data is such that we didn't find this necessary, given the large uncertainties present in the data we analyse in this study. Future studies should be able to allow for scaling factors in each band when improved data becomes available for this exoplanet. The S$_{cal}$ factor is directly correlated to the retrieved radii and can act to help the retrievals to maintain a physically sensible and higher radii instead of purely accounting for possible calibration imperfections, creating a degeneracy. This behaviour is likely exacerbated by the trend of retrievals deriving small radii \citep{Burningham_2021}. Our experience of this factor with the data sets used in this study is that it commonly acted to scale down the model (S$_{cal}$<1) which can then be counterbalanced by a higher radius, especially given the priors we applied in the case of 51 Eri b. This is why we placed a Gaussian prior on this parameter when also employing one on the radius parameter, in an effort to restrict this degeneracy and the ability for the retrieval to simply use S$_{cal}$ to retrieve our set radius prior. This will likely be a continued issue for retrievals going forward, where scaling factors designed for flux calibration and possible variability considerations could mask the documented inability of models to derive expected radii values, especially when using flat priors.

Unlike previous studies, we were able to fit the spectral profile of 51 Eri b without clouds, This is an interesting and important result as previous studies all employed cloud models within grid modelling, often based on more rigidly parameterised temperature-pressure profile assumptions (e.g. radiative-convective equilibrium) and chemistry. We acknowledge and stress, however, that our ability to fit the data with a preference for a cloudless modelling may be due to our flexible temperature-pressure profile being able to mimic and account for the presence of an unmodelled cloud. Our result matches with that from \citet{Burningham2017} and \citet{Molliere_2020}, where synthetic data of cloudy L dwarfs was successfully fit due to the use of a flexible temperature-pressure profile. \citet{Molliere_2020} also showed that when an incorrect cloud model was employed to fit synthetic cloudy data the retrieval determines a preference for a cloudless fit. This may be indicative that the power law deck and slab clouds were insufficient for analysis of this target and future work on 51 Eri b should include a more diverse set of cloud modelling. We attempted to explore a more physically motivated cloud prescription, compared to the the power law paramterisation, using the cloud parameterisation outlined in \citet{Lee_2013_HR8799b_retrieval} and employed in \citet{Lavie_2017_HELIOS}. However, the data quality for 51 Eri b is such that these cloud parameters could not be constrained to a useful extent and thus this analysis was inconclusive and omitted from this article.
The degenerate ability for a flexible temperature-pressure profile to account for clouds in the absence of any cloud modelling within a retrieval may be negated in the future by employing data across a wider wavelength range, when such data becomes available. Retrieval analysis including clouds will be further explored in further work when improved data becomes available for 51 Eri b, such as the further photometric points from JWST Program $\#$1412 which may assist in breaking model degeneracies.

Our ability to fit the 51 Eri b data without clouds may have also been assisted by the free chemistry nature of the retrieval, where grid-model are often much more constrained based on employment of coarse parameter sampling, solar abundance ratios and chemical equilibrium. In the case of abundances, for example, exoplanets have been shown to possess a variety of chemical compositions, often deviating from norms seen in our own solar system. For example, exoplanets can possess C/O ratios much higher than that present in our solar system \citep{Madhusudhan_2012_55_Cancri_e, Moses_2013}. This is further shown by the super-stellar C/O ratio we retrieved for 51 Eri b. This parameter allows us to hypothesize possible formation pathways. Due to 51 Eri b's large retrieved mass, measured orbital separation and retrieved C/O ratio, we suggest this may hint at formation via gravitational instability \citep{jr:Vigan2017}. Further spectral observations of 51 Eri b, using instruments such as GRAVITY, may help further constrain the C/O ratio and permit a more in depth analysis of possible formation scenarios for this exoplanet.

Overall, we suggest the best approach is testing and exploring a suite of free-retrieval setups along with self-consistent modelling, as performed in this study, when characterising self-luminous objects. Ideally, when further data becomes available from instruments aboard JWST, GPI2 and SPHERE+, the results derived from these different approaches should converges to agreement.

\section{Summary}

We introduce TauREx3 which we have modified to be suitable for directly imaged objects, and apply it to the benchmark brown dwarf GJ 570D and the cool exoplanet 51 Eri b.

We discuss issues with the Na+K cross sections when applied to T dwarf spectra. The retrievals converged to a high mass and radius, likely due to biases introduced by the methods used to compute these cross sections. This issue was overcome by splitting the retrieval into two parts. Part 1 retrieved the mass, radius, distance and $S_{cal}$ using the 1.2-2.5$\mu$m data, while part 2 retrieved the chemical profile of the atmosphere using the 0.85-2.5$\mu$m data. This allowed for more plausible results.

We compared our GJ 570 D results with other studies that performed retrieval analyses of this object \citep{Line_2015, Burningham2017, Kitzmann_2020}. The different analysis of GJ 570D, across various retrieval codes, shows an encouraging stability of most parameters, especially relating to the atmospheric chemistry as well as the temperature-pressure profile. We therefore successfully demonstrate TauREx3's suitability for brown dwarf emission analysis. 

We also carried out free chemistry and cloudless retrieval analyses on all published spectroscopy observations of 51 Eri b, while comparing our results to previous studies that used grid modelling. The main results of our 51 Eri b retrieval analysis are:

\begin{itemize}
  \item Our retrievals result in excellent fits to the observations without requiring cloud extinction, deriving a higher Log(Ev) when compared to retrievals including power law clouds. This is in contrast to the cloudy atmosphere conclusions made in all previous studies \citep{Macintosh2015, Samland_2017, Rajan_2017} who employed grid model fitting. However, this could be due to our flexible temperature-pressure profile being able to account for un-modelled clouds with this behaviour also being seen in \citet{Burningham2017} and \citet{Molliere_2020}.
  \item We confirm and constrain the presence of H$_{2}$O and CH$_{4}$.
  \item We find tentative evidence of NH3 in the atmosphere of 51 Eri b, to a $\sim$2.7 sigma confidence. Further observations are required to confirm this.
  \item We retrieve a super-solar C/O ratio, and a solar consistent [M/H] for 51 Eri b.
  \item Our surface gravity values are consistent with both classical hot-start and cold-start planetary thermal evolution models from \citet{2008_Fortney}.
  \item We demonstrate the importance of the $K$-band observations for constraining the effective temperature and temperature-pressure profile.
  \item Our highest Log(Ev) retrieval literature consistent radius values of $1.18^{+0.12}_{-0.12}$R$_{\rm Jup}$ and $1.09^{+0.11}_{-0.11}$R$_{\rm Jup}$ for our two data sets. This is despite not employing cloud modelling, something previous studies struggled to do. 
  \item Our analysis highlights strong similarities between the retrieved molecular mixing ratios and temperature-pressure profiles of 51 Eri b and GJ 570D. The slight gradient differences in temperature-pressure profiles is attributed to possibly accounting for an un-modelled cloud structure in the case of 51 Eri b's retrieval by adopting a more isothermal gradient. 
  \item Our retrieved $npoint$ temperature-pressure profiles for 51 Eri b adopts a much more isothermal profile compared to the adiabatic profile employed in the unsuccessful \texttt{ATMO} 2020 grid model fit. This more isothermal profile, again, could account for the impact of an unmodelled photospheric cloud structure, or alternatively could be indicative of diabatic convection triggered by the $\mathrm{CO/CH_4}$ chemical transition \citep{Tremblin_2016, Tremblin_2019} in the atmosphere of 51 Eri b.
  \item Our retrieved super-stellar C/O ratio, coupled with our retrieved mass and previously measured orbital separation, hints at a possible formation pathway of gravitational instability for 51 Eri b. However, this conclusion is tentative and higher quality data is required for a more thorough analysis of the possible formation history of 51 Eri b.
  When including clouds along with a less flexible temperature-pressure profile, our retrievals derived a strong preference for the inclusion of clouds. However, the Log(Ev) never surpassed that of the cloudless retrieval when employing the more flexible $npoint$ temperature-pressure profile.
  Table \ref{table_51_Eri_b_cloud_evs} shows that the conclusions of our retrievals, in relation to cloudy vs cloud-free and patchy clouds vs uniform clouds, can be biased by model setup and which data are employed.  This points to the need for diverse and rigours testing of retrieval model setups and data combinations, particularly when analysing high-contract imaging data with challenging SNRs.
  
\end{itemize}

\section*{Acknowledgements}

NW thanks the referee of this paper for providing very constructive and insightful feedback, helping to improve this resulting manuscript. NW thanks Ben Burningham for the discussion and assistance with implementing cloud models.  NW thanks Jackie Faherty for comments provided on this article. NW also thanks the BDNYC retrieval group for useful discussions regarding this work. NW acknowledgements postgraduate funding from the UK's Science and Technology Facilities Council (STFC). IW acknowledgements funding from the UK's Science and Technology Facilities Council (STFC) and from the European Research Council (ERC) under the European Union's Horizon 2020 research and innovation programme (grant agreement No 758892, ExoAI). Furthermore, IW acknowledges funding by the Science and Technology Funding Council (STFC) grants: ST/K502406/1, ST/P000282/1, ST/P002153/1 and ST/S002634/1. NS acknowledges funding from the OCAV-PSL project. The authors wish to acknowledge the very significant cultural role and reverence that the summit of Mauna Kea has always had within the indigenous Hawaiian community. We are most fortunate to have the opportunity to use observations conducted from this mountain. Some of the data used herein were obtained at the W. M. Keck Observatory, which is operated as a scientific partnership among the California Institute of Technology, the University of California and the National Aeronautics and Space Administration. The Observatory was made possible by the generous financial support of the W. M. Keck Foundation. This publication makes use of data products from the Two Micron All Sky Survey, which is a joint project of the University of Massachusetts and the Infrared Processing and Analysis Center/California Institute of Technology, funded by the National Aeronautics and Space Administration and the National Science Foundation. This work used the Cirrus UK National Tier-2 HPC Service at EPCC (http://www.cirrus.ac.uk) funded by the University of Edinburgh and EPSRC (EP/P020267/1)

\section*{Data Availability}

GJ 570D data can be accessed via the SpeX Prism Library (see footnote 1). Data used for 51 Eridani can be accessed via supplementary materials from \citet{Rajan_2017} and \citet{Samland_2017}.




\bibliographystyle{mnras}                      

\setlength{\bibsep}{0pt plus 0.1ex}
\scriptsize\bibliography{books,journal}       



\appendix

\section{APPENDIX FIGURES}

Here we present our posterior distributions for the \texttt{ATMO} 2020 grid modelling fits of properties of GJ 570D and 51 Eri b. We also present the individual posterior distributions for our retrieval analysis of different datasets.

\addtocounter{figure}{-1}
\begin{figure*}
    \centering
    \begin{subfigure}[m]{0.5\textwidth}
        \centering
        \includegraphics[height=3in]{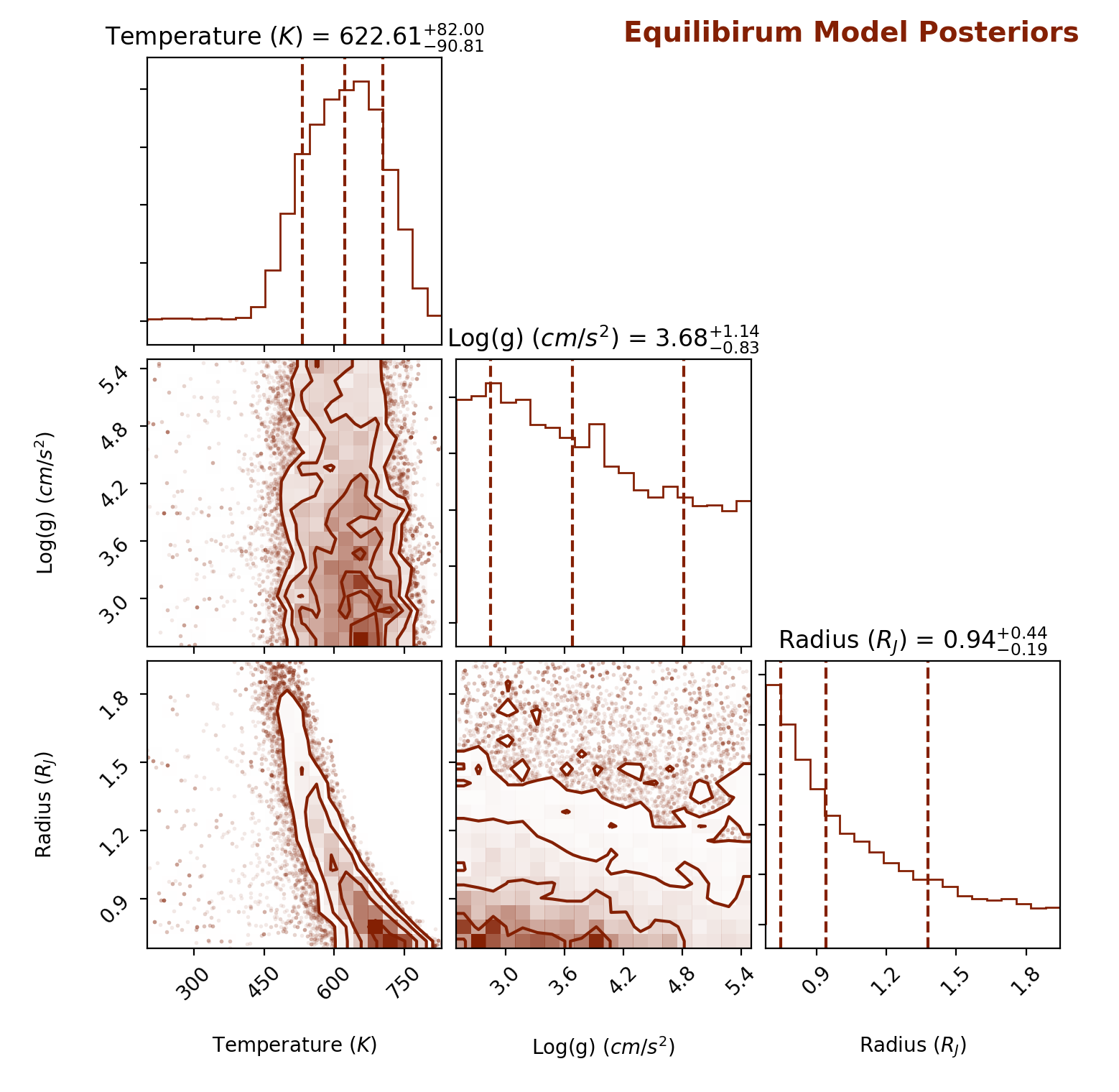}
        \caption{(a)}
    \end{subfigure}%
    ~ 
    \begin{subfigure}[m]{0.5\textwidth}
        \centering
        \includegraphics[height=3in,]{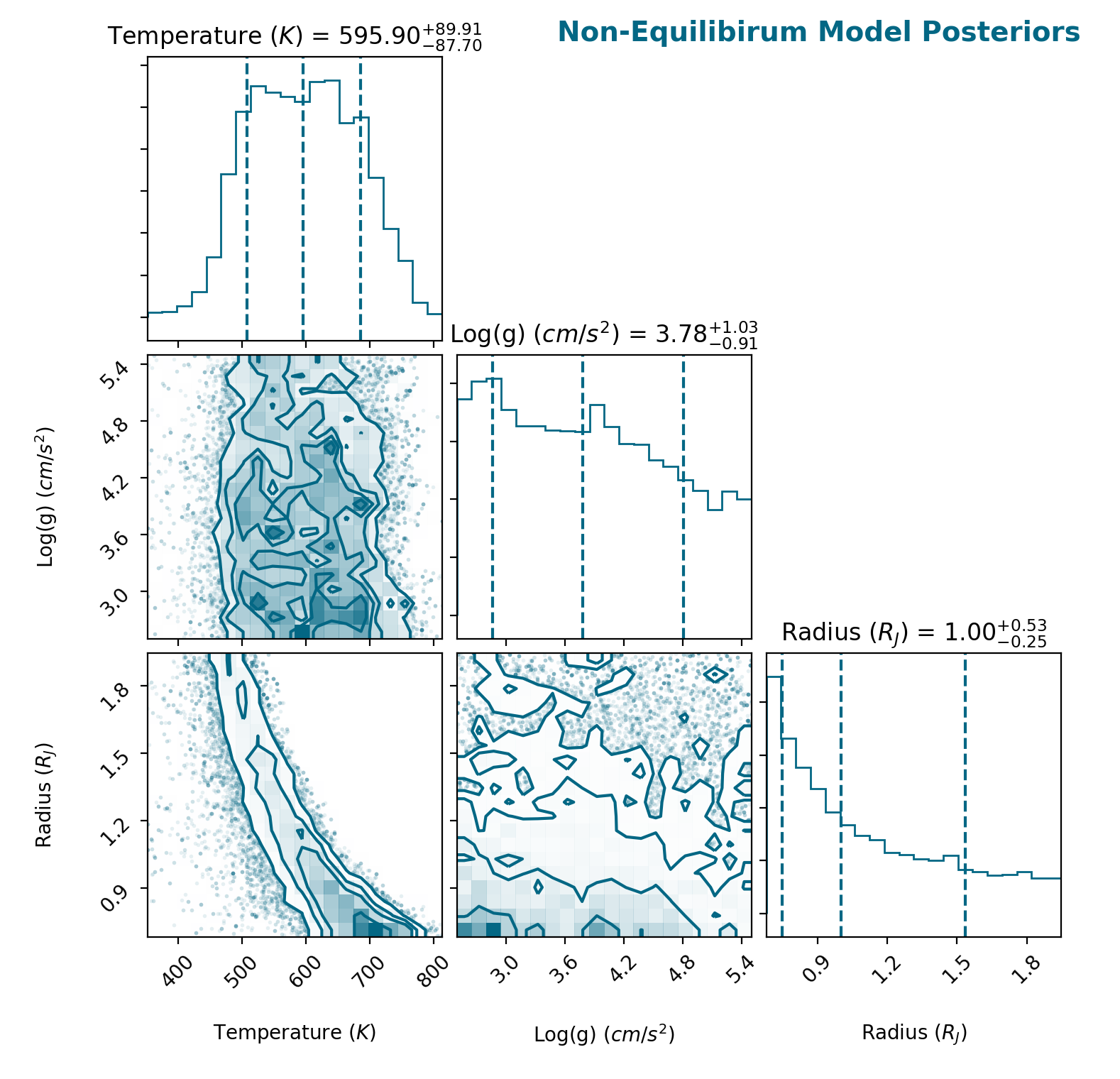}
        \caption{(b)}
    \end{subfigure}
    \caption{ATMO posteriors plots for 51 Eri b. Left: Equilibrium chemistry model. Right: Non-equilibrium chemistry model.}
\label{fig:ATMO_corner_plots_51Erib}
\end{figure*}

\addtocounter{figure}{-1}
\begin{figure*}
    \centering
    \begin{subfigure}[m]{0.5\textwidth}
        \centering
        \includegraphics[height=3in]{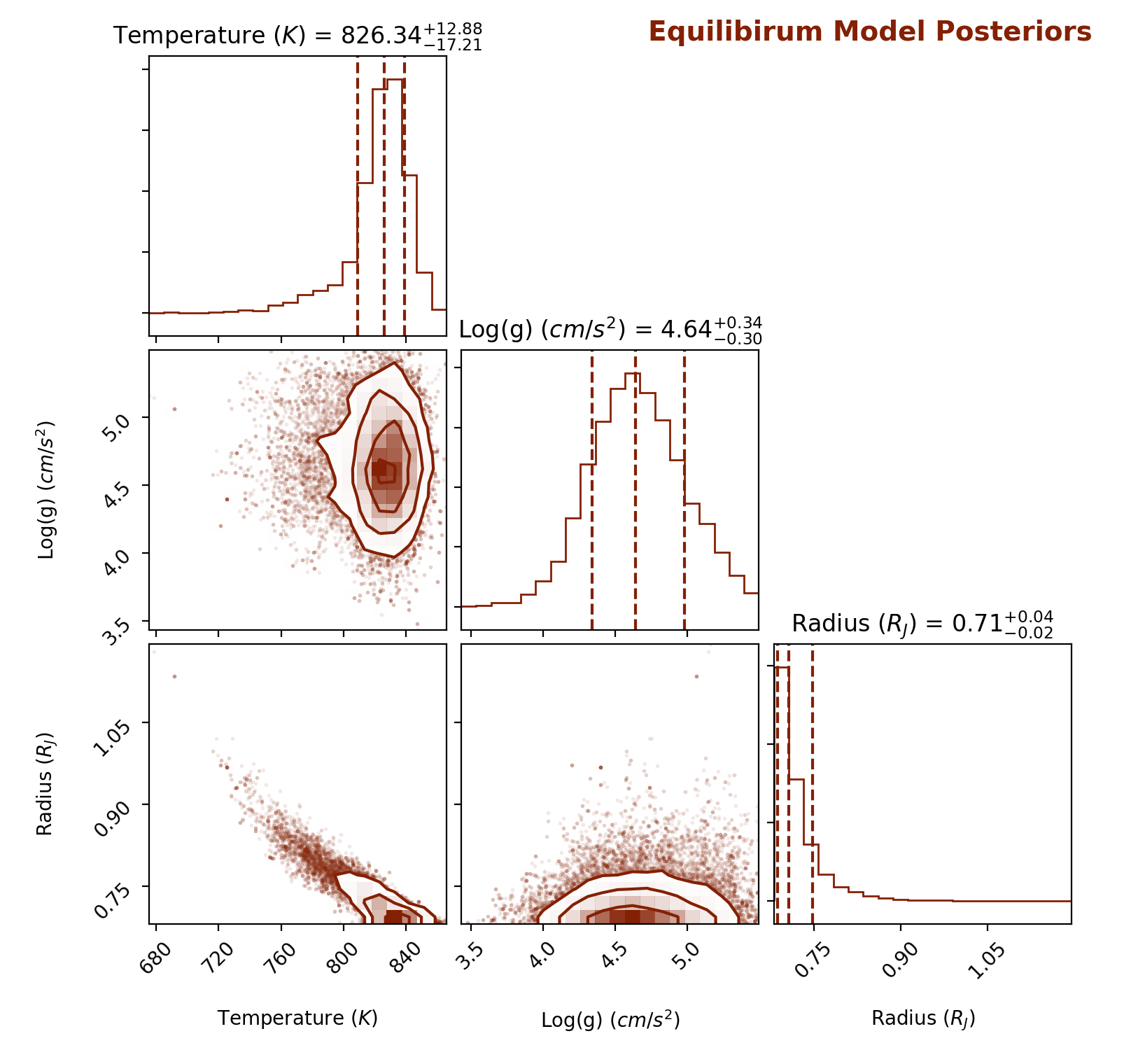}
        \caption{(a)}
    \end{subfigure}%
    ~ 
    \begin{subfigure}[m]{.5\textwidth}
        \centering
        \includegraphics[height=3in,]{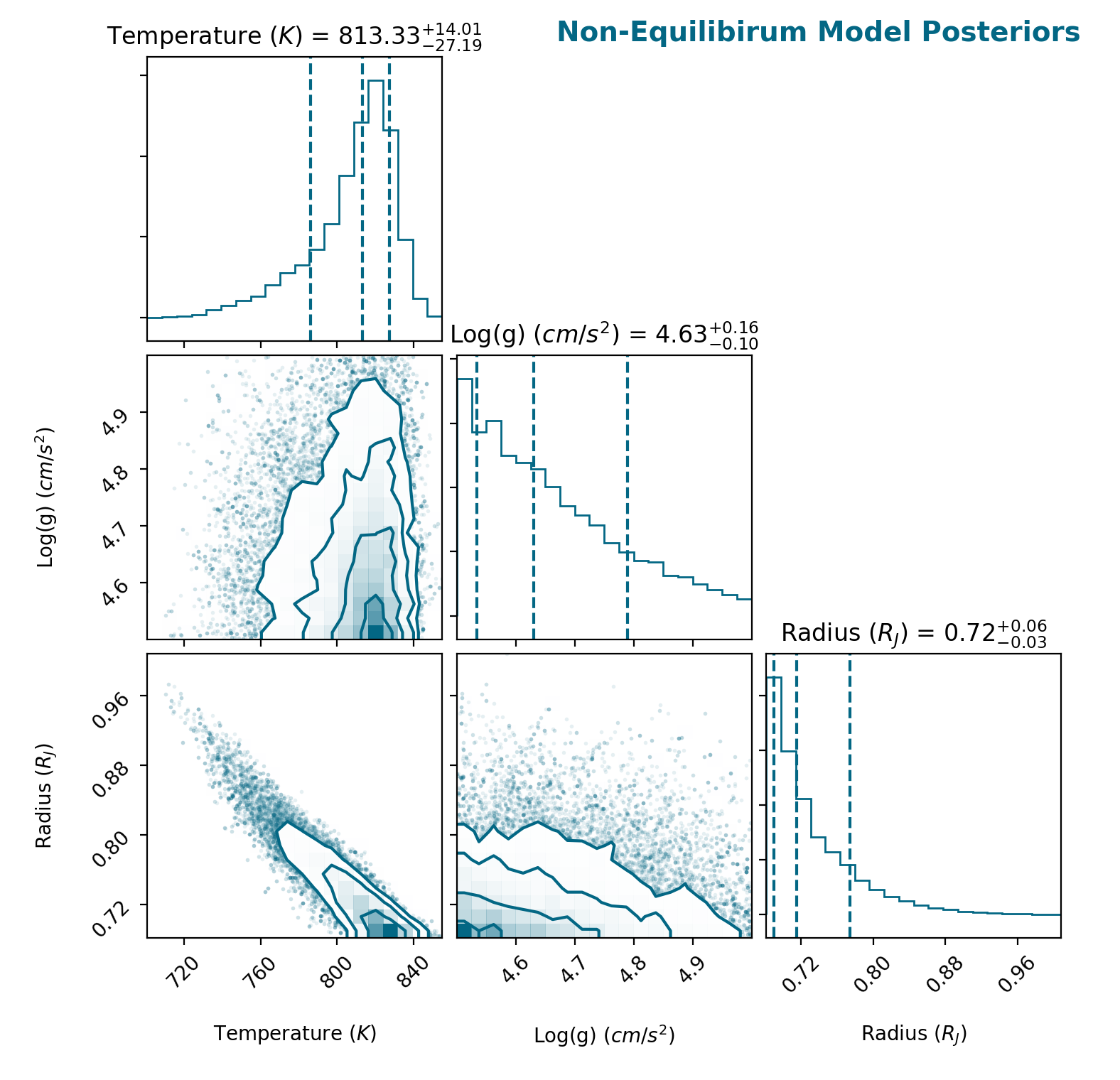}
        \caption{(b)}
    \end{subfigure}
    \caption{ATMO posteriors plots for GJ 570D. Left: Equilibrium chemistry model. Right: Non-equilibrium chemistry model.}
\label{fig:ATMO_corner_plots_GJ570D}
\end{figure*}

\begin{figure*}
\centering
\includegraphics[width=1\textwidth]{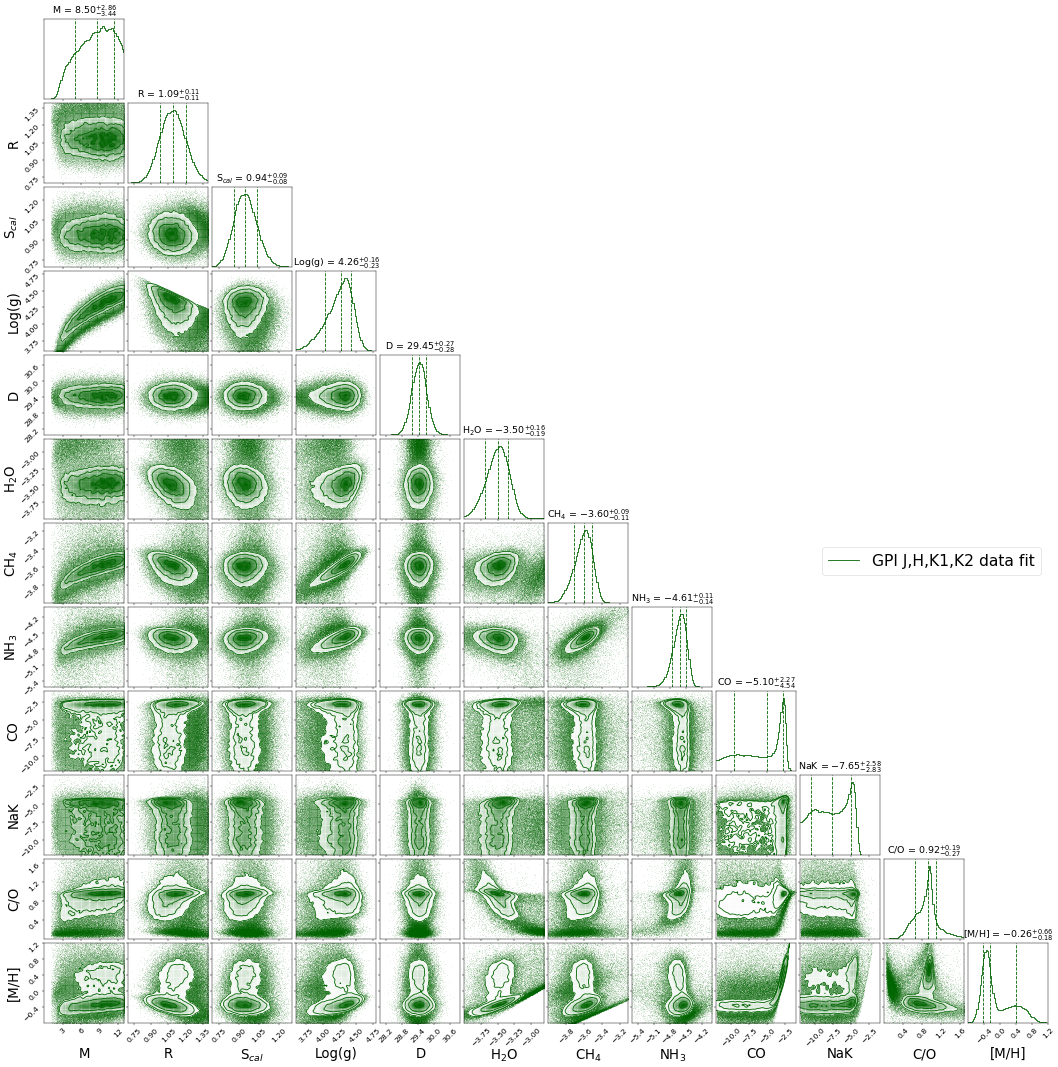}
\caption{51 Eri b posteriors for GPI $J$, $H$ and $K$ band data. Log(g), C/O and [M/H] posteriors are inferred parameters, while all the other parameters are sample as part of the retrieval.}
\label{fig:51Erib_GPI_posteriors}
\end{figure*}

\begin{figure*}
\centering
\includegraphics[width=1\textwidth]{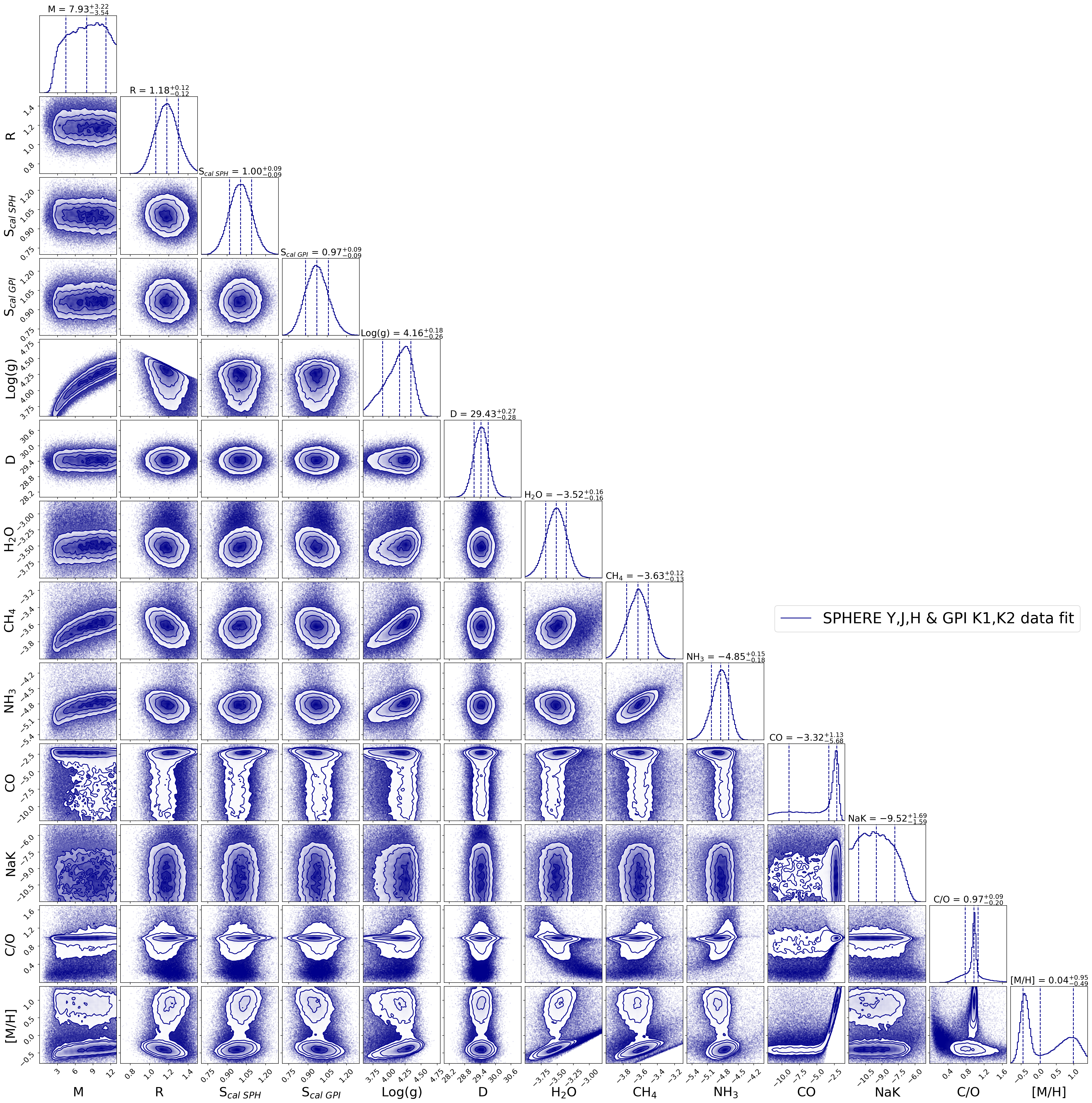}
\caption{51 Eri b posteriors for SPEHRE $Y$, $J$, $H$ and GPI $K$ band data. Log(g), C/O and [M/H] posteriors are inferred parameters, while all the other parameters are sample as part of the retrieval.}
\label{fig:51Erib_SPHERE_GPI_posteriors}
\end{figure*}

\begin{figure*}
\centering
\includegraphics[width=1\textwidth]{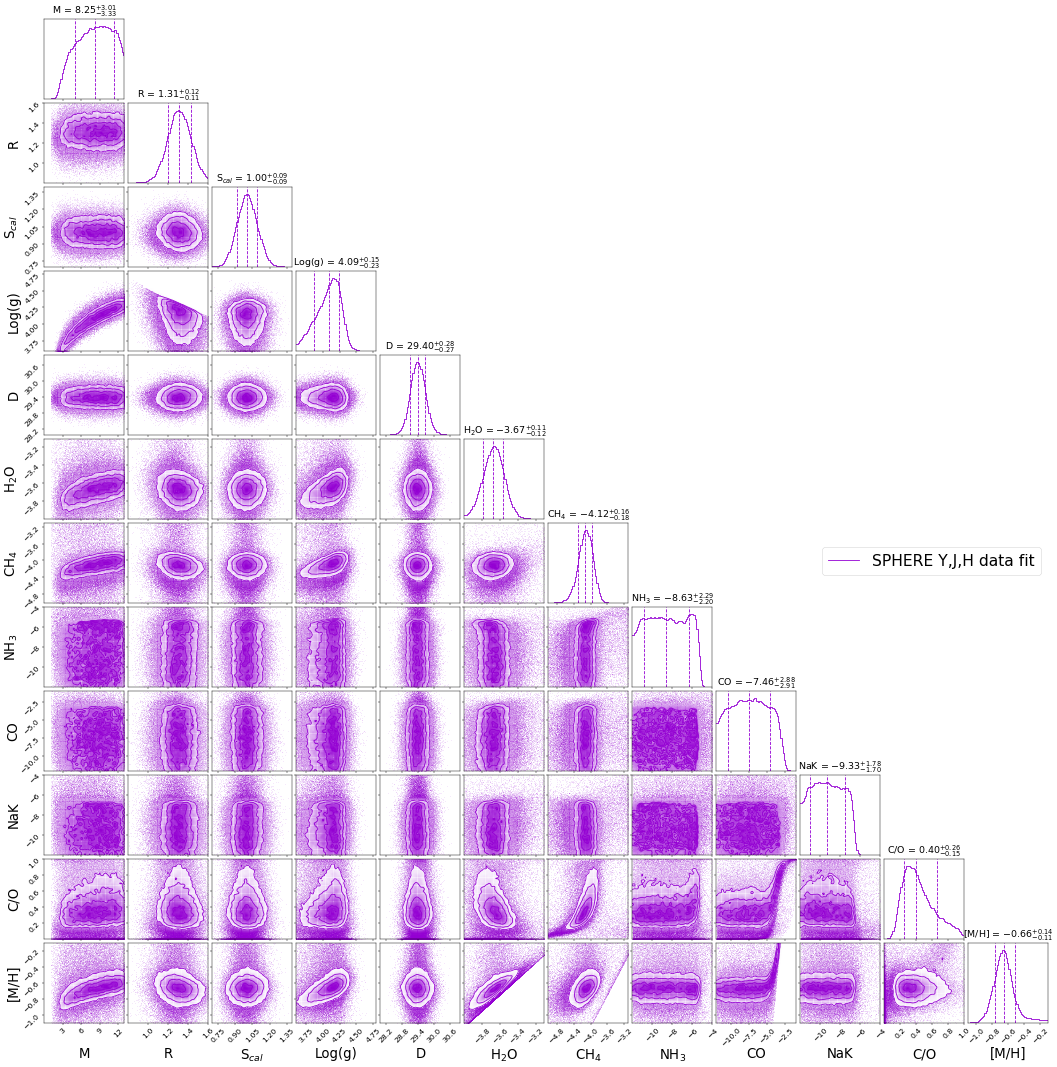}
\caption{51 Eri b posteriors for $Y$, $J$ and $H$ band data. Log(g), C/O and [M/H] posteriors are inferred parameters, while all the other parameters are sample as part of the retrieval.}
\label{fig:51Erib_SPHERE_only_posteriors}
\end{figure*}

\bsp	
\label{lastpage}
\end{document}